\documentclass[prb,twocolumn,nobalancelastpage]{revtex4-1}
\pdfoutput=1

\usepackage{amsmath,amsfonts,amssymb,color,graphicx,blkarray,mathtools}%

\usepackage[sort&compress]{natbib}
\setcitestyle{numbers,square}
\usepackage{hyperref}
\hypersetup{
        colorlinks=true,
}

\newcommand{\ket}[1]{\vert#1\rangle}
\newcommand{\bra}[1]{\langle#1\vert}

\newcommand{\lham}{\blacktriangleright}
\newcommand{\rham}{\blacktriangleleft}

\newcommand{\tr}{\intercal}  
\newcommand{\sep}{|}
\newcommand{\contr}{\triangleright}  

\newcommand{\mtx}[1]{\left(\begin{matrix}#1\end{matrix}\right)}

\DeclareMathOperator{\Tr}{Tr}
\DeclareMathOperator{\Pfaffian}{Pf}

\newcommand{\md}[1]{#1}

\newcommand{\Ymatrix}[2]{\begin{pmatrix}&#1\\#2\end{pmatrix}}
\newcommand{\Ysmatrix}[2]{\big(\begin{smallmatrix}&#1\\#2\end{smallmatrix}\big)}

\begin{document}

\title{Matrix product state algorithms for Gaussian fermionic states}
\author{Norbert Schuch}
\affiliation{Max-Planck-Institute of Quantum Optics, Hans-Kopfermann-Str.~1, 85748 Garching, Germany}
\affiliation{Munich Center for Quantum Science and Technology, Schellingstra{\ss}e 4, 80799 M{\"u}nchen, Germany}
\author{Bela Bauer}
\affiliation{Station Q, Microsoft Corporation, Santa Barbara, California 93106 USA}

\begin{abstract}
While general quantum many-body systems require exponential resources to be simulated
on a classical computer, systems of non-interacting fermions can be simulated exactly using polynomially
scaling resources. Such systems may be of interest in their own right, but also occur as
effective models in numerical methods for interacting systems, such as Hartree-Fock, density functional
theory, and many others. Often it is desirable to solve systems of many thousand constituent particles,
rendering these simulations computationally costly despite their polynomial scaling.
We demonstrate how this scaling can be improved by adapting methods based on matrix product states,
which have been enormously successful for low-dimensional interacting quantum systems,
to the case of free fermions. Compared to the case of interacting systems, our methods achieve an exponential
speedup in the entanglement entropy of the state.
We demonstrate their use
to solve systems of up to one million sites with an effective MPS bond dimension of $10^{15}$.
\end{abstract}

\maketitle

\tableofcontents

\section{Introduction}

Non-interacting or weakly interacting fermions play a crucial role in many aspects of condensed matter
theory and quantum chemistry. For example, using density functional theory, many materials 
can be accurately approximated by weakly interacting fermions; Hartree-Fock theory is an excellent
starting point for quantum chemical calculations; and transport in electronic nanodevices is often captured
by non-interacting models. While the computational effort in all these applications scales polynomially with the size of the system
that must be considered, the desire to obtain a realistic description -- reached,
for example, by including many orbitals in a complex material or a microscopic multi-band
model for a nanostructure -- means that computing properties of a non-interacting fermion model can
nevertheless become a computational bottleneck.

For interacting quantum systems, tensor networks have in many cases become the tool of choice
for numerical simulations. They exploit the locality of entanglement in the low-energy states of
local quantum Hamiltonians for a more compact classical description of these states. The
prototypical example are matrix-product states~\cite{fannes1992,ostlund1995}, which form
the basis of the density matrix renormalization group (DMRG)~\cite{white1992,white1992-1}.
Since its inception, DMRG has become the standard method to solve one-dimensional quantum
systems~\cite{schollwoeck2005,schollwoeck2011}. Many extensions of matrix-product states
to higher-dimensional systems have been
suggested~\cite{sierra1998,nishino1998,nishino2000,nishio2004,verstraete2004,jordan2008,vidal2007-1,vidal2008};
for recent reviews, see Refs.~\onlinecite{orus2014,orus2014-1,bridgeman2017}.

In this paper, we will show that it is possible to apply the
benefits of tensor networks to non-interacting fermions to obtain
significantly more efficient numerical methods for the simulation of very large free
fermion systems, allowing us to simulate systems far beyond what is
possible with existing methods.  While tensor network
states have previously been described for systems of free
fermions~\cite{kraus2010,schuch2012,dubail2015,haegeman2013,fishman2015,evenbly2016,haegeman2018},
thus far they have not been used primarily as numerical tools. In this
work, we will describe practical tensor network algorithms to compute the
many-body ground state as well as the dynamics of many-body states for
free-fermion systems.  Our methods capture the full equal-time Green's
function of the system, from which any physical observable can be
computed. While exact methods for this property scale cubically with the
system size, our approach scales linearly with the length of a
quasi-one-dimensional system and thus outperforms exact methods
significantly.

We work within the framework of Gaussian fermionic matrix product states
(GFMPS)~\cite{schuch2012}. These states are constructed using Gaussian
fermionic states, which are the most general class of states for which a
Wick's theorem holds, i.e.\ whose equal-time correlation functions are
characterized fully through the equal-time single-particle Green's
function. We will review this formalism in Sec.~\ref{sec:gcm}, where we
also introduce the required building blocks for Gaussian tensor network
algorithms, such as contraction and decomposition of tensors. We introduce
the specific case of GFMPS in Sec.~\ref{sec:GFMPS}, where we also discuss
the key technical tool of our work: a \emph{canonical form} for
GFMPS.  This crucial technical step will allow us to generalize many
well-established numerical algorithms for matrix-product states, such as
single- and two-site DMRG as well as time evolution approaches such as
TDVP~\cite{haegeman2011,haegeman2016} to the Gaussian setting; this will
be discussed in Secs.~\ref{sec:optim} and~\ref{sec:GFMPS-opt}. Finally, in
Sec.~\ref{sec:numerics}, we numerically demonstrate the performance of our
algorithms.

\subsection{Quadratic fermion systems}

To set the stage, consider a lattice system of $N$ sites, each comprised of some
number $n_s$ of fermionic modes. For our purposes, it will be convenient to work
in a basis of self-adjoint -- so-called Majorana -- fermions $c_i$, with
$c_i^\dagger = c_i$ and $\{c_i,c_j\}=2\delta_{ij}$. A fermionic system
comprised of conventional complex fermions created by some operator
$a_i^\dagger$ can always be rewritten in such a Majorana basis by taking
$c_{2i-1} = a_i+a_i^\dagger$, $c_{2i} = -i(a_i-a_i^\dagger)$. 
 We will be interested in quadratic Hamiltonian, i.e.,
Hamiltonians of the form
\begin{equation} \label{eqn:Hmaj}
    \mathcal H = -i \sum_{ij} H_{ij} c_i c_j + E\ ,
\end{equation}
with $E$ a constant offset.
We denote the total number of Majorana modes, which must be even in a
physical system, by $M=2\sum n_s$. Thus, $H$ is a real matrix of size $M\times
M$, which we can choose w.l.o.g.\ anti-symmetric, $H^\tr=-H$,
and we will do so in the following.

To compute properties of such a system, one can fully diagonalize the matrix $H$ to obtain all single-particle eigenvalues and eigenvectors, from which all other observables can be constructed. In the absence of any special structure of $H$, this will scale as $\mathcal{O}(N^3)$. If, on the other hand, only a few low-lying single-particle eigenstates (or a few states near a particular energy, such as the Fermi energy) are desired and the matrix $A$ is sufficiently sparse (as is the case for short-range hopping models), one can use sparse Krylov-space diagonalization methods such as Lanczos or shift-and-invert methods. These are generally expected to scale linearly in $N$ and the number of non-zero elements per row of $A$, as well as exhibit some dependence on spectral properties such as the energy gap. However, it should be emphasized that having computed only a few low-lying states, physical observables such as the density or more generally equal-time Green's function cannot be computed.

There are many other methods that overcome this limitation. A very efficient method to compute the spectral density of some operator (as well as certain dynamical correlation functions) in a time that scales only linearly with the number of modes is the kernel polynomial method (KPM)~\cite{weisse2006}.
There is also a large variety of Green's-function-based methods. To compute the frequency-dependent Green's function, the recursive Green's function (RGF) approach~\cite{thouless1981,lewenkopf2013} can be used. For a one-dimensional system, this method scales linearly with the length; for two- or three-dimensional systems, the method can be applied by taking the system to be a one-dimensional collection of "slices". However, to obtain equal-time properties, a frequency integral must be taken.
Finally, the non-equilibrium Green's function approach~\cite{datta2000nanoscale} incorporates systems driven out of equilibrium, for example by an applied voltage in a nanodevice. For a recent review and a related wavefunction-based approach, see Ref.~\onlinecite{gaury2014}.
Finally, in the context of density functional theory, several approximate methods to solve the Kohn-Sham equation in linear time have been developed, see e.g. Ref.~\onlinecite{goedecker1999}.
In contrast to these methods, our GFMPS method is based on the \emph{many-body} wavefunction for a low-energy or time-dependent state of the system, and thus allows the straightforward calculation of arbitrary expectation values without having to integrate over frequencies. Furthermore, our approach gives access to quantities such as the entanglement spectrum and entanglement entropy of the state.

\subsection{Gaussian tensor networks}

The power of matrix-product states is most easily understood from the properties of the Schmidt
decomposition. For a quantum state $|\psi\rangle$ on a tensor product space
$\mathcal{H}_A \otimes \mathcal{H}_B$, it is given by
\begin{equation}
|\psi\rangle = \sum_{\alpha=1}^M s_\alpha |A_\alpha\rangle |B_\alpha\rangle,
\end{equation}
where $s_\alpha \geq 0$ are real, and the \emph{Schmidt vectors} $|A_\alpha\rangle$, $|B_\alpha\rangle$ are orthonormal bases for
$\mathcal{H}_A$ and $\mathcal{H}_B$, respectively.
The von Neumann entanglement entropy between $A$ and $B$ is given by
$S = -\Tr ( \rho_A \log \rho_A ) = -\sum s_\alpha^2 \log s_\alpha^2$, and is bounded by $S \leq \log(M)$.
Thus, coarsely speaking, the number of terms that are relevant grows exponentially with the
bipartite entanglement of the state.

Given that low-energy states of local Hamiltonians are weakly entangled in the sense that they
exhibit area law scaling of the entanglement
entropy~\cite{eisert2010,hastings:arealaw,arad:rg-algorithms-and-area-laws,huang:area-law},
one can exploit the properties of the Schmidt decomposition to construct more compact representations 
of such states. In particular, it is natural to assume that the Schmidt values $s_\alpha$ decay very
quickly with $\alpha$ in such weakly entangled states~\cite{white1998,verstraete2006,hastings2007,schuch2008,hastings:arealaw,arad:rg-algorithms-and-area-laws,huang:area-law}. This directly leads to the construction of 
matrix-product states, 
which can be obtained by recursively performing a Schmidt decomposition
between all sites of a one-dimensional system and truncating
the sum on each bond to the $D$ largest Schmidt values $s_\alpha$.
The computational cost of such a state will scale exponentially with the bipartite entanglement entropy.

In systems of non-interacting fermions, the Schmidt decomposition and thus also the bipartite entanglement take
a simpler form. This is due to the fact that reduced density matrix for a
subsystem is itself a Gibbs state of a quadratic
Hamiltonian~\cite{botero2003,vidal2003,peschel2003,botero2004}. Thus, even its many-body spectrum can be
obtained through single-particle modes and their eigenvalues, and the entanglement can be fully
characterized through \emph{mode-wise} entanglement~\cite{botero2003,botero2004} between single-particle
modes in each part of the bipartite system. In other words, the Schmidt decomposition of a Gaussian
state can itself be expressed entirely in terms of Gaussian states. Crucially, each mode can contribute
a fixed amount to the bipartite entanglement, and the maximum entanglement is thus linear in the number
of modes rather than logarithmic in the number of terms in the Schmidt decomposition. This constitutes
an exponential compression of the Schmidt decomposition in the Gaussian case.

To exploit this computationally, one can construct a Gaussian MPS (GFMPS) in an analogous fashion to the
interacting case by recursively applying the Schmidt decomposition on each bond of a given Gaussian state.
Owing to the structure of Gaussian states, such a GFMPS will scale only polynomially rather than exponentially
with the amount of bipartite entanglement.

This exponential speedup also benefits more general tensor networks. In the context of a conventional
tensor network, a tensor is nothing but a high-dimensional array of numbers, which are enumerated by an array of indices.
The number of indices is commonly referred to as ``tensor rank'' $R$ (not to be confused with the
matrix rank), and each index runs over integers $1,\ldots,D$, where $D$ is referred to as the bond dimension.
A quantum state is expressed through such a tensor network by identifying some indices of the tensors
with physical degrees of freedom, and contracting (summing) over the remaining ones in some predetermined
pattern.

To understand the case of a Gaussian tensor network, it is convenient to think of a rank-$R$ tensor
of bond dimension $D$ as a quantum state on the tensor product of $R$ $D$-dimensional Hilbert spaces.
One can now consider the case where this state is Gaussian, i.e. it satisfies a Wick theorem and can
thus be described completely through the expectation values of quadratic operators.
Such a representation is exponentially
more compact, i.e. requires only $\mathcal{O}(R \log D)$ rather than $\mathcal{O}(D^R)$ numbers.
Using techniques that we will discuss in more detail below, one can generalize the contraction of
tensors and other important operations on tensor networks to this framework.

\section{Covariance matrix formalism for Gaussian tensor networks}
\label{sec:gcm}

We will now introduce Gaussian tensor networks
on a more technical level. This will establish the essential tools that we will use for the
construction of Gaussian MPS in Sec.~\ref{sec:GFMPS} as well as specific optimization algorithms
in Sec.~\ref{sec:optim}.
We first review the fermionic covariance matrix
formalism. We follow the conventions of Ref.~\onlinecite{Bravyi2017}; for further details,
see also Ref.~\onlinecite{bravyi2004}. Throughout this section, we will assume a system of $2n$
Majorana fermions described by a Hamiltonian of the form~\eqref{eqn:Hmaj}.

\subsection{Covariance matrix formalism}

Gaussian states are states that are fully characterized
through their equal-time two-point correlation functions. A convenient formalism
for describing these is the covariance matrix, which is given by
\begin{equation}
\label{eq:cm-def}
\gamma_{ij} = \tfrac{\mathrm{i}}{2}\mathrm{tr}\big(\rho [c_i,c_j]\big)\ .
\end{equation}
The covariance matrix $\gamma$ is real and antisymmetric, and satisfies
$\gamma^2\ge-\openone$; for pure states,
\begin{equation} \label{eq:pure-state-cond}
\gamma^2=-\openone\ .
\end{equation}
In the following, we discuss some
further properties of quadratic Hamiltonians and Gaussian states.

\paragraph{Energy of a state and normal form.}
The energy of a state $\rho$ under the Hamiltonian $\mathcal H$ [Eq.~\eqref{eqn:Hmaj}] is 
\begin{equation}
\label{eq:ham-energy}
\begin{aligned}
\mathrm{tr}[\mathcal H\rho]&=
    -\sum H_{ij}\:\mathrm{i}\:\mathrm{tr}(\rho c_i c_j)+E
\\
    &=-\sum H_{ij}\gamma_{ij}+E 
\\
&= \mathrm{tr}(H\gamma)+E\ .
\end{aligned}
\end{equation}
Any real antisymmetric matrix $A$ (such as $H$ or $\gamma$) can be brought
into a canonical form by an orthogonal transformation,
\begin{equation}
\label{eq:can-form-antisym}
    O A O^\tr = \bigoplus_{k=1}^n \Ymatrix{\lambda_k}{-\lambda_k}\ ,
\end{equation}
with $\lambda_k\ge0$. If $A$ is of odd size, additionally a single
$0$ block appears. Note that $\pm \lambda_k$ are
exactly the
eigenvalues of the Hermitian matrix $iA$.

We can use this to bring a Hamiltonian of the form Eq.~\eqref{eqn:Hmaj}
into a normal form
\begin{equation}
    \mathcal H = -\frac{\mathrm{i}}{2}\: \sum_{k=1}^n \epsilon_k [\eta_{2k-1}, \eta_{2k}]\ ,
\end{equation}
where $\eta_k$ are new canonical Majorana operators given by $\eta_k =
\sum_i O_{ki} c_i$,  with $O H O^\tr = \bigoplus_{k=1}^n
\Ysmatrix{\epsilon_k/2}{-\epsilon_k/2}$.  The $\epsilon_k$ correspond exactly
to the non-negative half of the single-particle eigenvalues of the
equivalent Bogoliubov-de Gennes
Hamiltonian.  The ground state of $\mathcal H$, which we will denote as
$|0\rangle$, is characterized by having $\mathrm{i} \eta_{2k-1} \eta_{2k} |0\rangle
= |0\rangle$ for all $k = 1, \ldots, n$, and thus has total energy $E_0 =
-\sum \epsilon_k$. Its covariance matrix in the basis of canonical
operators $\eta_k$ is thus given by
\begin{equation}
    \tilde{\gamma}_0 = \tfrac{i}{2}\bra 0 [\eta_i, \eta_j] \ket0 
= \bigoplus \Ymatrix{1}{-1}\ .
\end{equation}
Using the orthogonal matrix $O$ that brings $A$ into the normal form, we
can obtain the covariance matrix in the basis of local, physical Majorana
operators $c_i$ as $\gamma_0 = O^\tr \tilde{\gamma}_0 O$.
Using Eq.~\eqref{eq:ham-energy}, we can verify that the ground state
energy is indeed $E_0=-\sum \epsilon_k$. 
Numerically, $\gamma_0$ can be determined by diagonalizing the
antisymmetric matrix $H$ (giving imaginary eigenvalues) and then replacing
them by the sign of the imaginary part. For other approaches,
see Ref.~\onlinecite{Wimmer2012}.

\paragraph{Expectation values and state overlaps.}
The expectation value of arbitrary operators in a Gaussian state can be
computed using Wick's theorem, which takes a particularly simple form in
the representation of Gaussian covariance matrices. Consider a product of
Majorana operators $C = \prod_{i \in \mathcal{I}} c_i$. The expectation
value of $C$ in a state $|\phi \rangle$ with CM
$\gamma$ is given by~\cite{Bravyi2017}
\begin{equation}
    \langle \phi |C| \phi \rangle = \Pfaffian (i \gamma_{\mathcal{I}} ),
\end{equation}
where $\gamma_{\mathcal{I}}$ is the submatrix of $M$ with row and column
indices in $\mathcal{I}$.

The square modulus of the overlap between two states is easily determined from the
covariance formalism~\cite{lowdin1955,Bravyi2017}.
Consider two many-body states $|A\rangle$ and $|B\rangle$ with corresponding covariance matrices
$\gamma_A$ and $\gamma_B$ and the same fermion parity, i.e.
$\Pfaffian(i \gamma_A) = \Pfaffian(i \gamma_B) = p$. Then,
\begin{equation}
\label{eq:overlap-main}
|\langle A|B \rangle|^2 = p 2^{-n} \Pfaffian(\gamma_A + \gamma_B).
\end{equation}

When the phase of the overlap is also desired, a complication arises from
the fact that the covariance matrix formalism does not capture the global
phase of the state -- clearly for some fermionic state $|\psi\rangle$, all
states $e^{i \phi} |\psi\rangle$ will have the same expectation values
of physical operators, and thus the same covariance matrix. To overcome this
problem, one can introduce a reference state $|C\rangle$ and observe that
$\langle C|A \rangle\langle A|B \rangle\langle B|C \rangle$ is invariant
under multiplying any of the three states by a phase. Therefore, it can
be computed from the covariance formalism; for explicit expressions, see
Ref.~\onlinecite{Bravyi2017}.

\paragraph{Composite systems.}

Consider a Gaussian state described by a covariance matrix $\gamma'$ on
$N$ modes, and $\gamma''$ on $M$ modes. The joint state $\gamma$ is
characterized by the direct sum of the covariance matrices,
\begin{equation}
    \gamma = \gamma' \oplus \gamma'' = \mtx{\gamma' &0 \\ 0&\gamma''}.
\end{equation}
To describe a general state of such a joint system, it is convenient to introduce
collective labels for the Majorana modes in each subsystem. For example, we can
denote the $N$ modes of the state $\gamma'$ as $a$, and the $M$ modes of the
other system as $b$. A general Gaussian state of the joint system thus takes the
block form
\begin{equation}
\label{eq:gen-bipartite-system}
    \gamma = \mtx{\gamma_{aa}&\gamma_{ab}\\-\gamma_{ab}^\tr&\gamma_{bb}}.
\end{equation}
where $\gamma_{aa}$ contains the correlations of the $a$ modes,
$\gamma_{ab}$ the correlations of $a$ modes with $b$ modes, etc.
We will occasionally write $|a|$ for the number of modes in $a$, i.e. $|a|=N$.
More generally, for multipartite systems, such as a tripartite system with
modes $a$, $b$, and $c$, we can express $\gamma$ in different partitions,
e.g.\
\begin{equation}
\gamma=\begin{pmatrix}
\gamma_{aa}&\gamma_{ab}&\gamma_{ac}\\
\gamma_{ba}&\gamma_{bb}&\gamma_{bc}\\
\gamma_{ca}&\gamma_{cb}&\gamma_{cc}
\end{pmatrix}
=\begin{pmatrix}
\gamma_{ab|ab} & \gamma_{ab|c}\\
\gamma_{c|ab} & \gamma_{c|c}
\end{pmatrix}\ ,
\end{equation}
where we use a vertical bar to separate row and column indices when
ambiguous; e.g., $\gamma_{ab \sep c}$ holds the correlations of the modes
$a$ and $b$ vs.\ $c$, i.e., 
$\gamma_{ab\sep c}=\left(\begin{smallmatrix}\gamma_{ac}\\
\gamma_{bc}\end{smallmatrix}\right)$.
We will generally follow this notation throughout this paper. Note
that due to the explicit labeling of the modes in this notation, the
actual ordering of the blocks in the matrix does not matter.

\paragraph{Partial traces.}

The dual operation to composing systems is the partial trace, i.e.
considering only the state of a subsystem; we will make frequent use of it
in this paper.  Given a system composed of $N+M$ modes $a$ and $b$ with
covariance matrix 
\begin{equation}
    \gamma = \mtx{\gamma_{aa}&\gamma_{ab}\\-\gamma_{ab}^\tr&\gamma_{bb}}\ ,
\end{equation}
the covariance matrix which describes the state of the $a$ modes (i.e., the system after
tracing out the $b$ modes) is given by $\gamma_{aa}$; this follows
immediately from the definition of the covariance matrix.

\subsection{Tensor networks and contraction}
\label{sec:contraction}

\paragraph{Formalism.}
Tensor networks are a way to describe multipartite quantum states
\begin{equation}
\ket{\Psi}=\sum_{i_1,\dots,i_N}c_{i_1,\dots,i_N}\ket{i_1,\dots,i_N}
\end{equation}
by rewriting $c_{i_1,\dots,i_N}=\sum_{\alpha_k} \prod
C_{\{i_s,\alpha_p\}}$, using tensors $C_{\{i_s\},\{\alpha_p\}}$ which depend
on only on a few (if any)  physical indices $\{i_s\}$ and auxiliary indices (or
\emph{entanglement degrees of freedom}) $\{\alpha_p\}$ each, 
where each auxiliary index is contained in exactly two tensors, 
and where the
auxiliary indices are contracted (i.e.\ summed over).  In the context of
fermionic tensor networks, it is convenient to instead consider each
$C_{\{i_s\},\{\alpha_p\}}$ as describing a
pure state (``fiducial state'') $\ket{C_{\{i_s\},\{\alpha_p\}}}=
\sum C_{\{i_s\},\{\alpha_p\}} \ket{\{i_s\},\{\alpha_p\}}$ 
of the physical \emph{and} virtual indices jointly, where the contraction is
achieved by projecting pairs of virtual indices onto a maximally entangled
state $\ket{\omega_{\alpha,\alpha'}}$: 
\begin{equation}
\label{eq:contraction-GH-def}
\ket\Psi = \left(\bigotimes \bra{\omega_{\alpha,\alpha'}}\right) 
\left(\bigotimes \ket{C_{\{i_s\},\{\alpha_p\}}}\right)\ .
\end{equation}
This framework is particularly suited for the case of fermionic tensor network states, as
it removes the requirement to specify an ordering of the fermionic tensors
(and bonds), since both $\bra{\omega_{\alpha,\alpha'}}$ and
$\ket{C_{\{i_s\},\{\alpha_p\}}}$ have fixed fermion parity, and all
tensors $C$ mutually commute (as they don't share fermionic operators),
and similarly all bonds $\omega$.

\begin{figure}[t]
\centering
\includegraphics[width=7cm]{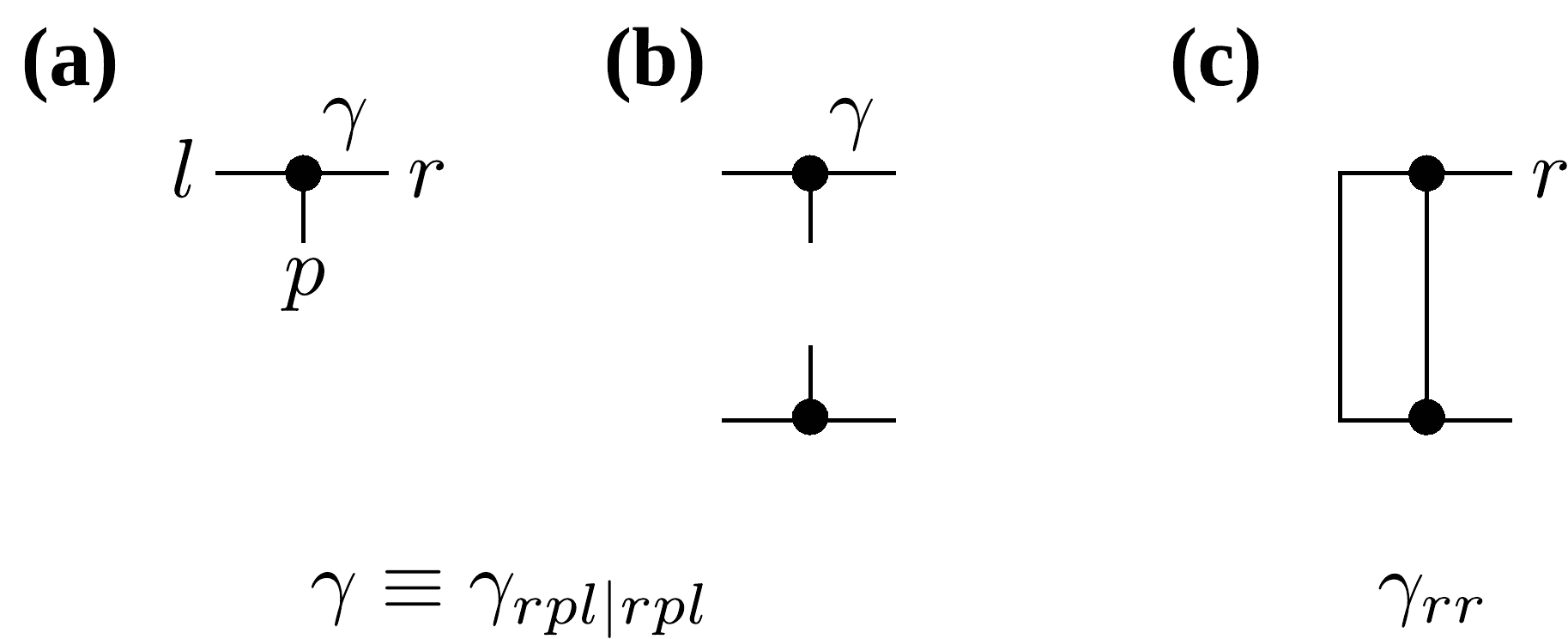}
\caption{
\textbf{(a)} Tensors are described by their covariance matrix $\gamma$, which
has blocks of rows/columns labeled by the different indices; the size of
each block equals the number of Majorana modes in the block.  \textbf{(b)} Since
$\gamma$ describes pure and mixed state on equal footing, we will
sometimes draw even pure state tensors with both ket and bra indices.
\textbf{(c)} This allows e.g.\ to naturally capture tracing of an index in
the graphical calculus.}
\label{fig:tndef}
\end{figure}

\paragraph{Contraction.}
In the case of Gaussian fermionic tensor networks, the fiducial states are
given by covariance matrices $\gamma$ which are indexed by the
corresponding physical and auxiliary modes.  The maximally entangled state
$\ket\omega$ (which can either be a product of maximally entangled states
between complex fermions, or a product of pairs of
Majorana fermions in their joint vacuum) is itself Gaussian, such that each
contraction gives rise to a new Gaussian state. The basic step is thus the
contraction of two Gaussian tensors (i.e.\ states). This corresponds to
integrating out the auxiliary degrees of freedom in a Gaussian integral,
and thus gives rise to a Schur
complement~\cite{Giedke2002,Eisert2002,fiurasek2002,bravyi2004}.

This leads to the following explicit expressions for contraction: Given two tensors
\begin{equation}
G=\left(\begin{matrix}G_{aa}&G_{ac}\\-G_{ac}^\tr&G_{cc}\end{matrix}\right)
\quad\mbox{and}\quad
H=\left(\begin{matrix}H_{bb}&H_{bc'}\\-H_{bc'}^\tr&H_{c'c'}
\end{matrix}\right),
\end{equation}
(possibly with blocked indices), the tensor resulting from contracting the
$c$ index of $G$ with the $c'$ index of $H$ is given by
\begin{align}
    \label{eq:contraction:schur}
K=\mtx{K_{aa}&K_{ab}\\-K_{ab}^\tr&K_{bb}}
=&\mtx{\!G_{aa}\\&\!\!H_{bb}}\! \\ \nonumber +\! \mtx{G_{ac}\\&\!\!H_{bc'}}\!
    &\mtx{\!G_{cc}&\openone\\-\openone&H_{c'c'}}^{\!\!-1\!}\!
\mtx{\!G_{ac}\\&\!\!H_{bc'}\!}^{\!\tr}\ .
\end{align}
Note that the contraction is oriented (swapping $c$ and $c'$ changes the
sign of the $\openone$'s); we will from now on use the notation
$c\triangleright c'$ to indicate the order.
The matrix inverse in Eq.~\eqref{eq:contraction:schur} can be carried
out block-wise using Schur complements.
Explicitly, 
\begin{multline}
\label{eq:sammlung:inv-w-id}
\begin{pmatrix}A&\openone\\-\openone&D\end{pmatrix}^{-1} 
= \\
\begin{pmatrix} (DA+\openone)^{-1} \\ & (AD+\openone)^{-1} \end{pmatrix}
\begin{pmatrix} D & -\openone \\ \openone & A \end{pmatrix}
\end{multline}
and as a special case which we will use later
\begin{equation}
\label{eq:sammlung:inv-w-0}
\begin{pmatrix}A&\openone\\-\openone&0\end{pmatrix}^{-1}
=
\begin{pmatrix} 0 & -\openone \\ \openone & A \end{pmatrix}\ .
\end{equation}

\begin{figure}[t]
\includegraphics[width=7cm]{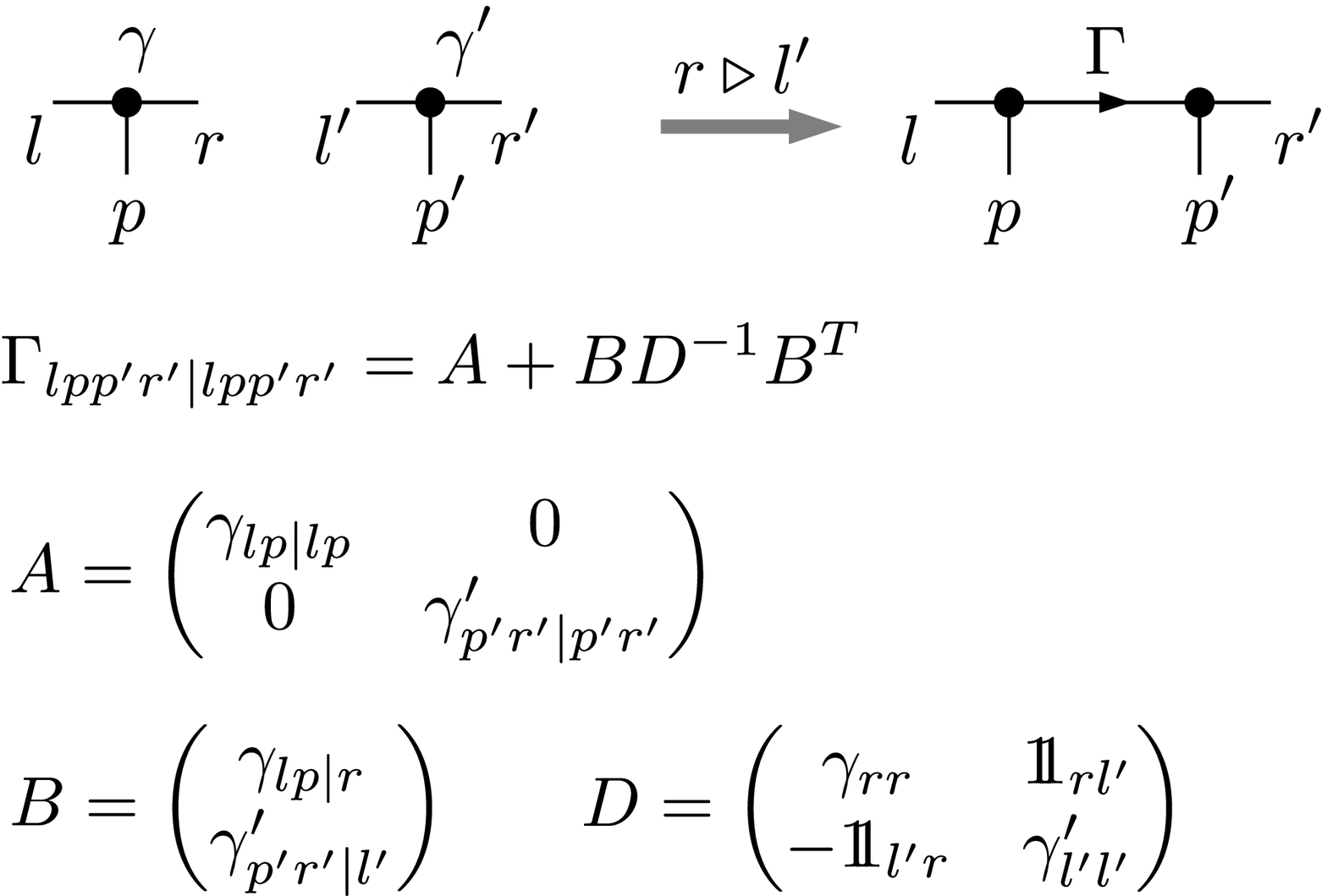}
\caption{
Contraction of two indices $r$ and $l'$ (contraction is oriented, denoted
as $r\contr l'$, and indicated by the arrow on the contracted index).  The
CM $\Gamma$ of the resulting state is obtained as a Schur complement, as
given by the formulas in the figure or Eq.~(\ref{eq:contraction:schur}).
(Here, $\openone_{xy}$ denotes the identity matrix betweeen modes $x$ and
$y$.)} \label{fig:contraction}
\end{figure}

For generalizations of this approach as well as the normalization of the
state after contraction, see App.~\ref{app:overlaps}.

\paragraph{Graphical calculus.} Tensor networks can be conveniently
expressed using a graphical calculus, where each tensor is described by a
box with one leg per index (where each index can contain any number of
Majorana modes), Fig.~\ref{fig:tndef}a. Contraction of indices is indicated by
connecting the corresponding legs; the orientation of the contraction will
be indicated by an arrow. This is shown in Fig.~\ref{fig:contraction}.

Note that the covariance matrix formalism (and thus Gaussian tensor
networks) does not distinguish between pure and mixed states.  Yet, we will
generally use the tensor notation for \emph{pure} states; when referring
to the density operator, we will instead draw the tensor twice, once for
the ``ket'' and once for the ``bra'' system (Fig.~\ref{fig:tndef}b).  In
particular, an essential operation will be tracing out part of a tensor,
which corresponds to contracting the ``ket'' and ``bra'' part of the
corresponding index, and which will be drawn as in Fig.~\ref{fig:tndef}c.

\subsection{Entanglement in Gaussian states and Schmidt decomposition}
\label{sec:schmidt}

A crucial step in the DMRG algorithm is the Schmidt decomposition, i.e.,
the decomposition of a bipartite entangled state $\ket\Psi$ in the form
$\ket\Psi = \sum_{k} \lambda_k \ket{\ell_k}\ket{r_k}$ with orthonormal
vectors $\{\ket{\ell_k}\}$ and $\{\ket{r_k}\}$ (see also Sec.~\ref{sec:svd}).

The procedure for fermionic Gaussian states is quite analogous. Consider a
pure bipartite Gaussian state
\begin{equation}
\label{eq:bip-state-svd}
    \gamma = \mtx{\gamma_{aa}&\gamma_{ab}\\-\gamma_{ab}^\tr&\gamma_{bb}}\ ,
\end{equation}
and assume for the moment that $\gamma_{ab}$ is square (i.e., the two systems have the same dimension) and has full rank; this corresponds to the case where all modes on either side are entangled.
Since $\gamma$ is pure, we have $\gamma^2=-\openone$, which implies
\begin{equation}
\label{eq:pure-blocks-relations}
\gamma_{aa}^2 = \gamma_{ab}^{\phantom\tr}\gamma_{ab}^\tr - \openone \ ,
\quad
\gamma_{bb}^2 = \gamma_{ab}^\tr\gamma_{ab}^{\phantom\tr} - \openone \ ,
\end{equation}
as well as $\gamma_{aa}\gamma_{ab}=-\gamma_{ab}\gamma_{bb}$, or equivalently
\begin{equation}
\label{eq:pure-blocks-offdiag-relations}
\gamma_{bb}=
-\gamma_{ab}^{-1}\gamma_{aa}\gamma_{ab}
\end{equation}
We now perform an SVD of $\gamma_{ab}$, $O \gamma_{ab} Q^\tr=
\Lambda_{ab}$, where $\Lambda_{ab}$ is diagonal with positive entries which are ordered descendingly, and
$O$ and $Q$ are real orthogonal, as they diagonalize the real symmetric
matrices
$\gamma_{ab}^{\phantom\tr}\gamma_{ab}^\tr$ and
$\gamma_{ab}^\tr\gamma_{ab}^{\phantom\tr}$, respectively.
 Eq.~\eqref{eq:pure-blocks-relations} then
immediately implies that also $O\gamma_{aa}^2O^\tr$
is diagonal with descending entries. Following
Eq.~\eqref{eq:can-form-antisym}, we find that 
\begin{equation}
\label{eq:O_gaa_O}
O\gamma_{aa}O^\tr =
\bigoplus_k\begin{pmatrix}&\lambda_k\\-\lambda_k\end{pmatrix}
\end{equation}
(possibly after rearranging rows of $O$ within degenerate blocks; by
performing the same permutation on $Q$, we can ensure that
$O\gamma_{ab}Q^\tr$ remains diagonal).
Therefore,
\begin{align}
Q\gamma_{bb}Q^\tr 
&= 
Q(-\gamma^{-1}_{ab}\gamma_{aa}\gamma_{ab})Q^\tr
\\
&= 
-(O\gamma_{ab}Q^\tr)^{-1}(O\gamma_{aa}O^\tr)(O\gamma_{ab}Q^\tr)
\end{align}
must be again of the form
\begin{equation}
Q\gamma_{bb}Q^\tr =
\bigoplus_k\begin{pmatrix}&-\lambda_k\\\lambda_k\end{pmatrix}\ ,
\end{equation}
where the fact that the entries $\lambda_k$ equal those in Eq.~(\ref{eq:O_gaa_O}) follow from  \eqref{eq:pure-blocks-relations}, as well as that
\begin{equation}
O\gamma_{ab}Q^\tr = 
\bigoplus_k\begin{pmatrix}\mu_k\\&\mu_k\end{pmatrix}
\end{equation}
with $\mu_k = \sqrt{1-\lambda_k^2}$.  Together, we find that the local
rotations $O\oplus Q$ transform $\gamma$ to
\begin{equation}
\label{eq:gaussian-schmidt}
(O\oplus Q)\gamma(O\oplus Q)^\tr = 
\bigoplus_k
W(\lambda_k)\ ,
\end{equation}
with 
\begin{equation}
\label{eq:gaussian-schmidt-Sblock}
W(\lambda_k)=\begin{pmatrix}
&\lambda_k& \mu_k\\
-\lambda_k & & & \mu_k\\
 -\mu_k & & & -\lambda_k\\
 & -\mu_k & \lambda_k 
\end{pmatrix},\ \:\mu_k=\sqrt{1-\lambda_k^2}\ ,
\end{equation}
where each block $W(\lambda_k)$ captures the correlations between two $a$
and two $b$ Majorana modes (in this order).  This is the fermionic
Gaussian equivalent of the Schmidt decomposition (and of the Williamson
normal form for bosonic Gaussian states), a result first derived by Botero
and Reznik~\cite{botero2004}.

From this form, the von Neumann entanglement entropy between the 
modes $a$ and $b$  is now easily found to be
\begin{equation}
    S_\mathrm{vN} = \sum_k H\left(\frac{1-\lambda_k}{2}\right) 
\end{equation}
with $H(p) = -p \log p - (1-p) \log (1-p)$.

In the more general case where $\gamma_{ab}$ does not have full rank (or
is not square), we can still perform an SVD of $\gamma_{ab}$, which in
that case will give rise to blocks where $\gamma_{ab}\gamma_{ab}^\tr$ (or
$\gamma_{ab}^\tr\gamma_{ab}$) is zero.
Eq.~\eqref{eq:pure-blocks-relations} implies that the corresponding blocks
of $\gamma_{aa}$ ($\gamma_{bb}$) decouple and are equal to
$\bigoplus\left(\begin{smallmatrix}&-1\\1\end{smallmatrix}\right)$, i.e.,
they describe a mode in a pure (vaccum) state, which is thus not entangled
to any other mode. In the general case, we thus obtain the form
\begin{multline}
\label{eq:gaussian-schmidt-general}
(O\oplus Q)\gamma(O\oplus Q)^\tr = \\
\bigg[ \bigoplus \left(\begin{smallmatrix}&-1\\1\end{smallmatrix}\right) \bigg]
\,\oplus\,
\bigg[ \bigoplus_k W(\lambda_k) \bigg]
\,\oplus\,
\bigg[ \bigoplus \left(\begin{smallmatrix}&-1\\1\end{smallmatrix}\right) \bigg]
,
\end{multline}
where the first (last) term in the sum lives only on the $a$ ($b$) modes, and the middle
term captures their correlations; as before, it is obtained from the SVD of
$\gamma_{ab}$, together with a suitable rearrangement of degenerate modes.

Note that while instead of performing the SVD of $\gamma_{ab}$, we could
have as well diagonalized $\gamma_{aa}$ and $\gamma_{bb}$. The approach
approach chosen here has two advantages: First, it doesn't require to match up
eigenvalues, and second, the singular values of $\gamma_{ab}$ relate to
the square roots of the eigenvalues of $\gamma_{aa}$, thus yielding a higher
accuracy (also in the basis transformations $O$ and $Q$) for weakly entangled modes.

\subsection{Tensor decompositions}

Tensor decompositions, such as the singular value decomposition (SVD) and the closely
related QR decomposition, are core ingredients in DMRG and other tensor network algorithms.  Their main
role is that they allow to decompose the state in terms of an
orthonormal basis; furthermore, they are an efficient way of obtaining
the Schmidt decomposition and thus information about the entanglement,
including the full entanglement spectrum. We will start by
introducing the Gaussian version of the SVD, and then discuss how to build
different related decompositions which are potentially easier or more
efficient to implement.

\begin{figure}
{\centering
\includegraphics[width=.95\columnwidth]{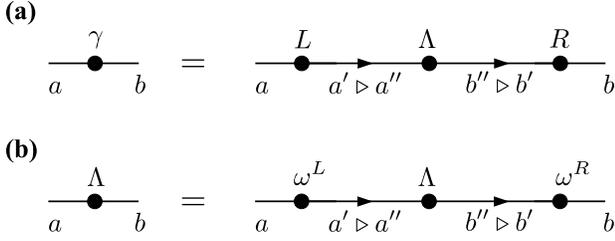}
}
\caption{
Gaussian fermionic SVD. \textbf{(a)} The SVD expresses a CM $\gamma$ as the
contraction of three CMs $L$, $\Lambda$, and $R$, where $L$ and $R$ are
isometries when contracting $a$ and $b$, respectively (i.e., $L_{a'a'}=0$,
$R_{b'b'}=0$), and $\Lambda=\bigoplus W(\lambda_k)$,
Eq.~\eqref{eq:gaussian-schmidt-Sblock}. \textbf{(b)}~The derivation
proceeds by using the Schmidt decomposition of $\gamma=(O\oplus
Q)^\tr\Lambda(O\oplus Q)$, rewriting $\Lambda$ as the contraction of itself
with two maximally entangled states $\omega^L$ and $\omega^R$, and
applying
$O$ and $Q$ to $a$ and $b$, respectively.  } \label{fig:gaussian-svd}
\end{figure}

\subsubsection{Singular value decomposition (SVD)} \label{sec:svd}

The SVD decomposes a matrix $M=UDV^\dagger$, where $U$ and $V$ are
isometries ($U^\dagger U=\openone$, $V^\dagger V=\openone$), and $D$ is
diagonal and full rank.  It is closely related to the Schmidt
decomposition, which for a state $\ket{\Psi}=\sum_{ij} M_{ij}\ket{i}\ket{j}$
is obtained from the SVD of $M$ as
\begin{gather}
    \mathclap{\ket\Psi=\sum d_k \ket{a_k}\ket{b_k}} \\
    d_k=D_{kk},\,\, \ket{a_k}=\sum U_{ik}\ket{i},\,\, \ket{b_k}=\sum V_{jk}^*\ket{j}\ .
\end{gather}
This can be viewed as an elementary tensor network: the state $\ket{\Psi}$ is given
as the contraction of tensors $U$, $D$ and $V^*$, where $U$ and $V^*$ have one auxiliary
and one physical index each, and $D$ has only auxiliary indices.

We have discussed the Gaussian version of the Schmidt decomposition in the previous
section, Sec.~\ref{sec:schmidt}.
Building on this, we will now describe how to view the Schmidt decomposition
as a tensor decomposition, i.e. how to view the state as the contraction of a
minimal tensor network of three tensors.
Specifically, the Gaussian SVD corresponds to the decomposition of the covariance matrix $\gamma$ of
a bipartite system $ab$ into the contraction of three covariance matrices $L$, $\Lambda$,
and $R$, as indicated in Fig.~\ref{fig:gaussian-svd}a (including the
labeling of the modes).  Here, the isometry
condition $U^\dagger U=\openone$ translates naturally to $L_{a'a'}=0$, which
is the CM of the maximally mixed state $\openone$, cf.\
Fig.~\ref{fig:tndef}c, and correspondingly $R_{b'b'}=0$;
furthermore, we require $\Lambda=\bigoplus W(\lambda_k)$ [cf.~Eq.~\eqref{eq:gaussian-schmidt}].

As  a starting point, consider the covariance matrix $\gamma$ written as
\begin{equation} 
\gamma=(O\oplus Q)^\tr \Lambda (O\oplus Q)
\end{equation}
with orthogonal $O$ and $Q$; note that at this point, we don't need to make
any assumption about $\Lambda$, i.e. the decomposition can be more general than Eqn.~\eqref{eq:gaussian-schmidt}.
We will generalize it to the case with
additional local degrees of freedom as in
Eq.~\eqref{eq:gaussian-schmidt-general} in a moment.
The first step is to rewrite $\Lambda$ as a contraction of three tensors
$\omega^L$, $\Lambda$, and $\omega^R$, as depicted in
Fig.~\ref{fig:gaussian-svd}b (note the relabelling of the indices of
$\Lambda\equiv \Lambda_{a''b''\sep a''b''}$) -- this is the equivalent
of contracting with the identity tensor, i.e.\ writing
$D=\openone\cdot D\cdot\openone$. The tensors $\omega^{L,R}$ are
given by
\begin{equation}
\omega^L \equiv \omega^L_{aa'\sep aa'}= 
\begin{pmatrix}0 & \openone_{aa'}\\
    -\openone_{a'a} & 0 \end{pmatrix}
\end{equation}
and
\begin{equation}
\omega^R \equiv \hat\omega_{b'b\sep b'b}= \begin{pmatrix}0 &
\openone_{b'b}\\ -\openone_{bb'} & 0 \end{pmatrix}\ .
\end{equation}
They describe maximally entangled states between $a$ and $a'$ and $b'$ and
$b$, respectively (corresponding to the identity tensor $\openone$). 
It is easy to check using Eqn.~\eqref{eq:contraction:schur}
that contracting $a'\triangleright a''$ returns $\Lambda$, and likewise for
$b''\triangleright b'$~\footnote{This can also be seen without calculation by noting that this amounts to a
postselected teleportation protocol (i.e.\ attaching a maximally entangled state
and projecting onto the maximally entangled state).}.

In a second step, we now take Fig.~\ref{fig:gaussian-svd}b and apply $O$
and $Q$ to the $a$ and $b$ system, respectively.  On the left-hand side,
this yields $(O\oplus Q)\Lambda(O\oplus Q)^\tr=\gamma$, while on the right-hand
side, $\omega^L$ and $\omega^R$ are transformed to
\begin{equation}
\label{eq:svd-omega-to-L}
L = 
(O^\tr\oplus\openone)\omega^L (O\oplus\openone) =
\begin{pmatrix}0 & O^\tr\\
    -O & 0 \end{pmatrix}\ ,
\end{equation}
and
\begin{equation}
\label{eq:svd-omega-to-R}
R = 
(\openone\oplus Q^\tr)\omega^R (\openone\oplus Q) =
\begin{pmatrix}0 & Q\\
    -Q^\tr & 0 \end{pmatrix}\ ,
\end{equation}
respectively.  Furthermore, the contraction and the application of $O$ and
$Q$ commute, as they act on different indices (and as can be also checked
explicitly).  We thus find that $\gamma$ can be written as the contraction
of $L$, $\Lambda$, and $R$, as in Fig.~\ref{fig:gaussian-svd}a, with $L$
and $R$ isometries. In the case where $\Lambda$ is the Schmidt
decomposition, this yields the Gaussian analogue of the SVD.

In the general case where the Schmidt decomposition contains additional
unentangled local degrees of freedom, as in
Eq.~\eqref{eq:gaussian-schmidt-general}, these modes need to be attached
to the $a$ and $b$ system of $\omega^L$ and $\omega^R$, respectively,
before applying the transformations in
Eqs.~(\ref{eq:svd-omega-to-L},\ref{eq:svd-omega-to-R}); the resulting $L$ and
$R$ are still isometries, and the decomposition in
Fig.~\ref{fig:gaussian-svd}a still holds with $\Lambda=\bigoplus
W(\lambda_k)$. Note that we can always regard a pair of decoupled modes
as $W(\lambda_k\equiv1)$ if we do not want to minimize the bond dimension
(i.e., the dimension of $\Lambda$).

\subsubsection{Truncation of bond dimension.}

An important scenario, in particular in $2$-site DMRG, is to truncate the
bond dimension.  This corresponds to carrying out an SVD $M=UDV^\dagger$
of a matrix $M$, and replacing it by a matrix $M'=(UW)(W^\dagger D
V^\dagger)$ with an isometry $W$ such that $WW^\dagger$ projects onto the
$\chi$ largest singular values of $D$.

In the Gaussian scenario, given a CM $\gamma$ this amounts to carrying out
the Schmidt decomposition Eq.~\eqref{eq:gaussian-schmidt-general}, and
keeping only the blocks $W(\lambda_k)$ with the $\kappa$ smallest
$\lambda_k$, while replacing the additional blocks with 
\begin{equation}
W(\lambda=1) \equiv \begin{pmatrix}
    &1\\-1\\&&&-1\\&&1
\end{pmatrix}\ ,
\end{equation}
i.e.\ decoupled modes.  Those modes are henceforth counted towards the
decoupled modes in Eq.~\eqref{eq:gaussian-schmidt-general}, and the
truncated SVD is obtained by applying the SVD construction,
Fig.~\ref{fig:gaussian-svd}a, with $\Lambda$ from the truncated Schmidt
decomposition.

\subsubsection{Variants: QR and related decompositions.}
\label{sec:other-decomp}

In most applications, a full SVD is not needed. For example,
to obtain the canonical form of an MPS (see below for details) 
one only needs to be able to rewrite a covariance matrix 
$\gamma\equiv\gamma_{ab\sep ab}$,  with $|a|\ge |b|$, 
as the contraction of an isometry and a tensor with equal-sized systems on both sides.
(In DMRG, $a$ corresponds to the left virtual + physical modes, while $b$ corresponds to the right virtual modes.)
To this end, we need to isolate modes in $a$ that are not entangled with any modes in $b$; we will denote
the modes that are unentangled as $a_0$ and the others as $a_1$, i.e. $|a|
= |a_0| + |a_1|$ and $|a_1| = |b|$. (We assume that all modes in $b$ are
entangled with modes in $a$, as happens in practice; the generalization is
straightforward.)
To achieve this, we find an $|a| \times |a|$ orthogonal transformation $O$ such that
\begin{equation}
(O\oplus\openone)\gamma(O\oplus\openone)^\tr =
\left( \begin{array}{c|cc}
\gamma_{a_0 a_0} &0 &0 \\ \hline 
0 &\gamma_{a_1 a_1} &X \\
0 &-X^\tr &\gamma_{bb}
\end{array} \right)
\end{equation}
where $\gamma_{a_0 a_0} = \bigoplus \left( \begin{smallmatrix} &-1 \\ 1 \end{smallmatrix} \right)$, and $X$ is $|b| \times |b|$.
Note that this is a weaker version of the decomposition of Eqn.~\eqref{eq:gaussian-schmidt}, where no transformation
is applied on the $b$ modes. To obtain this form, it is sufficient to choose $O$ such that 
\begin{equation}
O\gamma_{ab} = \begin{pmatrix}0\\X\end{pmatrix}
\end{equation}
with $X$ a square matrix; this can e.g.\ be achieved by determining the
kernel of $\gamma_{ab}^\tr$ (or the image of $\gamma_{ab}$). The fact that this
implies $\gamma_{a_0 a_1} = 0$ can be verified by invoking purity of the state.
Once such an $O$ is found, following the SVD construction of Fig.~\ref{fig:gaussian-svd}a
with $Q=\openone$ and
$\Lambda\equiv \left( \begin{smallmatrix} \gamma_{a_1 a_1} &X \\ -X^\tr &\gamma_{bb} \end{smallmatrix} \right)$
gives the desired result. For the case where $|a|\le |b|$, the equivalent transformation is obtained by acting with $O$ on the $b$ modes.  Note that for $|a|=|b|$, $O$ can be chosen trivial.

Similarly, in order to obtain a decomposition which allows to truncate the bond dimension (such as in 2-site DMRG), it
is sufficient to determine $O$ such that $\gamma_{aa}$ is diagonalized;
this allows to determine which modes decouple, and yields the $\lambda_k$
of the $W(\lambda_k)$ which can be used to identify and disentangle the least entangled modes, without the need to determine $Q$.

\section{Gaussian fermionic MPS (GFMPS)}
\label{sec:GFMPS}

\subsection{Construction}

\begin{figure}[t]
{\includegraphics[width=0.9\columnwidth]{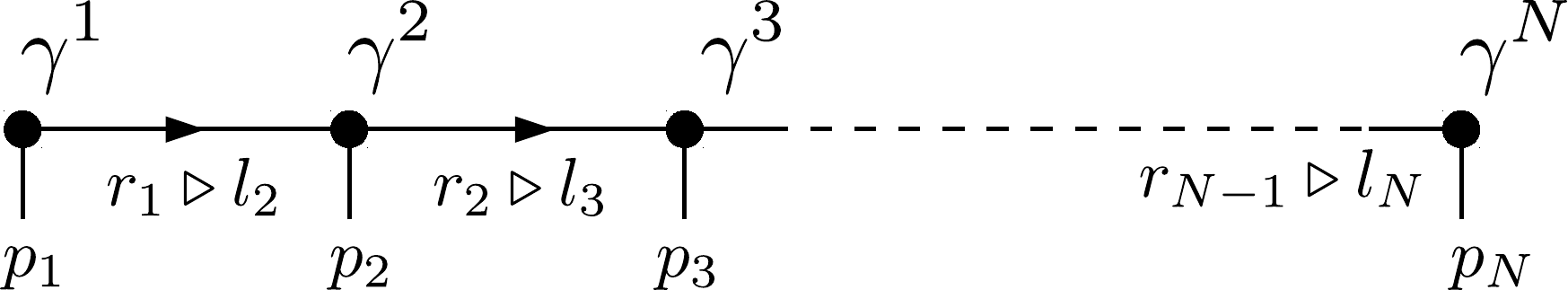}}
\caption{
Construction of a Gaussian fermionic MPS: To each site, we associate a
tensor $\gamma^s_{l_sp_sr_s\sep l_sp_sr_s}$ (with $l_1$ and $r_N$
trivial); the state of the ``physical modes'' $p_1,\dots,p_N$ is obtained
by contracting the ``virtual modes'' $r_s\contr l_{s+1}$.
}
\label{fig:GFMPS-construction}
\end{figure}

For the remainder of this manuscript, we will focus on the Gaussian fermionic
version of matrix-product states. A matrix product state (MPS) is a particular
type of tensor network state in one dimension, where the sites of the physical system
are arranged on a chain and one tensor is associated with each physical site.
The tensor is connected to its left and right neighbors.
In the following discussion, we will focus on the case of open boundary conditions.
For a chain of $N$ sites, an MPS consists of $N$ rank-3 tensors. In the
case of Gaussian fermionic MPS (GFMPS), it is thus fully specified by a set of covariance matrices
\begin{equation}
\gamma^s_{l_sp_sr_s\sep l_sp_sr_s},
\end{equation}
where $s=1,\dots,N$ denotes the sites of the chain (from left to right),
$p_s$ corresponds to the physical modes at each site, and the dimensions
$|r_s|=|l_{s+1}|$ of the virtual modes $r_s$, $l_{s+1}$
are the \emph{Majorana bond mode number} of the bond $(s,s+1)$. As in conventional
MPS, the bond mode number -- analogous to the bond dimension -- can be increased
to systematically increase the class of variational states, and an exact description
is recovered for $|l_s| \sim \mathcal{O}(N)$. As noted in the Introduction, the bond
dimension $M$ in a conventional MPS is related to the Majorana bond number $\chi$ used here
by $M = \sqrt{2}^\chi$. For open boundary conditions,
there are no left modes on the left-most tensor and no right modes on the right-most tensor,
i.e. $|l_1|=0$ and $|r_N|=0$. It is therefore often convenient to omit these and set
$\gamma^1 = \gamma^1_{p_1 r_1 \sep p_1 r_1}$ and $\gamma^N =
\gamma^N_{l_Np_N\sep l_Np_N}$. When clear from the context, we will often omit the
site subscripts to the modes and e.g.\ write $\gamma^s_{lpr|lpr}$.

A description of the state on $N$ sites is now
obtained by arranging the $\gamma^s$ on a line and contracting the
adjacent virtual indices $\md{r_s}\triangleright\md{l_{s+1}}$, as depicted
in Fig.~\ref{fig:GFMPS-construction}, which yields a CM describing the
physical modes $\md{p_1},\dots,\md{p_N}$.
Note that on physical grounds, it is clear that the state obtained through
the above construction is independent of the order of the contractions,
even though this is not evident from the description of the contraction in
the CM formalism through Schur complements.

Owing to the properties of the Gaussian formalism, a GFMPS always
describes a normalized state up to an overall global phase. This fact has
the potential to complicate practical MPS algorithms somewhat: in the
conventional formalism, physical observables such as the energy are
generally bilinear in the individual tensor elements. Thus, optimization
at least over individual tensors manifestly is a quadratic and thus
well-behaved problem.  Since GFMPS are a special case of MPS, this must in
principle also be true for GFMPS, but it is not apparent in the Gaussian
formulation. However, with some extra steps the usual structure can be
exposed and the conventional and well-tested algorithms, such as single-
and two-site DMRG optimization as well as related time-evolution
algorithms,
can be formulated. A central role is played by the canonical form of the MPS. While this
form is in practice very useful for the stability and performance of conventional MPS algorithms,
it is not strictly required; in the Gaussian case, on the other hand, we will find it to be essential.
In the following, we will first introduce the canonical form of GFMPS, and 
in the following sections will proceed to discuss
some important algorithms, such as efficient computation of the total energy, single- and two-site
optimization, and finally the time-dependent variational principle.

\subsection{Canonical form}
\label{sec:canform}

\paragraph{Definition.}

\begin{figure}[b]
{\includegraphics[width=0.8\columnwidth]{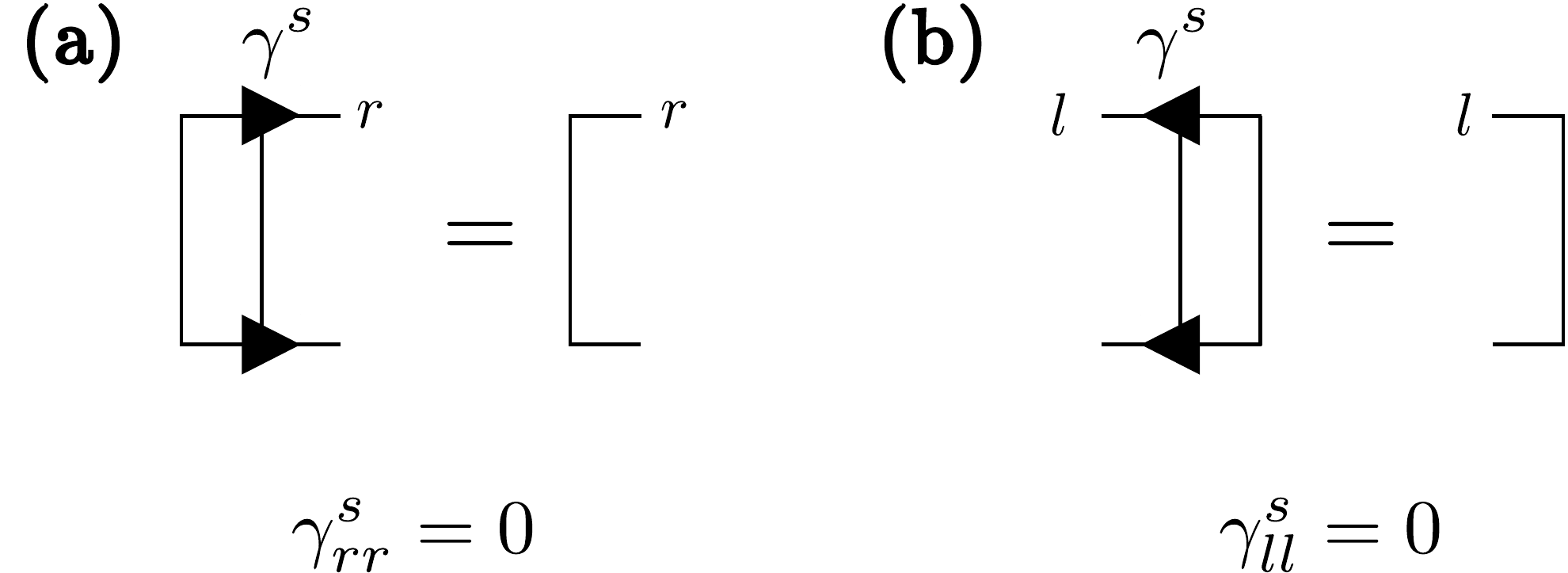}\hspace*{2em}}
\caption{
(a) Left- and (b) Right-canonical form of a tensor $\gamma^s$, and the
corresponding condition on $\gamma^s$.
}
\label{fig:can-form}
\end{figure}

The first step in defining the canonical form of an MPS is to choose a
center ``working'' site $s_0$ with respect to which the canonical form is
defined. Then, all tensors to the left of $s_0$ are brought into 
left-canonical form, and the ones to its right into right-canonical
form. The definitions of these are given by $\gamma_{rr}^s=0$ (left-canonical form, $s<s_0$) and $\gamma_{ll}^s=0$ (right-canonical form, $s>s_0$), 
as illustrated in Fig.~\ref{fig:can-form}.
This definition precisely correspond to that in the 
conventional MPS formalism: There, an MPS tensor is said to be
left-canonical if the contraction with its own adjoint over the left and
physical index yields the identity matrix (see tensor diagram of
Fig.~\ref{fig:can-form}a), or equivalently, the tensor -- seen as a map
from right to physical and left index -- is an isometry.  The contraction
corresponds to tracing out the left and physical system, and thus for GFMPS,
the left-canonical form amounts to require that the reduced density
matrix of the $r$ modes is $\gamma^s_{rr}=0$.  Correspondingly, for the
right-canonical one has $\gamma^s_{ll}=0$. Note that this is precisely the 
isometry condition we used in the derivation of the Gaussian SVD.

\paragraph{Elementary move.}

The elementary move in establishing and keeping the canonical form is the
following: Given a site $s$ such that all tensors left of it are in
left-canonical form, transform $\gamma^s$ into left-canonical form by
changing $\gamma^s$ and $\gamma^{s+1}$, without changing the state
described by the GFMPS. We focus on the left-canonical form, but the
corresponding procedure for the right-canonical form is completely analogous.

\begin{figure}
\includegraphics[width=0.3\columnwidth]{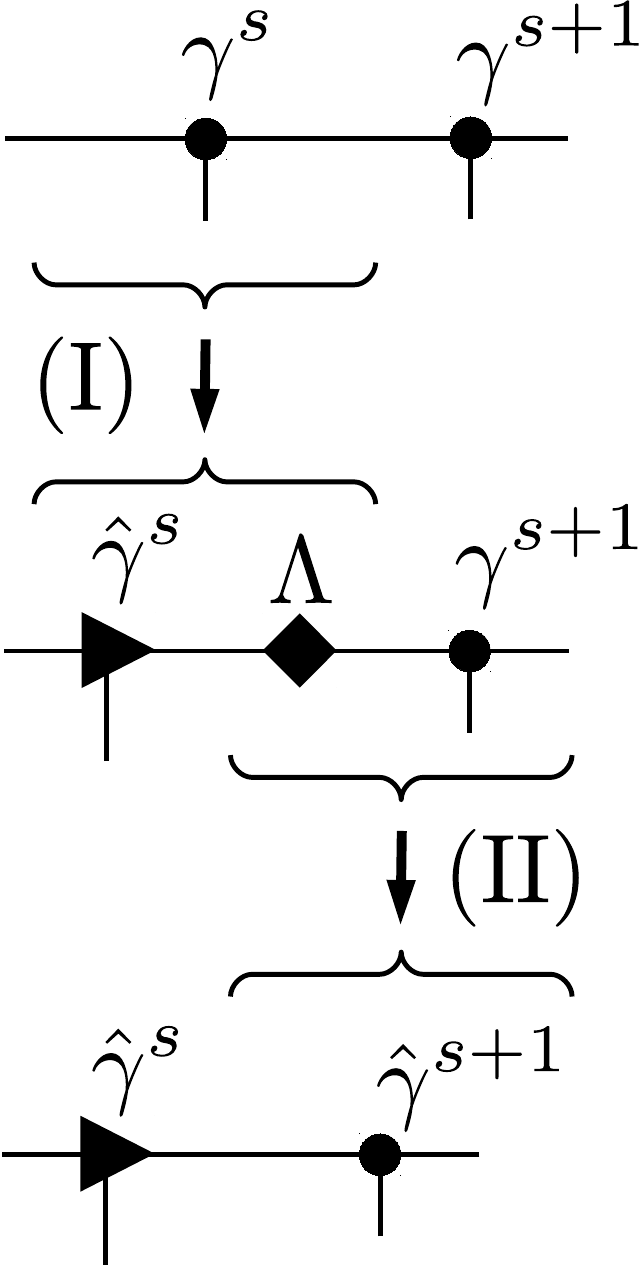}
\caption{
Bringing a tensor into left-canonical form.  In Step (I), $\gamma^s$ is
split into a left-canonical tensor $\hat\gamma^s$ and a tensor $\Lambda$
on the bond.  In Step (II), $\Lambda$ is absorbed in the tensor
$\gamma^{s+1}$ to the right, yielding $\hat\gamma^{s+1}$.
}
\label{fig:can-form-elem-move}
\end{figure}

We proceed in two steps, as shown in Fig.~\ref{fig:can-form-elem-move}: In step (I),
$\gamma^s$ is rewritten with two tensors, namely a new tensor
$\hat\gamma^s$ in left-canonical form, contracted with a new tensor
$\Lambda$ on its right virtual bond.  In step (II), $\Lambda$ is
contracted with $\gamma^{s+1}$, giving a new tensor $\hat\gamma^{s+1}$.

Step (I) is implemented by blocking $(lp)\equiv a$ and performing the SVD
of $\gamma^s_{ar\sep ar}$ as described in Sec.~\ref{sec:svd}, or one the 
simplified versions in Sec.~\ref{sec:other-decomp}, which yields a decomposition of
$\gamma^s$ into an isometry $L\equiv \hat\gamma^s$ and $\Lambda$, as
desired.  (If a decomposition with non-trivial $R$ [cf.\ Fig.~\ref{fig:gaussian-svd}(a)], such as the SVD,
is used, $\Lambda$ and $R$ need to be contracted.)
Specifically, it is sufficient to choose an $O$ such that
$O\gamma^s_{ar}=\left(\begin{smallmatrix}0\\Y\end{smallmatrix}\right)$
with $Y$ square, define
\begin{equation}
\left(
\begin{array}{c|cc}X & 0 & 0 \\
\hline
0 & 
\\
0 &\multicolumn{2}{c}{\smash{\raisebox{.5\normalbaselineskip}{$\ \Lambda$}}} 
\end{array}
\right)
\equiv
    (O\oplus\openone)\gamma^s (O\oplus\openone)^\tr
\end{equation}
(on the l.h.s., the first two blocks correspond to $a$ and the third block
to $p$), and let
\begin{equation}
\hat\gamma^s\equiv 
(O\oplus \openone)^\tr \left(
\begin{array}{c|cc}X & 0 & 0 \\
\hline
0 & 0 & \openone \\
0 & -\openone & 0 \end{array}
\right)
(O\oplus\openone)\ .
\end{equation}
For the sizes of $X$ and $\Lambda$ and more details on the procedure,
see Sec.~\ref{sec:other-decomp}.

In the second step, $\Lambda$ and $\gamma^{s+1}$ are contracted as indicated in
Fig.~\ref{fig:can-form-elem-move}, using Eq.~\eqref{eq:contraction:schur}.

In order to obtain the right-canonical form, the same steps have to be
followed; the only core difference is the inverse contraction order, which
in particular implies that the signs of the $\pm\openone$ in the SVD and
the contraction have to be reversed.
 
\paragraph{Obtaining and updating the canonical form.}
Given this elementary step, the GFMPS can be kept in canonical form at all times
using the same procedures that are used for conventional MPS and described in detail
in, e.g., Ref.~\onlinecite{schollwoeck2011}. In particular, given an initial GFMPS on $N$ sites
that is not in canonical form, it can be brought into canonical form by sweeping from one end
to the other, i.e. in $N$ elementary operations. All practical optimization or time evolution
algorithms for MPS perform local operations in each step and sweep over the system from one
end to the other (see also Sec.~\ref{sec:GFMPS-opt}). Therefore, the state can be kept in
canonical form by moving the center $s$, i.e. the site such that all tensors left (right) of it are
left-canonical (right-canonical), by only a single site, requiring only local operations.

\section{Energy computation for GFMPS}
\label{sec:optim}

We now explain how to compute and optimize the energy of a GFMPS for a
given Hamiltonian. We start with a simple elementary step, which takes a
two-site Hamiltonian and a two-site GFMPS in canonical form, and expresses
the energy of the GFMPS as a linear function of one of the GFMPS tensors
through an effective single-site Hamiltonian.  This will allow us to
derive an iterative formula for the energy which is valid for any
Hamiltonian, and which can moreover be used to optimize the energy of a
given working site.  Subsequently, we show how this formula can be
specialized to different cases, such as local Hamiltonians, to obtain more
efficient ways to compute the energy. 

For sake of completeness, we recall from the Introducion
[Eqs.~(\ref{eqn:Hmaj},\ref{eq:ham-energy})]
that a general Gaussian Hamiltonian is of the form
\begin{equation} \label{eqn:Hmaj-rep}
\mathcal H = -\mathrm{i} \sum H_{ij}c_i c_j + E\ ,
\end{equation}
(with $H$ an antisymmetric matrix and $E$ a constant offset), and 
the energy of a state $\rho$ with CM $\gamma$ is given by
\begin{equation} \label{eq:ham-energy-2}
\mathrm{tr}[\mathcal H\rho] = 
\mathrm{tr}(H\gamma)+E\ .
\end{equation}
We will henceforth denote this Hamiltonian by $(H,E)$.

\subsection{Energy in GFMPS: Elementary step}

\begin{figure}
\includegraphics[width=6cm]{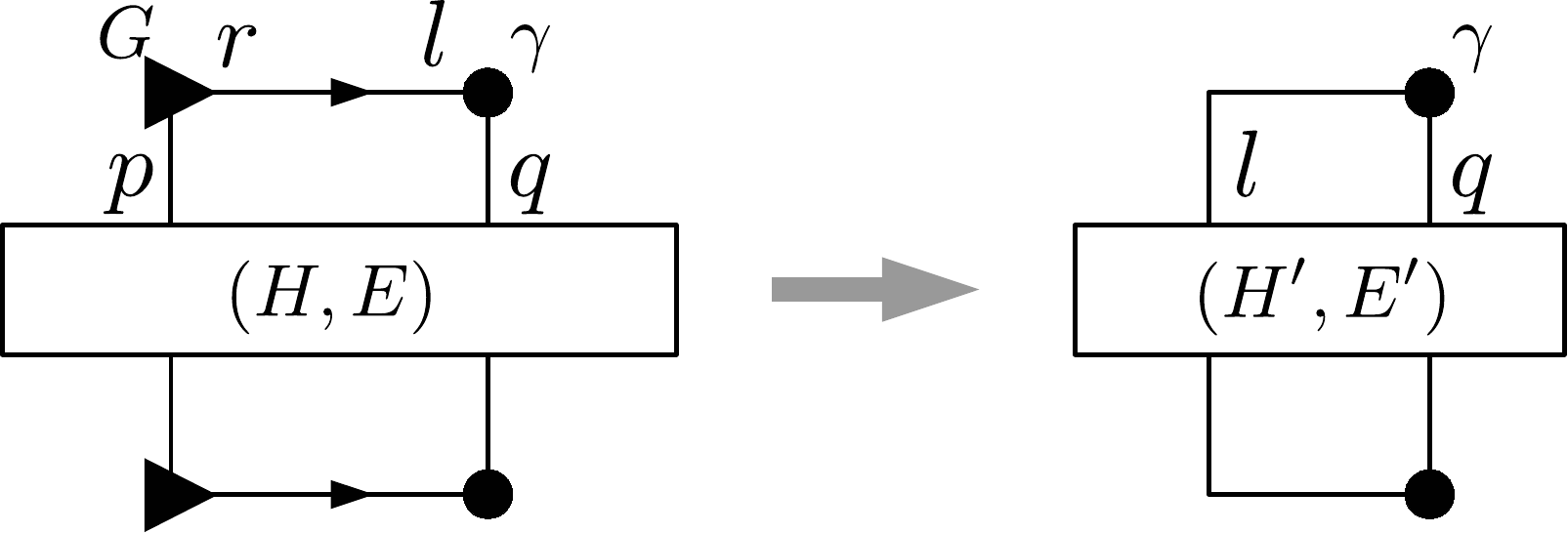}
\caption{Elementary step for evaluating energies in MPS: Dependence of the
energy on one tensor $\gamma$ (see text).}
\label{fig:ham-elementarystep}
\end{figure}

Let us first consider the scenario depicted in
Fig.~\ref{fig:ham-elementarystep}: We are given two  tensors
$G_{pr\sep pr}$ and $\gamma_{lq\sep lq}$, where $G_{rr}=0$, i.e. it is in left-canonical form,
and $r\triangleright l$ is contracted, and consider a Hamiltonian $(H,E)$ acting
on $pq$. Following Eq.~\eqref{eq:contraction:schur} for the contraction,
using Eq.~\eqref{eq:sammlung:inv-w-0} for the inverse, and rearranging
some terms, we find that the contraction yields a CM
\begin{equation}
\Theta=\mtx{G_{pp}&0\\0&0} + 
    \mtx{G_{pr}&0\\0&\openone}
    \mtx{\gamma_{ll}&-\gamma_{ql}^\tr\\\gamma_{ql}&\gamma_{qq}}
    \mtx{G_{pr}^\tr&0\\0&\openone}\:.
\end{equation}
We can thus rewrite
\begin{align}
\nonumber
&\mathrm{tr}[H\Theta] \\
    & = 
    \mathrm{tr}\Big[H \mtx{G_{pp}&0\\0&0}\Big]
    + \mathrm{tr}\Big[H
    \mtx{G_{pr}&0\\0&\openone}
    \gamma
    \mtx{G_{pr}^\tr&0\\0&\openone}\Big]
\\
    &= 
    \mathrm{tr}[H_{pp}G_{pp}]
    + \mathrm{tr}[H'\gamma]\ ,
\end{align}
with 
\begin{equation}
\label{eq:ham:hprime}
H'=\mtx{G_{pr}^\tr&0\\0&\openone} H \mtx{G_{pr}&0\\0&\openone}\ ,
\end{equation}
where the size of the identities is $|q| \times |q|$, and of course $|r|=|l|$,
and therefore $H'$ has dimensions $(|l|+|q|)\times(|l|+|q|)$.
The effective Hamiltonian acting on the $\gamma$ system -- comprised of the modes in $l$ and $q$ -- is thus obtained
from the old Hamiltonian through 
\begin{equation}
(H,E)\mapsto (H',E'=E+\mathrm{tr}[H_{pp}G_{pp}])
\end{equation}
with $H'$ from Eq~\eqref{eq:ham:hprime}.

Note that if the right tensor were canonical and we instead contracted $l\triangleright r$,
the signs of the $\openone$'s in Eq.~\eqref{eq:ham:hprime} had to be reversed.

\subsection{Energy in GFMPS: Iteration formula}

\begin{figure}
\includegraphics[width=7cm]{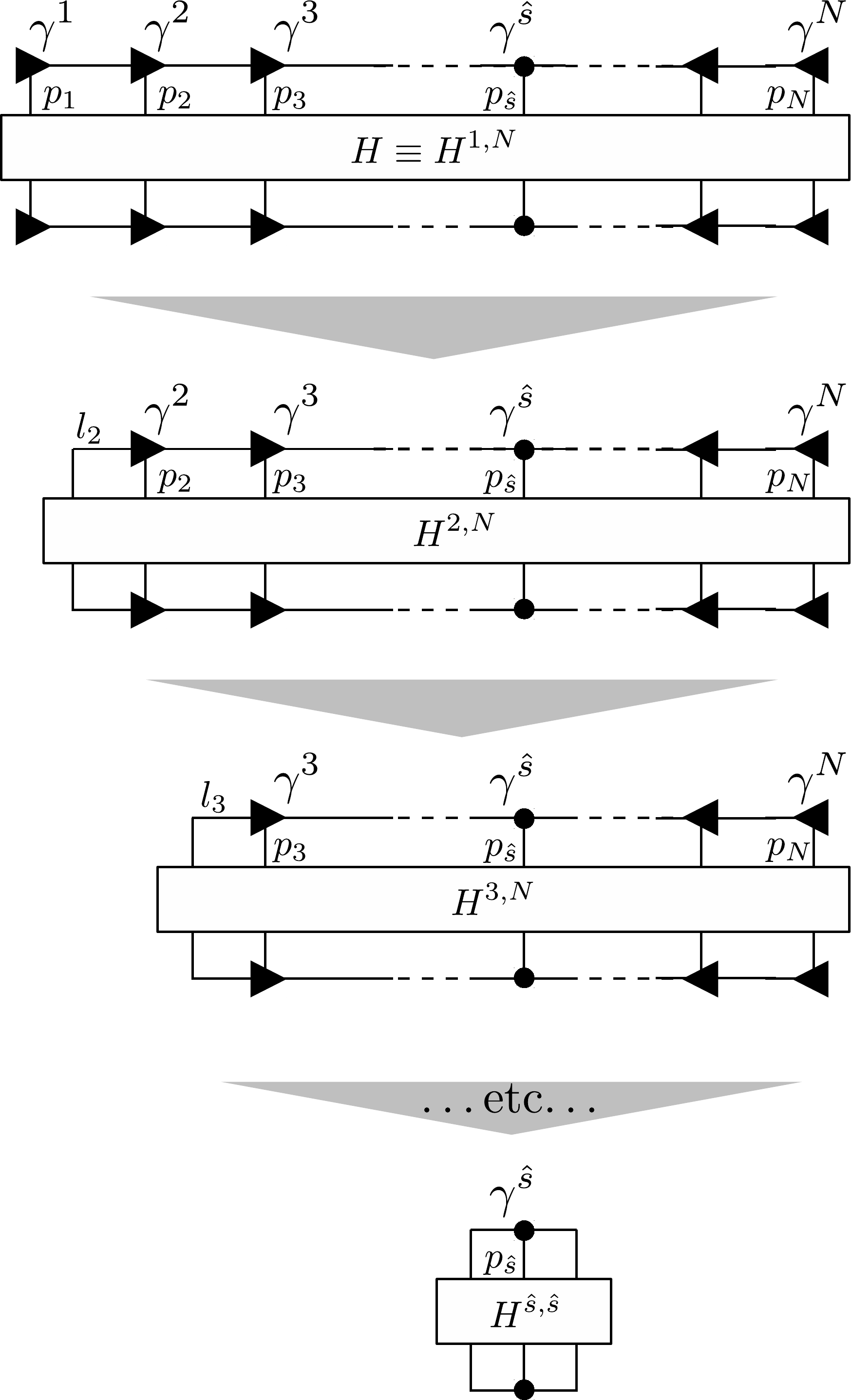}
\caption{Scheme for the contraction of the Hamiltonian. We proceed
stepwise from both ends (here shown from the left end), which gives an
effective Hamiltonian for successively smaller patches which also acts on
the dangling virtual bonds at the boundary, finally yielding an effective
Hamiltonian for $\gamma^{\hat s}$.
}
\label{fig:ham-fulliteration}
\end{figure}

We are now ready to derive how the energy of a 
given GFMPS for some Hamiltonian $(H^{1,N},E^{1,N})$, acting on $p_1,\dots,p_N$,
depends on the GFMPS tensor at a specific site  $\hat s$. 
It is based on iterated application of the elementary step above, and is illustrated  in Fig.~\ref{fig:ham-fulliteration}.

Consider an MPS in canonical form around $\hat s$, with $H$
acting on all physical sites $p_1,\dots,p_N$. For the description, we assume
the generic case $1 < \hat s < N$, but the procedure is easily applied to any $\hat s$.
By applying the procedure derived in the last section to $\gamma^1$ (corresponding to $G$ above)
vs.\ the remaining tensors (seen as one blocked tensor $\gamma$), we obain an effective Hamiltonian
\begin{equation}
H^{2,N}
\stackrel{l_2\leftarrow r_1} = (\gamma^{1}_{p_1r_1}\oplus\openone)^\tr H^{1,N}
(\gamma^1_{p_1r_1}\oplus\openone)\ ,
\end{equation}
where $H^{2,N}$ is acting on $l_2,p_2,\dots, p_N$, and we re-label $r_1$
to $l_2$, as indicated on top of the equality sign.
We can repeat this procedure
\begin{equation}
\label{eq:ham:iteration-from-left}
H^{s+1,N}
\stackrel{l_{s+1}\leftarrow r_s} =
(\gamma^{s}_{l_sp_s|r_s}\oplus\openone)^\tr H^{s,N}
(\gamma^s_{l_sp_s|r_s}\oplus\openone)\ ,
\end{equation}
until we reach site $\hat s$. 

Similarly, we can contract $H$ from the right, using
\begin{equation}
H^{\hat s,s-1}
\stackrel{r_{s-1}\leftarrow l_s} = (\gamma^{s}_{p_sr_s|l_s}\oplus
(-\openone))^\tr H^{\hat s,s}
(\gamma^{s}_{p_sr_s|l_s}\oplus (-\openone))
\end{equation}
-- note the $-\openone$ due to the opposite contraction order -- until we
reach $\hat s$. 

In words, each step removes one site from the MPS and instead updates the
way in which the Hamiltonian acts on the dangling virtual leg at the
boundary (this update rule thus involves the correlations between this
dangling leg and the physical leg in the bulk).  This is illustrated for
the left boundary in Fig.~\ref{fig:ham-fulliteration}: In order to update
$H$, we have to take the Hamiltonian part corresponding to the
left+physical leg which are being removed (site $s$), and transform it
with $\gamma^s_{lp|r}$ which relates those legs to the new dangling leg
(and correspondingly from the right).

In order to compute the total energy, we must also keep track of the $E$
contribution. It is updated as
\begin{equation}
\label{eq:ham:update-constant}
\begin{aligned}
E^{s+1,N}&=\mathrm{tr}[H^{s,N}_{l_sp_s|l_sp_s}\gamma^s_{l_sp_s|l_sp_s}]
+ E^{s,N}\\
E^{\hat s,s-1}&=\mathrm{tr}[H^{\hat
s,s}_{p_sr_s|p_sr_s}\gamma^s_{p_sr_s|p_sr_s}] + E^{\hat s,s}
\end{aligned}
\end{equation}
with $E^{2,N}=\mathrm{tr}[H^{1,N}_{p_1p_1}\gamma^1_{p_1p_1}]=
\mathrm{tr}[H_{p_1p_1}\gamma^1_{p_1p_1}]$, and accordingly
$E^{\hat s,N-1}=\mathrm{tr}[H_{p_Np_N}\gamma^N_{p_Np_N}]$.

We are thus left with a single-site Hamiltonian
$H^{\hat s,\hat s}$ acting on $l_{\hat s}p_{\hat s}r_{\hat s}$, such that
the total energy is
\begin{equation}
\label{eq:ham:total-effective-ham}
E_\mathrm{tot}(\gamma^{\hat s}) = \mathrm{tr}[H^{\hat s,\hat s}\gamma^{\hat
s}] + E^{\hat s,\hat s}\ .
\end{equation}

\subsection{Local Hamiltonians}

Let us now consider the special case of local Hamiltonians -- for
simplicity, we consider nearest-neighbor Hamiltonians -- and explain how
the evaluation of the Hamiltonian in this case can be accomplished in
linear time in the system size. Moreover, just as in conventional DMRG, it
is possible to store intermediate terms at each cut such that the effort
required to update this information when moving by one site is independent
of system size.

In the following, it will be convenient to also carry an $l_1$ and $r_N$
label, with the corresponding spaces being zero-dimensional (in the CM).
In any step of the computation, the Hamiltonian will be of the banded form
\begin{equation} \label{lh_ham_1}
 H^{s,N} = \ 
\begin{blockarray}{ccccc}
l_s & p_{s} & p_{s+1} & p_{s+2} & \dots \\
\begin{block}{(c|c|c|c|c)}
A^s_\lham & B^s_\lham & & & \phantom\ddots \\
\BAhline
-(B^s_\lham)^\tr &  D^s_\lham+ h_{11} & h_{12} & & \phantom\ddots \\
\BAhline
  & -h_{12}^\tr & h_{22}+k_{11} & k_{12} & \phantom\ddots \\
\BAhline
   &             & -k_{12}^\tr   & k_{22} + \dots & \ddots\\
\BAhline
    & &   & \ddots & \ddots \\
\end{block}
\end{blockarray}\;\; .
\end{equation}
Here, $A^s_\lham$, $B^s_\lham$, and $D^s_\lham$ summarize the information from all sites to the
left of $s$, and $h$ ($k$) is the two-site Hamiltonian term acting on sites
$p_{s},p_{s+1}$ ($p_{s+1},p_{s+2}$), respectively. One update step
Eq.~\eqref{eq:ham:iteration-from-left} maps this to 
\begin{equation} \label{lh_ham_2}
 H^{{s+1},N} = \ 
\begin{blockarray}{cccc}
l_{s+1} & p_{s+1} & p_{s+2} & \dots \\
\begin{block}{(c|c|c|c)}
A^{s+1}_\lham & B^{s+1}_\lham  & & \phantom\ddots \\
\BAhline
-{(B^{s+1}_\lham)}^\tr  & D^{s+1}_\lham+k_{11} & k_{12} & \phantom\ddots \\
\BAhline
             & -k_{12}^\tr   & k_{22} + \dots & \ddots\\
\BAhline
 &   & \ddots & \ddots \\
\end{block}
\end{blockarray}\;\; .
\end{equation}
with the update rule
\begin{equation}
\label{eq:ham:locham-update-rule}
\begin{aligned}
A^{s+1}_\lham&=\gamma_{lr}^\tr A^s_\lham \gamma_{lr} + 
     \gamma_{pr}^\tr(D^s_\lham+h_{11})\gamma_{pr}\\
    & \hspace*{4em} 
	    + \gamma_{lr}^\tr B^s_\lham\gamma_{pr} -\gamma_{pr}^\tr{(B^s_\lham)}^\tr\gamma_{lr} 
\\
B^{s+1}_\lham & = \gamma_{pr}^\tr h_{12}\\
D^{s+1}_\lham & = h_{22}
\end{aligned}
\end{equation}
where $\gamma\equiv \gamma^s$. 
We thus see that (\emph{i}) the block-diagonal form of $H^{s,N}$ is
preserved, (\emph{ii}) the initial Hamiltonian is of this form with 
$A^1_\lham=0$, $B^1_\lham=0$, $D^1_\lham=0$, and (\emph{iii}) updating the information about the
Hamiltonian only requires local updates.

Analogously, we can also keep the integrated Hamiltonian terms for all sites right of $s$ through the update rule
\begin{equation}
\label{eq:ham:locham-update-rule-right}
\begin{aligned}
A^{s-1}_\rham & = 
    \gamma_{rl}^\tr A^s_\rham\gamma_{rl} + 
    \gamma_{pl}^\tr(h_{22}+D^s_\rham)\gamma_{pl} 
    \\
    & \hspace*{4em} 
    + \gamma_{rl}^\tr B^s_\rham\gamma_{pl} 
    - \gamma_{pl}^\tr{(B^s_\rham)}^\tr\gamma_{rl}
\\
B^{s-1}_\rham & = \gamma_{pl}^\tr h_{12}^\tr\\
D^{s-1}_\rham & = h_{11}
\ ,\\
\end{aligned}
\end{equation}
where again $\gamma=\gamma^s$, and $h$ acts on sites $p_{s-1},p_{s}$ (with $h_{11}$ corresponding to $p_{s-1}$, etc.),
starting with $A^{N}_\rham=B^N_\rham=D^N_\rham=0$.

For a state with working site $\hat s$, we find that 
\begin{equation} \label{eqn:Hss}
\hat H^{\hat s,\hat s} = 
\begin{pmatrix}
A^{\hat s}_\lham & B^{\hat s}_\lham & 0 \\
-{(B^{\hat s}_\lham)}^\tr & 
    D^{\hat s}_\lham + D^{\hat s}_\rham & -{(B^{\hat s}_\rham)}^\tr \\
0 & B^{\hat s}_\rham & A^{\hat s}_\rham 
\end{pmatrix}
\end{equation}
with mode ordering $lpr$. Similarly, we can compute $E^{\hat s, \hat
s}=E^{\hat s}_\lham + E^{\hat s}_\rham$, where $E^1_\lham=E^N_\rham=0$,
and
\begin{equation} \label{eq:E-left}
\begin{aligned}
E^{s+1}_\lham &= E^s_\lham +
\mathrm{tr}[H^{s,N}_{l_sp_s|l_sp_s}\gamma^s_{l_sp_s|l_sp_s}] 
\\ 
&= 
E^s_\lham +
\mathrm{tr}\big[\begin{pmatrix}
    A^s_\lham&B^s_\lham \\ -{(B^s_\lham)}^\tr & D_\lham^s +
h^s_{11}\end{pmatrix} \gamma^s_{lp|lp} \big]
\end{aligned}
\end{equation}
and accordingly
\begin{equation} \label{eq:E-right}
E^{s-1}_\rham = E^s_\rham +
\mathrm{tr}\big[\begin{pmatrix}
    h^{s-1}_{22} + D_\rham^s &-{(B^s_\rham)}^\tr \\ B^s_\rham & 
    A^s_\rham 
    \end{pmatrix} \gamma^s_{pr|pr} \big]
\end{equation}
where $h^s_{11}$ ($h^s_{22}$) are the left (right) part of the Hamiltonian
term acting between sites $s$ and $s+1$.
Overall, the dependence of the energy on $\gamma^{\hat s}$ is then given
by
\begin{equation}
    E_\mathrm{tot}(\gamma^{\hat s}) = \mathrm{tr}[H^{\hat s,\hat s}\gamma^{\hat s}] + E^{\hat s,\hat s}\ .
\end{equation}
Since all update rules for the $A$, $B$, $D$, and $E$ are local, they can be updated (and thus the energy dependence on $\gamma^{\hat s}$ computed) at a cost independent of the system size when moving the working site.

The generalization beyond nearest-neighbor Hamiltonians follows the same pattern
illustrated above, but the Hamiltonian of Eq.~\eqref{lh_ham_1} will have additional
off-diagonal terms. In particular, the first row/column, which contains
coupling terms between sites $q < s$ and $p \geq s$, will be modified as follows:
(\emph{i}) The terms $A^s_\lham$ and $B^s_\lham$ will contain additional contributions from
terms between $q < s$ and $s$, and (\emph{ii}) there will be additional off-diagonal blocks similar to
$B^s_\lham$ for coupling terms between $q < s$ and $p > s$.
The update rule~\eqref{eq:ham:locham-update-rule}
is adapted by applying the rule for $B^{s+1}_\lham$ for all similar off-diagonal
terms, and including additional terms in $A^{s+1}_\lham$.
The update for the constant term can be adapted similarly.
The total number of blocks that need to be treated in each step depends on the
total number of Hamiltonian terms between sites $q < s$ and $p \geq s$.

\subsection{Energy minimization\label{sec:energy-minimization-eigenvalue}}

A core ingredient in DMRG algorithms is the optimization of the energy as a function of a single tensor $\gamma^{\hat s}\equiv \gamma$. As we have seen, this energy dependence can be summed up in an effective Hamiltonian $H^{\hat s,\hat s}\equiv H$ and constant offset $E^{\hat s,\hat s}\equiv E$.
In order to minimize
$E(\gamma)=\mathrm{tr}(H\gamma)+E$,
we need to fill  all modes with negative energy. This can be done by 
by going to the eigenbasis of $H$,
\begin{equation}
H\cong \bigoplus_i \mtx{0 & e_i\\-e_i & 0}
\end{equation}
and choosing
\begin{equation}
\gamma\cong \bigoplus_i \mtx{0 & 1\\-1 & 0}\ .
\end{equation}
Numerically, this can be done by diagonalizing the antisymmetric
matrix $H$ (giving imaginary eigenvalues) and then replacing them by their
sign, $\gamma=\mathrm{sign}(H)$, or by following the numerical approaches
discussed in Ref.~\onlinecite{Wimmer2012}.

\section{GFMPS optimization algorithms}
\label{sec:GFMPS-opt}

We now describe some commonly used MPS algorithms specifically in the
context of GFMPS. Most importantly, we will describe how to carry out the
DMRG algorithm in its formulation as a variational method over GFMPS.  In
addition, we will discuss how to implement the time-dependent variational
principle (TDVP), which is suitable for both time evolution and ground
state simulations, with GFMPS, as well as a translational invariant
version of those methods.

\subsection{Gaussian fermionic DMRG}

The DMRG algorithm can be naturally described as a variational method over
the family of MPS with a given bond dimension.  The key idea is to sweep
through the system and sequentially optimize a small number (usually one
or two) GFMPS tensors in each step.  Each such optimization is a quadratic
problem and can thus be efficiently solved.
In the case where more than one tensor is optimized, the state can be
brought back into MPS form using the singular value decomposition. While
local convergence does not guarantee global convergence of the energy,
this algorithm is empirically found to perform extremely well.

\subsubsection{The algorithm}

Gaussian fermionic DMRG, i.e. DRMG for a quadratic fermionic Hamiltonian
and with GFMPS as a variational family, follows closely the conventional
way of doing DMRG with MPS, see
e.g.~\cite{schollwoeck2005,schollwoeck2011}. It consists of an
initialization step, followed by a number of sweeps; each sweep in turn
consists of a right- and a left-sweep. For concreteness, we will consider
the case of a nearest-neighbor Hamiltonian, but the same steps equally
apply to more complex Hamiltonians (with suitable modifications regarding
the computation of effective Hamiltonians, cf.\ Sec.~\ref{sec:optim}). The
notation follows the one used in the previous sections. In addition, we
will denote the complete set of left and right boundary terms that enter
the effective Hamiltonian at position $s$ by $\mathcal
B_\lham^s=\{A^s_\lham,B^s_\lham,D^s_\lham,E^s_\lham\}$ and $\mathcal
B_\rham^s=\{A^s_\rham,B^s_\rham,D^s_\rham,E^s_\rham\}$, respectively.  We
begin by describing the variant where only one tensor is optimized at a
time ("single-site DMRG").

\noindent
\emph{(i) Initialization.---}%
Fix a bond mode number for each bond $(s,s+1)$, $s=1,\dots,N-1$, and  
choose a random (pure) initial tensor $\gamma^s$ for each site $s$.
Let $\mathcal B^N_\rham=\{\varnothing,\varnothing,\varnothing,\varnothing\}$, where by $\varnothing$ we denote a $0 \times 0$ matrix. \\
For $\hat s=N,N-1,\dots,2$ do:
\begin{enumerate}
\item Bring $\gamma^{\hat s}$ into right-canonical form (Sec.~\ref{sec:canform}).
\item Compute 
$\mathcal B^{\hat s-1}_\rham$ as given by Eq.~(\ref{eq:ham:locham-update-rule-right}), and store the result.
\end{enumerate}
After the initialization, we are left with an GFMPS with all tensors in right-canonical form.

\noindent
\emph{(ii) Right-sweep.---}%
Let $\mathcal B^1_\lham=\{\varnothing,\varnothing,\varnothing,\varnothing\}$.\\
For $\hat s=1,\dots,N-1$ do:
\begin{enumerate}
\item Compute $H^{\hat s,\hat s}$ from $\mathcal B^{\hat s}_\lham$  using Eq.~\eqref{eqn:Hss} (or a suitable modification, based on Sec.~\ref{sec:optim}, for longer ranged Hamiltonians.)
\item Choose $\gamma^{\hat s}$ such as to optimize the energy $\mathrm{tr}[H^{\hat s, \hat s}\gamma^{\hat s}]$.
\item Bring $\gamma^{\hat s}$ into left-canonical form (updating $\gamma^{\hat s+1}$ correspondingly).
\item (Re-)compute  $\mathcal B^{\hat s+1}_\lham$. 
\end{enumerate}

\noindent\emph{(iii) Left-sweep.---}%
The left-sweep works analogous to the right-sweep, just that one iterates
over $\hat s=N,N-1,\dots,2$, and steps 3 and 4 are replaced by bringing
$\gamma^{\hat s}$ into right-canonical form and updating $\mathcal B^{\hat
s-1}_\rham$,  respectively.

To find the ground state, one performs several left- and right-sweeps until convergence is reached. Convergence is typically tested for using the ground state energy (see also our discussion in Sec.~\ref{sec:numerics}).

A straightforward generalization of this algorithm is to the "two-site DMRG" algorithm, which optimizes
two adjacent tensors in each step. This coincides with the original DMRG algorithm due to
White~\cite{white1992,white1992-1}. Given a GFMPS in canonical form around
a center site $\hat s$ during a sweep to the right, the tensors on sites
$\hat s, \hat{s}+1$ are optimized by first using
the steps laid out in Sec.~\ref{sec:optim} to obtain an effective Hamiltonian $H^{(\hat{s},\hat{s}+1),(\hat{s},\hat{s}+1)}$.
One then finds the ground state of this Hamiltonian on the modes $l_{\hat s}, p_{\hat s}, p_{\hat{s}+1}, r_{\hat{s}+1}$.
Using the singular value decomposition of Sec.~\ref{sec:svd} and absorbing the middle tensor
$\Lambda$ into the right tensor, one can put the state back into canonical GFMPS form with
a new center site $\hat{s}' = \hat{s}+1$; therefore, no extra steps
are required to keep the state canonical.
When sweeping to the left, the same steps are performed on the sites $\hat{s} -1, \hat{s}$,
and in the final SVD $\Lambda$ is absorbed into the left tensor.
During the singular value decomposition, the bond dimension can be truncated as appropriate.

In conventional DMRG, the single-site algorithm needs to be augmented with the technique described
in Ref.~\onlinecite{white2005}. This addresses two issues of the single-site algorithm, namely that
the number of states on a bond is difficult to adapt dynamically, and a
tendency to become stuck
in local minima. While this technique does not have a simple generalization to the GFMPS case, we
find in practice that for a fixed bond dimension, the convergence of single- and two-site DMRG
on GFMPS is very similar; furthermore, given the much lower cost of the GFMPS-based algorithms,
it is often not necessary to dynamically adapt the bond number.

\subsubsection{Scaling and efficiency\label{sec:GFDMRG-scaling}}

All steps in the GFDMRG algorithm can be carried out at a cost of
$\mathcal{O}\left((\chi+p)^3\right)$, where $\chi$ is the number of Majorana modes
on each bond and $p$ the number of physical Majorana modes on each site. In practical
numerical implementations, the most costly step will usually be the inversion required
to contract two tensors, which can be optimized using the block formulas of
Eqs.~(\ref{eq:sammlung:inv-w-id},\ref{eq:sammlung:inv-w-0}). The scaling
of a single sweep with system size is $\mathcal{O}(N)$. The number of
sweeps required to reach convergence heavily depends on the underlying
problem and cannot in general be bounded.

Additional cost is incurred if the Hamiltonian is not nearest-neighbor. In particular,
the number of additional blocks that need to be included when computing the effective
Hamiltonian at site $s$ depends on the number of Hamiltonian terms that couples sites
$p < s$ and $q \geq s$. Therefore, the scaling depends crucially on details of the
Hamiltonian. In a simple case, such as periodic boundary conditions leading to a single
additional term between the left and the right end of the system, the additional cost is
constant, while in the most general case of coupling between all sites an additional
factor $N^2$ could be incurred.

In contrast to conventional DMRG algorithms, the performance can be improved
significantly by blocking a group of $n_b$ sites together.
To understand this, consider a system of $N$ sites with $p$ Majorana modes on each site, and a GFMPS description with
Majorana bond number $\chi$.
In this case, the matrices of the MPS will be of size $(2\chi + p) \times (2\chi + p)$, and
the computational cost of a sweep will scale as $L (2\chi + p)^3$.
If we group $n_b$ sites into one, we instead obtain a scaling of $(L/n_b) (2\chi + n_b p)^3$.
In general, one has $\chi \gg p$, and optimal performance will be achieved for $n_b > 1$.
For local Hamiltonians, the optimal choice is $n_b = \chi/2$. For non-local Hamiltonians,
an additional advantage can arise because the Hamiltonian may become less long-ranged
if sites are blocked together.

Finally, in the discussion so far we have considered general Gaussian states rather than
states with a fixed particle number. In principle, Gaussian states with fixed
particle number can be described by matrices that are smaller by a factor of two, thus
potentially leading to a speedup of order $2^3$.

We will perform a more detailed analysis of the performance for specific models in
Sec.~\ref{sec:runtime}.

\subsection{TDVP}

The time-dependent variational principle (TDVP) can be adapted
to use MPS~\cite{haegeman2011} as variational manifold to approximately solve 
the time-dependent Schr\"odinger equation both in real and imaginary time. 
In Ref.~\onlinecite{haegeman2016}, it is shown that TDVP can be re-formulated as a
modification of the DMRG method, with only a  few modifications.  In the following,
we discuss how to modify the GFDMRG algorithm described above to obtain TDVP
for GFMPS following the recipe of Ref.~\onlinecite{haegeman2016}.

To perform the evolution for some sufficiently small timestep $t$
(where real time evolution corresponds to real $t$, and imaginary time evolution to $t=i\tau$, $\tau>0$),
the steps 2, 3, and 4 in \emph{(ii) Right-sweep} are replaced by the following sequence:
steps :
\begin{enumerate}
\item[2a.] 
Evolve $\gamma^{\hat s}$ under $H^{\hat s,\hat s}$ for time $t$. \item[3a.] 
Perform Step (I) of the elementary move to bring $\gamma^{\hat s}$ into left-canonical form, see Sec.~\ref{sec:canform} and   Fig.~\ref{fig:can-form-elem-move}, i.e. bring $\gamma^{\hat s}$ into left-canonical form $\hat\gamma^{\hat s}$ by splitting off the SVD $\Lambda$, but do not absorb $\Lambda$ in $\gamma^{\hat s+1}$ yet.
\item[4.] (Re-)compute $\mathcal B^{\hat s+1}_\lham$, using the new (left-canonical) $\hat \gamma^{\hat s}$. (Since $\mathcal B^{\hat s+1}$ does not depend on $\gamma^{\hat s+1}$, this is already the correct $\mathcal B^{\hat s+1}$.)
\item[2b.] Compute the effective Hamiltonian $H_\Lambda$ for $\Lambda$, which describes the dependence of the energy on $\Lambda$:
\begin{equation}
\label{eq:Heff-Lambda}
H_\Lambda=\begin{pmatrix} 
A^{\hat s+1}_\lham & B^{\hat s+1}_\lham \gamma_{pl}^{\hat s+1} \\
{-(B^{\hat s+1}_\lham\gamma_{pl}^{\hat s+1})}^\tr & A^{\hat s}_\rham
\end{pmatrix}\ ,
\end{equation}
and evolve $\Lambda$ with $H_\Lambda$ for time $-t$. 
\item[3b.] Absorb the evolved $\Lambda$ into $\gamma^{\hat s+1}$.
\end{enumerate}
A core ingredient, used in Step 2a and 2b, is to evolve a state $\rho$ with covariance matrix $\Gamma$ with a
time-independent Hamiltonian $\mathcal H = \mathrm{i} \sum H_{ij} c_i c_j$ either in real or imaginary time. 
In the covariance matrix formalism, the Schr\"odinger equation in real time takes the form
\begin{equation}
\label{eq:time-evol-cm}
\Gamma(0)\mapsto \Gamma(t)=O(t)\Gamma(0)O(t)^\tr\ ,
\end{equation}
with $O(t)=\exp(4Ht)$~\footnote{This can be shown by observing that Eq.~\eqref{eq:time-evol-cm} is equivalent to 
$c_k(t)=\sum O(t)_{kl} c_l(0)$ and thus  $\dot c_k = 4\sum H_{kl} c_l$, and on the other hand comparing it
with the evolution $\dot c_k = \mathrm{i}[\mathcal H,c_k]$ of $c_k$ in the
Heisenberg picture.}.
For imaginary time, this is replaced by integrating~\cite{kraus2010-ghf}
\begin{equation}
\dot\Gamma(t) = -4(H+\Gamma(t)H\,\Gamma(t))\ .
\end{equation}

For details on the implementation, such as the optimal choice of
time discretization as well as the generalization to a two-site version of
the algorithm that allows a dynamical choice of the bond dimension, we
refer the reader to Ref.~\cite{haegeman2016}. Note that in the case where
$t\to\mathrm{i}\:\infty$, we recover the DMRG algorithm.

\subsection{Infinite systems}

It was already pointed out in Ref.~\onlinecite{white1992} that DMRG is in
principle suitable for infinite systems. While this approach initially
enjoyed some success, for a long time DMRG for finite systems was
considered more accurate. More recently, however, DMRG for infinite
systems has been reinterpreted in the language of matrix product states,
which has allowed for accurate and efficient methods for infinite,
translationally invariant MPS to be
developed~\cite{mcculloch2008infinite,crosswhite2008,vanderstraeten2019,schollwoeck2011}.
The underlying idea of this reinterpretation is to consider an MPS made up
of a unit cell of $K$ sites that repeats indefinitely; in the simplest
case, $K=1$, the tensor is the same on each site of the lattice. It turns
out that most MPS algorithms can be generalized to such an infinite,
translationally invariant ansatz by reformulating them in terms of
eigenvectors of transfer operators along the MPS.

While non-interacting systems can be solved directly in the thermodynamic
limit by expressing the problem in momentum space,
such an approach may scale unfavorably with the number of sites in the unit cell and 
does not give access to real-space correlation functions without solving for all
momenta and transforming back to real space. Therefore, for a large number of sites
in the unit cell, a GFMPS approach directly in the thermodynamic limit will be
advantageous.

To illustrate how  GFMPS can be generalized to the infinite
case, we will limit ourselves to the iDMRG algorithm, and
restrict the discussion to the aspects which need to adapted in the GFMPS
case. However, similar generalizations are also possible for other
infinite MPS methods, such as VUMPS~\cite{vanderstraeten2019}.
We begin by briefly reviewing the conventional iDMRG algorithm of Ref.~\onlinecite{white1992}, where we specialize to the case of a two-site Hamiltonian $H$. Our description follows closely Ref.~\onlinecite{schollwoeck2011}, which we refer to for details. The algorithm proceeds in the following steps:
\begin{enumerate}
\item Compute the covariance matrix $\Gamma^{n=1}$ corresponding to the ground state of $H$ on two sites, and perform an SVD such that $\Gamma$ is given as the contraction of three tensors $A^n$, $\Lambda^n$ and $B^n$, where $A^n$ and $B^n$ are left- and right-canonical MPS tensors, respectively.
\item Extend the lattice by two sites in the middle and consider an MPS where the tensor immediately to the left (right) of the newly inserted sites is given by $A^n$ ($B^n$). Follow the steps outlined in Sec.~\ref{sec:optim}, especially Eqns.~\eqref{eq:ham:locham-update-rule}-\eqref{eqn:Hss} (adapted for two center sites), to obtain the effective Hamiltonian for the two sites newly added in the center. Use Eqns.~\eqref{eq:E-left},\eqref{eq:E-right} to obtain the constant part of the energy.
\item Find the ground state $\Gamma^{n+1}$ of this new effective Hamiltonian and again perform a singular value decomposition to obtain $A^{n+1}$, $\Lambda^{n+1}$ and $B^{n+1}$. Return to step (2).
\end{enumerate}
All of the steps outlined here can be performed using techniques already discussed. Note that this approach effectively simulates a finite system of length $2n$, where $n$ is the number of iterations performed. The energy obtained as the sum of the constant term and the effective Hamiltonian for the center sites is variational for this system of $2n$ sites. One generally finds that for sufficiently large $n$, the tensors converge.

One can now take the tensors obtained in the last step, and use them to form an infinite, translationally invariant MPS~\cite{mcculloch2008infinite,schollwoeck2011}. Let $A^n$, $\Lambda^n$, and $B^n$ be the tensors obtained in the last iteration of the infinite DMRG algorithm, and $\Lambda^{n-1}$ the central tensor from the previous iteration. We then define, analogous to Ref.~\onlinecite{vidal2007tebd},
\begin{align}
\Gamma^A &= \left( \Lambda^{n-1} \right)^{-1} \contr A^n &\lambda^A &= \Lambda^n \\
\Gamma^B &= A^n \contr \left( \Lambda^{n-1} \right)^{-1} &\lambda^B &= \Lambda^{n-1}.
\end{align}
(In the following, we use the notation $A\contr B$ to denote the contraction of
tensors; which indices are to be contracted follows from the structure of
the tensor network.)
Here, $\Lambda^{-1}$ denotes a square rank-2 tensor such that $\Lambda
\contr \Lambda^{-1}$ yields the maximally entangled state between the two
sets of modes. From the SVD, $\Lambda$ is of the form $\Lambda = \bigoplus
W(\lambda_k)$ (see Eq.~\eqref{eq:gaussian-schmidt-Sblock}); for a tensor
of this form, the inverse of this sense is given by exchanging the left
and right modes (i.e., mapping $\lambda_k\to-\lambda_k$). This can be verified explicitly using
Eqn.~\eqref{eq:contraction:schur}.

If the state is well-converged and the bond dimension large enough,
$\gamma^{L,1}=\lambda^B \contr \Gamma^A \contr \lambda^A \contr \Gamma^B$
and $\gamma^{L,2} = \lambda^A \contr \Gamma^B
\contr \lambda^B \contr \Gamma^A$ will be approximately left-canonical,
and $\gamma^{R,1}=\Gamma^A \contr \lambda^A \contr \Gamma^B \contr
\lambda^B$ and $\gamma^{R,2}=\Gamma^B \contr \lambda^B \contr \Gamma^A
\contr \lambda^A$ will be approximately right-canonical.
Expectation values of local
observables, such as the energy on a bond,
can be obtained from the observation that the reduced density
matrix for the two-site unit cell AB is given by $\left(\lambda^B \contr
\Gamma^A \contr \lambda^A \contr \Gamma^B \contr \lambda^B \right)_{p^A
p^B | p^A p^B}$, where the physical indices in the subscript refer to the
physical indices of $\Gamma^A$ and $\Gamma^B$.
Note that for the average energy, one must compute the energy for the AB and BA
unit cells separately and average.

\section{Numerical results}
\label{sec:numerics}

Below we test our numerical implementation of the algorithms presented in the previous sections for three relevant
cases: We first study a one-dimensional system that exhibits non-trivial spatial correlation functions, but can be exactly solved, thus
providing a simple comparison. Then we turn to a two-dimensional system, which we solve on quasi-two-dimensional cylinder
geometries, to assess how accurately our approach works beyond a strictly one-dimensional system and when
it performs better than exact approaches.
Finally, we study charge transport through a quantum point contact in a simple geometry at finite bias voltage
to demonstrate the capability of TDVP to compute dynamical properties of nanostructures.

While the models discussed in this section preserve particle number, we use the Majorana formalism described
throughout this manuscript without exploiting particle number conservation.

\subsection{The resonant level model}

\begin{figure}
  \includegraphics{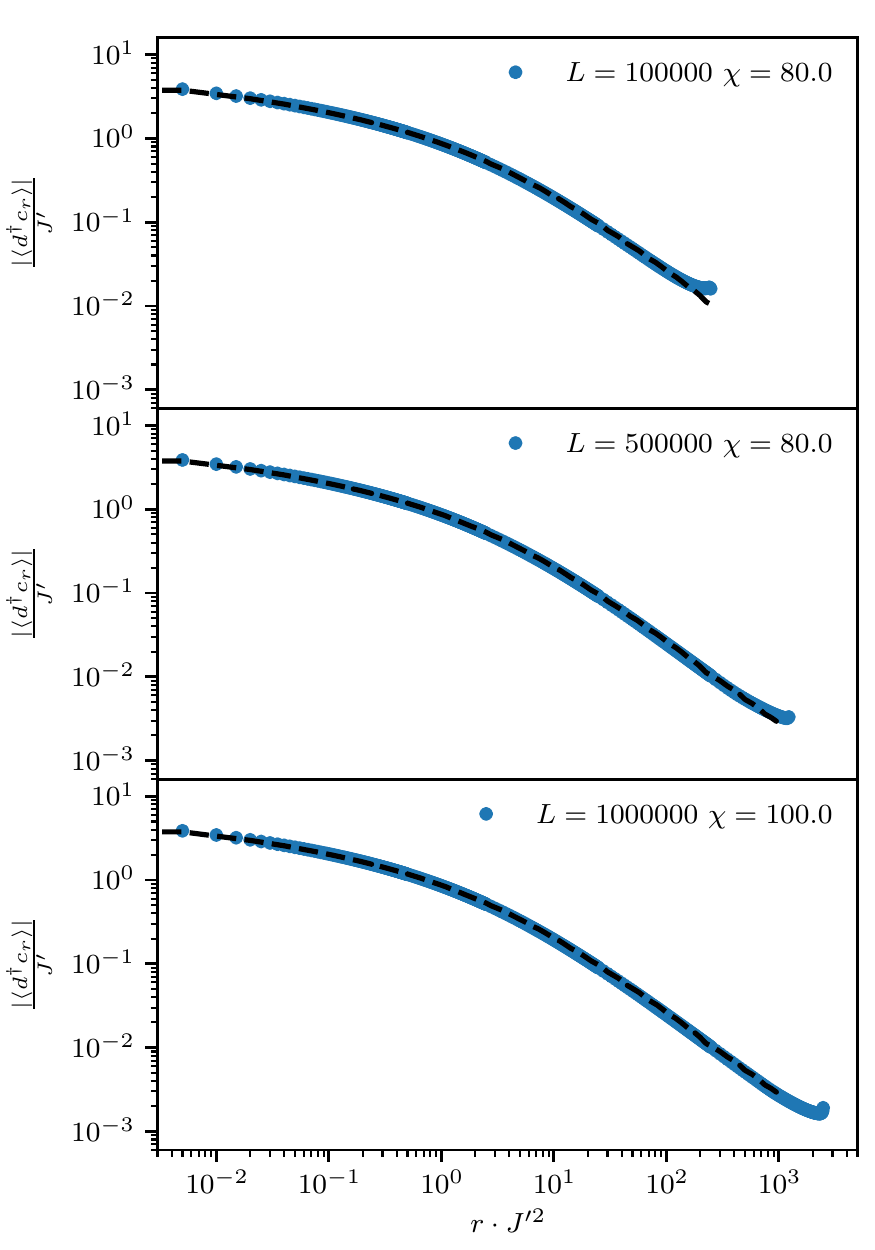}
  \caption{Rescaled real-space correlation function $\langle d^\dagger c_r
\rangle$ of the ground state of Eqn.~\eqref{eqn:HRLM} as a function of
(rescaled) distance $r \cdot J'^2$. Other parameters are $J'=0.05J$ and
the number of sites per block $n_b=\chi/2$ (see
Sec.~\ref{sec:GFDMRG-scaling}). $\chi$ denotes the maximum
Majorana bond number, and $L$ is the size of the system including the
impurity. Dashed lines indicate a fit to the exact
solution~\eqref{eqn:Cr}. \label{fig:RLM} }
\end{figure}

First, we consider a one-dimensional system which still exhibits non-trivial real-space correlation functions:
the resonant-level model, which appears as the non-interacting limit of a number of
impurity models (see, e.g., Refs.~\onlinecite{anderson1961,toulouse1970,schlottmann1980,filyov1981}).
Here, we follow the discussion of Ref.~\onlinecite{ghosh2014}. We consider a fermionic impurity with a corresponding
creation operator $d^\dagger$ coupled to a bath of fermions, which here is represented by a periodic chain of $L$
fermions with associated creation operators $c_i^\dagger$. The Hamiltonian is given by
\begin{eqnarray} \label{eqn:HRLM}
H &=& J' \left( d^\dagger c_1 + c_1^\dagger d \right) + J T_{\rm chain} \\
T_{\rm chain} &=& \sum_i^{L-1} \left( c_i^\dagger c_{i+1} + c_{i+1}^\dagger c_i \right) + \left( c_L^\dagger c_1 + c_1^\dagger c_L \right)
\end{eqnarray}
Here, $T$ is the kinetic energy operator on the chain and $J$ sets the bandwidth; we set the unit of energy to $J=1$ and fix the system to half filling, leaving $J'$ as the only free parameter aside from system size.

The quantity of interest is the correlation between the impurity site and the sites in the bath, $C(r) = \langle d^\dagger c_r \rangle / J'$.
Owing to the non-interacting nature of the problem, this correlation function can be computed exactly in the thermodynamic limit (assuming a constant density of
states around the Fermi energy) 
and at zero temperature is found to be given by
\begin{eqnarray} \label{eqn:Cr}
C(r) &=& A f(B J'^2 r) \\
f(\kappa) &=& -\pi \int_{-\infty}^{\infty} dx \frac{x \cos(\kappa x) + \sin(\kappa x)}{(x^2+1)(1+\delta(x))}.
\end{eqnarray}
The constants $A$ and $B$ are independent of $L$ and $J'$, but depend on the microscopic choice for the bath such as the Fermi velocity and density of states at the Fermi energy,
and are taken as fit parameters here; however, they can be computed exactly for certain choices of the bath~\cite{ghosh2014}.
The relevant universal behavior is that $C(r)$ shows a characteristic crossover from short-distance to long-distance
behavior around $\kappa = B J'^2 r \approx 1$; for $\kappa \ll 1$, $f(\kappa) \sim \ln(\kappa) + \gamma$, while for $\kappa \gg 1$,
$f(\kappa) \sim 1/\kappa$.

In the previous literature, this crossover was illustrated by collapsing many datasets for different values of $J'$, each covering a different
range of $\kappa$. Since we are able to study systems almost three orders of magnitude larger than previously possible, we can exhibit the full scaling form
in a \emph{single} dataset. This is shown in Fig.~\ref{fig:RLM}.
Here, $\chi$ denotes the maximum number of Majorana modes on the bonds of the MPS. This corresponds to a conventional bond dimension
of $D(\chi)=\sqrt{2}^\chi$; for $\chi=80$, this works out to $D \approx 10^{12}$, and for $\chi=100$ to $D \approx 10^{15}$. The dashed
line indicates the exact solution of Eqn.~\eqref{eqn:Cr} with $A$, $B$ chosen as best fit to the MPS data. We see that agreement is essentially
exact up to the largest distances, which are weakly affected by finite-size effects.

\subsection{Quasi-two-dimensional systems}

\begin{figure}
    \centering
    \includegraphics{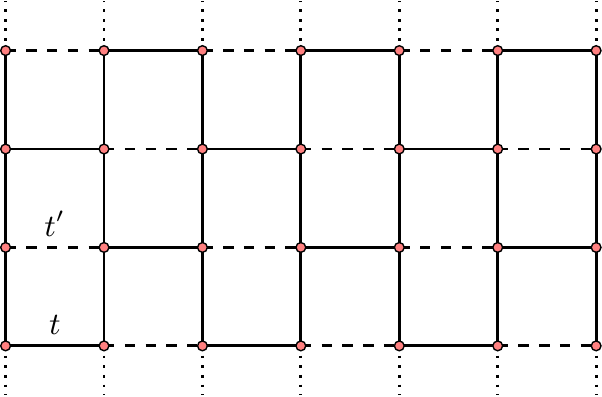}
    \caption{Lattice used to define the 2d Hamiltonian Eqn.~\eqref{eqn:H2d}. Solid lines indicate bonds with strength $t$, and dashed
    lines indicate bonds with strength $t'$. For $t'=t$, the square lattice is recovered, while for $t'=0$ the brickwall representation of the
    honeycomb lattice is obtained. The dotted lines indicate periodic boundary conditions in one direction, effectively wrapping
    the system onto a cylinder.}
    \label{fig:lattice}
\end{figure}

For a more challenging test, we turn to a model of spinless fermions hopping on a two-dimensional lattice,
\begin{equation} \label{eqn:H2d}
H_{\rm hop} = - \sum t_{xy}^{x'y'} \left( c_{xy}^\dagger c_{x'y'} + c_{x'y'}^\dagger c_{xy} \right)
\end{equation}
where $c_{xy}^\dagger$ creates a fermion on the $x$'th ($y$'th) site in the horizontal (vertical) direction, and
the hopping elements $t_{xy}^{x'y'}$ are chosen to give rise to the brickwall lattice as sketched in Fig.~\ref{fig:lattice}.
There are three particular limits of this model that we are interested in: (1) The limit where $t'=t$, where the system simply becomes
the square lattice; (2) the limit of $t'=0$, where it becomes the brickwall lattice, whose connectivity equals that of the honeycomb lattice;
and (3) the limit of $t=0$, where it becomes a lattice of isolated dimers.

From the point of view of a matrix-product state description, the most relevant property of the system is its entanglement structure,
which differs significantly between these three limits. Most easily understood is the case of isolated dimers: in this case, the spectrum
is fully gapped with a gap of $2t$, and there is only entanglement between the sites that together form a dimer, and no entanglement
(or correlations) otherwise. The state thus has a very simple MPS description. Upon adding the $t$ term and interpolating to
the square lattice limit $t=t'$, the gap decreases and finally the system becomes gapless at $t=t'$. Correspondingly, the entanglement
increases. The square lattice hosts a gapless state with one-dimensional Fermi surface. This leads to a logarithmic
violation of the area law~\cite{wolf2006,gioev2006,barthel2006,li2006} and thus represents the most difficult case for a tensor network state approach. Finally, in the
case of the brickwall/honeycomb lattice, the system is gapless, but with Dirac points instead of Fermi surfaces.
In this limit, the entanglement follows an area law despite the gapless nature of the system.

In the following, we will focus on the case of the honeycomb/brickwall lattice, and defer a discussion of the
other cases to Appendix~\ref{app:squarelimits}.
Due to the area law in the honeycomb/brickwall case, the maximum entropy that must be captured in the MPS
scales linearly in its width $W$. We therefore choose $\chi$ proportional to the width of the system $W$.

For the numerical simulations using DMRG methods, we study this lattice on cylinders of circumference $W$ and length $L$, i.e.
with a total of $W \cdot L$ sites. We map the sites to a chain following each rung (of length $W$) from bottom to top. We have found that the
single-site and two-site optimization algorithm generally lead to comparable results, and thus use the single-site algorithm
throughout. We perform at least 5 sweeps and terminate when the absolute change in energy between sweeps drops below $10^{-3}$.

\subsubsection{Convergence of real-space correlation functions}

\begin{figure}
    \centering
    \includegraphics{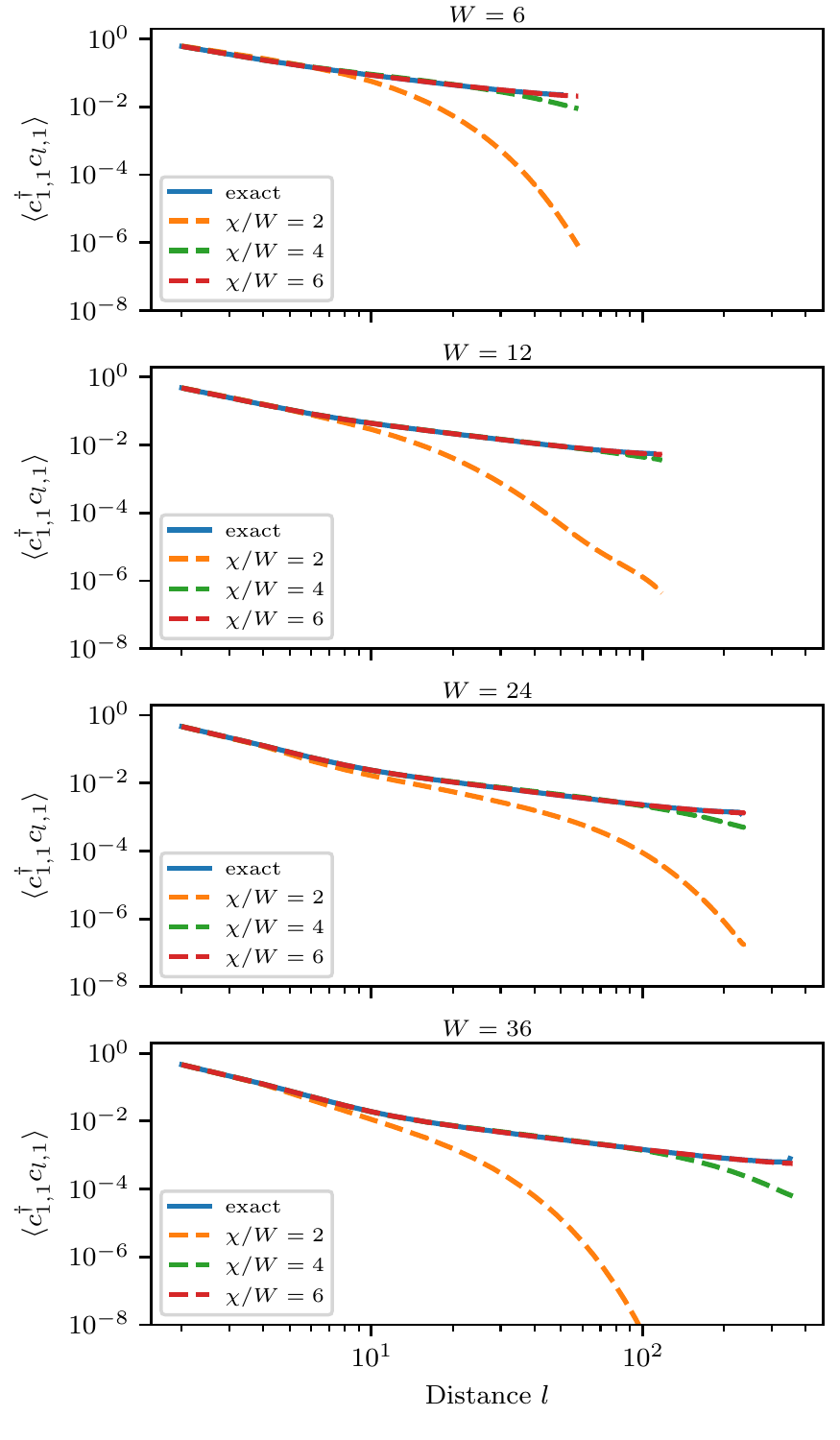}
    \caption{Real-space correlation functions for the ground state of Eqn.~\eqref{eqn:H2d} with $t=1$, $t'=0$.
    Correlations are shown between two sites of the same (odd) vertical position and different horizontal
    positions, i.e. probing the decay of correlations along the cylinder.
    These results are obtained using the single-site optimization algorithm with $W$ sites blocked into one
    for systems of aspect ratio 10, i.e. $L=10W$. Exact results are obtained from a full diagonalization of the
    same system.}
    \label{fig:honeycombcorr}
\end{figure}

To characterize how well the MPS ground state approximates the exact one, we study the real-space
correlations. Fig.~\ref{fig:honeycombcorr} shows the decay of the equal-time Green's function
along the long direction of the cylinder in the limit of a brickwall/honeycomb lattice, $t'=0$.
Here, we scale the Majorana bond number with the width of the cylinder, showing $\chi = 2W,4W,6W$.
This choice is motivated by the area law.

It is well-known that for finite bond dimension, correlations in an MPS decay exponentially at
sufficiently long distances~\cite{fannes1992}. At shorter distances, the correlations can approximate
a polynomial decay. This is clearly borne out in the data of Fig.~\ref{fig:honeycombcorr}: for small
Majorana bond number $\chi$, the correlations exhibit an artificial and unphysical exponential decay, while for
sufficiently large $\chi$ the exact polynomial decay $\mathcal{O}(1/l)$ is recovered to
high accuracy.

\subsubsection{Performance comparison against exact methods}
\label{sec:runtime}

\begin{figure}
    \centering
    \includegraphics{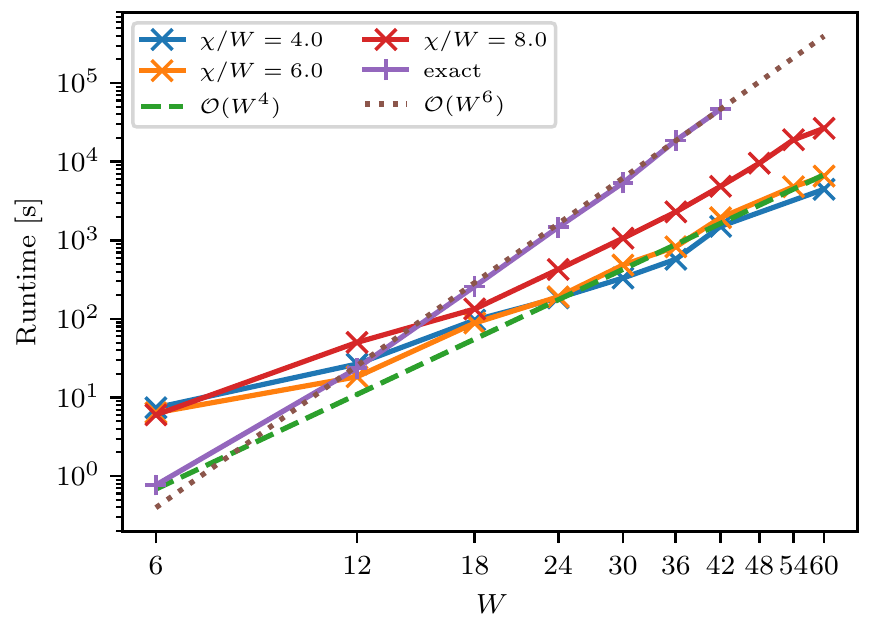}
    \caption{Run time to a converged solution for the ground state of Eqn.~\eqref{eqn:H2d} with $t=1$, $t'=0$
    for aspect ratio $L=20W$, i.e. the largest system has $60^2 \cdot 20 = 72{,}000$ sites. The dashed
    line indicates a best fit of the GFMPS data for $\chi/W=6$ to $aW^4$, with $a$ a fit parameter,
    and the dotted line is a best fit of the exact method to $aW^6$.
    These simulations are performed on 4 cores of an Intel(R) Xeon(R) E5-2690 v3 @ 2.60GHz.}
    \label{fig:scaling}
\end{figure}

We now examine the runtime of the MPS algorithm for quasi-one-dimensional systems compared to
exact, full diagonalization methods. Of course absolute run-times
are implementation-specific and, in particular in the case of MPS methods, may depend sensitively
on details such as the choice of initial states, convergence criterion and many others. More pertinent
is therefore an analysis of the scaling of the method.

Exact methods based on fully diagonalizing the hopping matrix of the underlying problem will scale
as the cube of the matrix size, i.e. in the case at hand as $\mathcal{O}(W^3 L^3)$. Estimating the
cost of the MPS method is more challenging since in general, we cannot make any strong assertions on the
convergence with the number of sweeps~\footnote{See Ref.~\onlinecite{landau2015polynomial} for an
MPS-based variational method where convergence can be guaranteed in certain cases.} or the bond dimension.
However, we can quantify the cost of a single sweep for fixed bond
dimension. As stated in Sec.~\ref{sec:GFDMRG-scaling},
the cost is $(N/n_b) (2\chi+n_b p)^3 D$, where $N$ is the number of sites, $p$ the number of Majorana
modes per site, $n_b$ the number of physical sites that are grouped into one site of the MPS,
$\chi$ the Majorana bond number, and $D$ the number of operators that must be tracked to accomodate
long-range hopping terms, which depends on the details of the Hamiltonian and $n_b$.
For the case of a quasi-1d system at hand, as we have
seen above, to achieve fixed accuracy we scale $\chi \sim W$, and block a number of sites $n_b = \chi/2$
for optimal performance; therefore, $n_b \sim W$. For this case, it turns out that $D$ is a constant
independent of $W$. Therefore, the overall scaling becomes $L W^3$. For the case where we fix
the aspect ratio, $L \sim W$, we thus
obtain a scaling of $\mathcal{O}(W^6)$ for the exact methods and $\mathcal{O}(W^4)$ for the MPS method.

The memory usage of the exact method scales as $(WL)^2$, while the GFMPS approach scales as
$(WL/n_b)(2\chi+n_bp)^2$. For the choices made above ($W \sim L,\chi,n_b$), this leads to the exact
method scaling as $W^4$ and the GFMPS approach scaling as $W^3$. In practice, given our choice
of aspect ratio, the memory requirements of the exact approach are much greater than the GFMPS
approach, and turn out to be the limiting factor.

We numerically confirm the scaling of the runtime in Fig.~\ref{fig:scaling}, where we show the
time to achieve a converged solution instead of the time to perform a single sweep. The GFMPS
approach shows polynomial scaling consistent with $W^4$ across a range of bond dimensions. This
is consistent with the empirical expectation that the number of sweeps scales at most very weakly
with the size of the system.

For the parameters chosen here and our implementation, the crossover where the GFMPS method
becomes faster than exact approaches is around $W=14$. This is most strongly influenced by
the aspect ratio: upon approaching the one-dimensional limit, the GFMPS approach will
become more and more favorable.

\subsection{Transport through a quantum point contact}

\begin{figure}
    \centering
    \includegraphics[width=\columnwidth]{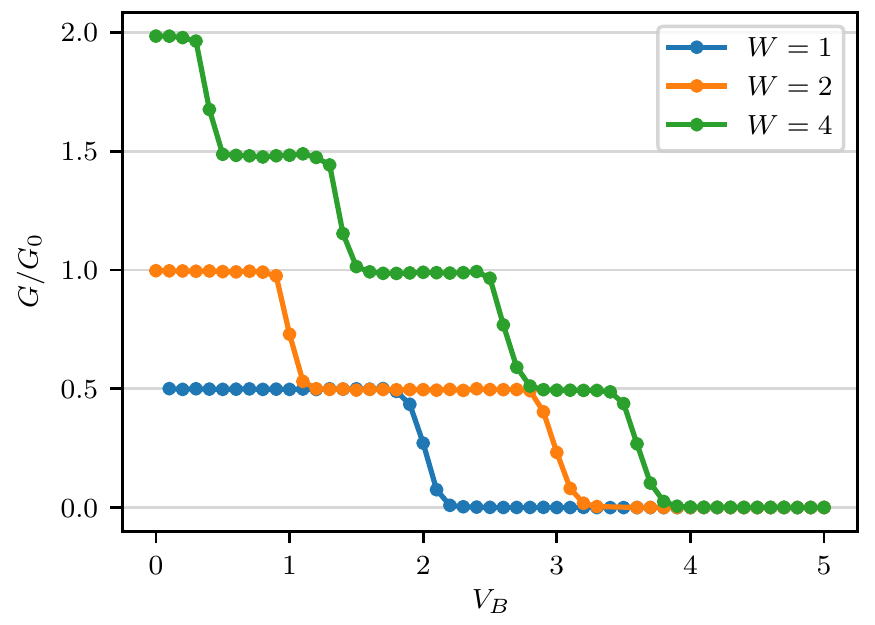}
    \caption{GFMPS simulation of the quantized conductance across a Gaussian potential barrier of height $V_B$ in a $W \times 500$ system. For details, see main text.}
    \label{fig:conductance}
\end{figure}

To illustrate the power of the time-dependent variational principle applied to GFMPS, we study a simple
transport problem, namely the quantization of conductance through a quantum point contact in a multi-channel
wire~\cite{nazarov2009quantum}. We model the system as a square lattice of
spinless fermions of size $W \times L$, i.e.\
as shown in Fig.~\ref{fig:lattice} with hopping Hamiltonian~\eqref{eqn:H2d} for $t'=t=1$ . We add an
on-site potential
\begin{equation}
H_{\rm pot} = \sum \left[ -\mu_0 + V_B \exp\left( -\frac{(x-L/2)^2}{d^2} \right) \right] n_{xy}
\end{equation}
where $n_{xy} = c_{xy}^\dagger c_{xy}$. This corresponds to a Gaussian potential barrier of height $V_B$
and width $d$ at the center of the system in $x$ direction, and independent of $y$.

We initialize the system in the ground state of $H_{\rm hop} + H_{\rm pot}$ using the single-site
DMRG algorithm and then apply a voltage $V$ by evolving under the Hamiltonian
$H_{\rm hop} + H_{\rm pot} + H_{\rm bias}$ with
\begin{equation}
H_{\rm bias} = \frac{V}{2} \sum_y \left( \sum_{x=1}^{L/2-4d} n_{xy} - \sum_{x=L/2+4d}^L n_{xy} \right).
\end{equation}
Here, we use the sites that are more than $4d$ away from the center of the barrier as leads and
apply a symmetric bias voltage, and consider the region between the leads as the quantum point contact.
We then meausure the charge in the left half of the system, $N_{\rm left} = \sum_y \sum_{x=1}^{L/2} c_{xy}^\dagger c_{xy}$,
as a function of time. Due to conservation of charge, we can compute the current flowing across the junction as $J(t)=dN_{\rm left}/dt$.
Alternatively, we could evaluate the charge operator across the junction.

After a transient time of order $d$, the current settles to a plateau value that will last until the charge transport
reaches the end of the leads after a time of order $L$. We compute the average current $\bar{J}$ over this plateau, and define the
conductance as $G = \bar{J}/V$. According to well-known theoretical calculations, in the limit of a smooth barrier the conductance
should be quantized as $G = N G_0/2$, where $N$ is the number of conducting \emph{spin} channels (i.e., for a usual spin-degenerate
system, $N=2K$ with $K$ the number of spinful channels, but here we consider a spinless system) and $G_0 = 2e^2/h$.
We use units where $e = \hbar = 1$, and thus $G_0 = 1/\pi$.

For $V_B=0$, we expect the system to have $W$ conducting channels and thus $G=W G_0/2$; as we increase the barrier height,
the number of channels is reduced one by one as the lowest-momentum channels are cut off first. We confirm this behavior in
Fig.~\ref{fig:conductance} for a system with $W=4$, $L=500$, $d=10$, and applied bias voltage $V=0.1$.
Our simulations are performed with a GFMPS of Majorana bond number $\chi=50$ and with $n_b = W$, and TDVP is run with a timestep of $dt=0.05$.
We also note that since the Hamiltonian is
time-independent after the quench, a time $t$ can be reached in one step for exact simulations while requiring $t/dt$ steps in TDVP;
however, one can easily generalize the TDVP simulation to a time-dependent Hamiltonian, as would arise for example when
computing the ac conductance, without incurring additional computational cost.

\section{Conclusions}

In this paper, we have studied how to apply the DMRG algorithm and other MPS-based
algorithms to the simulation of systems of non-interacting
fermions. By combining the advantages of the exponentially compressed description 
of non-interacting fermionic states in terms of second moments and the
efficient description of states with bounded (area-law) entanglement
through Matrix Product States, we were able to simulate systems far
beyond what is possible with either method alone.  The central insight of
our GFDMRG algorithm is the use of a suitable canonical form, which allows
to express the minimization of the energy as a function of a given tensor 
as an eigenvalue problem, much like in conventional DMRG.  This enabled us
to realize the variational algorithm for non-interacting fermions in close
analogy to the conventional MPS-based formulation of DMRG, as well as to
generalize the method to other MPS-based algorithms such as TDVP for
time evolution or iDMRG for infinite systems. 

The computational cost of the GFDMRG method scales as $L\chi^3$ per sweep,
with $L$ the length of the chain and $\chi$ the number of modes per bond.
Due to the exponential compression of the representation in terms of
second moments, the effective bond dimension is $D=2^{\chi/2}$.  This
allows us to simulate even systems with an amount of entanglement that grows
faster than an area law, such as a Fermi surface or states after a quantum
quench, efficiently and to high accuracy for
very large system sizes.  We have demonstrated the power of our method for
both one-dimensional and quasi-2D systems, and obtained results for up to
one million lattice sites and effective bond dimensions of $10^{15}$.

An interesting follow-up question is how to generalize the method to
two-dimensional systems using Projected Entangled Pair States (PEPS).  A
key point to be addressed is that in our 1D algorithm, the canonical form
serves a more central role than in conventional MPS simulations: While in
the latter case, the canonical form is primarily used to stabilize the
method, in our case it is the key ingredient which allows us to express
the dependence of the energy on a tensor as an eigenvalue problem in the
first place. Therefore, generalizing the method to 2D requires either to
understand how to solve the energy optimization problem without using a
canonical form, or consider more specialized two-dimensional tensor networks
that have a canonical form.

\acknowledgements

NS acknowledges support by the European Research Council (ERC) under the
European Union's Horizon 2020 research and innovation programme through
the ERC Starting Grant WASCOSYS (No.~636201), and by the Deutsche
Forschungsgemeinschaft (DFG) under Germany's Excellence Strategy (EXC-2111
-- 390814868).

\appendix

\section{Real-space correlations in the presence of a Fermi surface}
\label{app:squarelimits}

\begin{figure}
    \centering
    \includegraphics{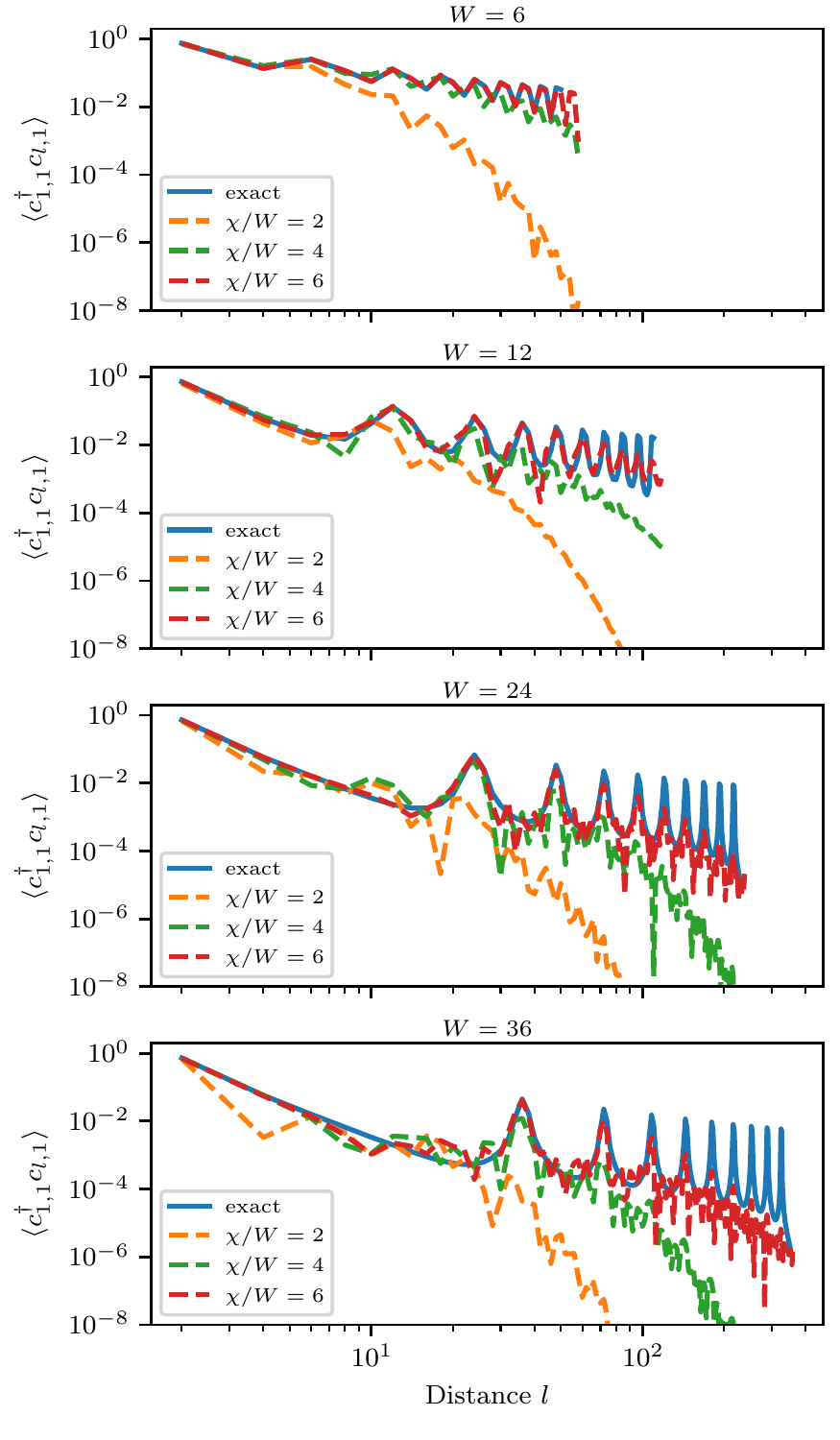}
    \caption{Convergence of real-space correlations functions for the ground
    state of Eq.~\ref{eqn:H2d} in the limit of a square lattice, $t=t'$.}
    \label{fig:squarecorr}
\end{figure}

In Fig.~\ref{fig:squarecorr}, we show the real-space correlation functions, analogous
to Fig.~\ref{fig:honeycombcorr}, for the case of fermions hopping on the square lattice,
i.e. $t=t'$. While these exhibit much more structure than the correlations in the honeycomb
case, which is due to the interplay of finite Fermi momentum $k_F$ and the quantization of momentum
around the periodic direction of the cylinder, they still
have an envelope that decays as a power-law. Importantly,
in contrast to the honeycomb case of Fig.~\ref{fig:honeycombcorr}, the system violates the
area law and thus an accurate description requires a $\chi$ that grows faster
than $W$. This can be seen in the data: while for $W=6$, $\chi/W=6$ is sufficient
to accurately describe the correlations, for $W=36$ the same $\chi/W$ can not capture
the correlations accurately beyond a distance of roughly $10^2$.

\section{Computation of overlaps of GFMPS}
\label{app:overlaps}

\begin{figure}
\includegraphics[width=6cm]{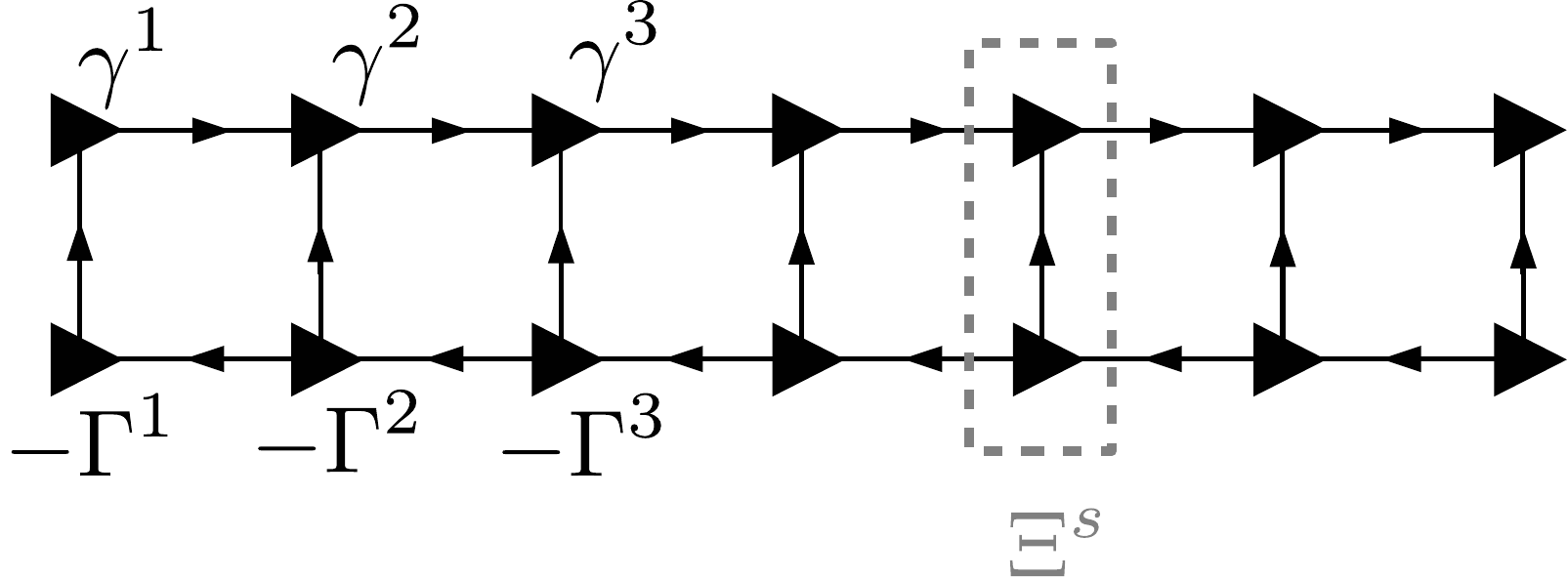}
\caption{Overlap of GFMPS. The overlap is obtained by contracting the GFMPS with tensors $\gamma^s$ and the particle-hole transformed GFMPS with tensors $-\Gamma^s$. The contraction can be carried out efficiently by contracting, say, from left to right, in each step contracting first the upper and then the lower MPS tensor to a left boundary tensor. In each step, the opposite contraction order in the lower part must be observed.}
\label{fig:overlap}
\end{figure}

In this appendix, we describe how to efficiently evaluate the absolute
value squared of the
overlap of two GFMPS. We focus on this instead of the overlap itself since
the phase of the overlap is not fully determined
by a Gaussian covariance matrix; to see this, note that the covariance matrix
of two Gaussian states $\ket{\phi}$ and $e^{i \theta} \ket{\phi}$ is the same.
For a more detailed discussion of this issue and possible solutions, see
Ref.~\onlinecite{Bravyi2017}.

We begin by slightly generalizing the contraction formalism introduced
in Sec.~\ref{sec:contraction}. In particular, Eq.~\eqref{eq:contraction:schur}
can be generalized to the contraction of two indices $b$, $c$ on a given
covariance matrix $\gamma$ even when there is correlation between the two,
i.e. $\gamma_{bc|bc} \neq 0$. This should be contrasted with the case of
Eq.~\eqref{eq:contraction:schur}, where the input covariance matrix is
the direct sum of two parts and the indices being contracted have no correlations.
Starting from the covariance matrix
$\gamma_{abc|abc}$, the CM after contracting $b\triangleright c$ is
given by 
\begin{equation}
\label{eq:schur-contraction-correlated}
K = \gamma_{aa} +
\gamma_{a|bc}\left(\gamma_{bc|bc}+\begin{pmatrix}0&\openone\\-\openone &
0 \end{pmatrix} \right)^{-1}\gamma_{a|bc}^\tr\ .
\end{equation}

Given two states $\rho$ and $\sigma$ with CMs $\gamma_\rho$ and
$\gamma_\sigma$, their overlap is
$\mathrm{tr}(\rho\sigma)=2^{-\chi/2}\sqrt{\det(\openone-\gamma_\rho
\gamma_\sigma)}$, where $\chi$ is the number of Majorana
modes~\cite{bravyi2004}, cf.\ also Eq.~\eqref{eq:overlap-main}. From this we can immediately infer that the
change in normalization induced by contracting two indices
$b\triangleright c$ of a state with CM $\gamma_{abc|abc}$ -- which is
equivalent to projecting $bc$ on a maximally entangled state -- is given by
\begin{equation}
\label{eq:normalization-from-contraction}
\mathcal N = 2^{-\chi/2}\sqrt{\left|\det\left[\gamma_{bc|bc} + 
\left(\begin{smallmatrix}
&\openone\\-\openone\end{smallmatrix}\right)\right]\right|}\
,
\end{equation}
with $\chi=|b|=|c|$ (this can be seen by multiplying with 
$\Big(\begin{smallmatrix}
&\openone\\-\openone\end{smallmatrix}\Big)$ in the determinant); note that
this exactly corresponds to the determinant of the inverse in the Schur complement in
the contraction Eq.~\eqref{eq:schur-contraction-correlated}.
This normalization also holds for the case $|a|=0$, i.e. when all indices are contracted.

In order to compute the overlap of two Gaussian states
with covariance matrices $\gamma$ and $\Gamma$, we can form the joint covariance matrix
$\left(\begin{smallmatrix}\gamma\\&-\Gamma\end{smallmatrix}\right)$ and
contract all indices using Eq.~\eqref{eq:normalization-from-contraction};
note that in that case $\mathcal N$ is independent of the direction of the contraction
(i.e., replacing $\openone\leftrightarrow-\openone$).
The minus sign reflects the fact that contracting as defined here involves
a particle-hole transformation on $\Gamma$ (corresponding
to particle number conservation of the maximally entangled state $\omega$
onto which the bond is projected).
To compute the overlap of two GFMPS -- cf.\ Fig.~\ref{fig:overlap} -- we can perform the same procedure iteratively
by contracting the two GFMPS together from left to right. In each step, the change
in normalization must be tracked according to Eq.~\eqref{eq:normalization-from-contraction}.
Furthermore, the particle-hole transformation on the second MPS requires
to reverse the direction of the contraction of its virtual indices. A convenient way to carry out the procedure is to perform the following contractions:
\begin{enumerate}
    \item $L^{(1)} = (-\Gamma^1) \contr \gamma^1$
    \item $L' = L^{(n)} \contr \gamma^{(n+1)}$
    \item $L^{(n+1)} = (-\Gamma^{(n+1)}) \contr L'$ and return to step 2.
\end{enumerate}
Here, in slight abuse of previous notation, we use the symbol $\contr$ to denote the
(directional) contraction of two tensors (rather than indices), where the correct
indices to contract are specified in Fig.~\ref{fig:overlap}. In each contraction, including
the very last one, the normalization factor $\mathcal N$ is tracked; the
overall normalization of the overlap is then given by the product of all
normalizations. We denote this overlap as
$\mathcal{N}(\gamma^s,\Gamma^s)$.

One can then obtain the normalized overlap of the two states as
\begin{equation}
|\langle\gamma|\Gamma\rangle|^2 = \frac
    {\mathcal{N}(\gamma^s,\Gamma^s)}
    {\sqrt{\mathcal{N}(\gamma^s,\gamma^s) \mathcal{N}(\Gamma^s,\Gamma^s)}}\ .
\end{equation}
The computation of the denominator is greatly simplified by transforming each GFMPS
into the left- or right-canonical form. If the left-canonical gauge is chosen for all tensors,
then $\gamma^s_{rr}=0$, yielding a normalization $2^{-\chi/2}$ for
the contraction of a bond with $\chi$ Majorana modes, and thus 
\begin{equation}
\mathcal{N}(\gamma^s,\gamma^s) = \prod_i 2^{-\chi^\gamma_{i,i+1}/2}
\end{equation}
with $\chi^\gamma_{i,i+1}$ the bond number at link $(i,i+1)$.

\bibliography{gmps}

\begin{thebibliography}{71}%
\makeatletter
\providecommand \@ifxundefined [1]{%
 \@ifx{#1\undefined}
}%
\providecommand \@ifnum [1]{%
 \ifnum #1\expandafter \@firstoftwo
 \else \expandafter \@secondoftwo
 \fi
}%
\providecommand \@ifx [1]{%
 \ifx #1\expandafter \@firstoftwo
 \else \expandafter \@secondoftwo
 \fi
}%
\providecommand \natexlab [1]{#1}%
\providecommand \enquote  [1]{``#1''}%
\providecommand \bibnamefont  [1]{#1}%
\providecommand \bibfnamefont [1]{#1}%
\providecommand \citenamefont [1]{#1}%
\providecommand \href@noop [0]{\@secondoftwo}%
\providecommand \href [0]{\begingroup \@sanitize@url \@href}%
\providecommand \@href[1]{\@@startlink{#1}\@@href}%
\providecommand \@@href[1]{\endgroup#1\@@endlink}%
\providecommand \@sanitize@url [0]{\catcode `\\12\catcode `\$12\catcode
  `\&12\catcode `\#12\catcode `\^12\catcode `\_12\catcode `\%12\relax}%
\providecommand \@@startlink[1]{}%
\providecommand \@@endlink[0]{}%
\providecommand \url  [0]{\begingroup\@sanitize@url \@url }%
\providecommand \@url [1]{\endgroup\@href {#1}{\urlprefix }}%
\providecommand \urlprefix  [0]{URL }%
\providecommand \Eprint [0]{\href }%
\providecommand \doibase [0]{http://dx.doi.org/}%
\providecommand \selectlanguage [0]{\@gobble}%
\providecommand \bibinfo  [0]{\@secondoftwo}%
\providecommand \bibfield  [0]{\@secondoftwo}%
\providecommand \translation [1]{[#1]}%
\providecommand \BibitemOpen [0]{}%
\providecommand \bibitemStop [0]{}%
\providecommand \bibitemNoStop [0]{.\EOS\space}%
\providecommand \EOS [0]{\spacefactor3000\relax}%
\providecommand \BibitemShut  [1]{\csname bibitem#1\endcsname}%
\let\auto@bib@innerbib\@empty
\bibitem [{\citenamefont {Fannes}\ \emph {et~al.}(1992)\citenamefont {Fannes},
  \citenamefont {Nachtergaele},\ and\ \citenamefont {Werner}}]{fannes1992}%
  \BibitemOpen
  \bibfield  {author} {\bibinfo {author} {\bibfnamefont {M.}~\bibnamefont
  {Fannes}}, \bibinfo {author} {\bibfnamefont {B.}~\bibnamefont
  {Nachtergaele}}, \ and\ \bibinfo {author} {\bibfnamefont {R.~F.}\
  \bibnamefont {Werner}},\ }\href {\doibase 10.1007/BF02099178} {\bibfield
  {journal} {\bibinfo  {journal} {Communications in Mathematical Physics}\
  }\textbf {\bibinfo {volume} {144}},\ \bibinfo {pages} {443} (\bibinfo {year}
  {1992})}\BibitemShut {NoStop}%
\bibitem [{\citenamefont {{\"O}stlund}\ and\ \citenamefont
  {Rommer}(1995)}]{ostlund1995}%
  \BibitemOpen
  \bibfield  {author} {\bibinfo {author} {\bibfnamefont {S.}~\bibnamefont
  {{\"O}stlund}}\ and\ \bibinfo {author} {\bibfnamefont {S.}~\bibnamefont
  {Rommer}},\ }\href {\doibase 10.1103/PhysRevLett.75.3537} {\bibfield
  {journal} {\bibinfo  {journal} {Phys. Rev. Lett.}\ }\textbf {\bibinfo
  {volume} {75}},\ \bibinfo {pages} {3537} (\bibinfo {year}
  {1995})}\BibitemShut {NoStop}%
\bibitem [{\citenamefont {White}(1992)}]{white1992}%
  \BibitemOpen
  \bibfield  {author} {\bibinfo {author} {\bibfnamefont {S.~R.}\ \bibnamefont
  {White}},\ }\href {\doibase 10.1103/PhysRevLett.69.2863} {\bibfield
  {journal} {\bibinfo  {journal} {Phys. Rev. Lett.}\ }\textbf {\bibinfo
  {volume} {69}},\ \bibinfo {pages} {2863} (\bibinfo {year}
  {1992})}\BibitemShut {NoStop}%
\bibitem [{\citenamefont {White}\ and\ \citenamefont
  {Noack}(1992)}]{white1992-1}%
  \BibitemOpen
  \bibfield  {author} {\bibinfo {author} {\bibfnamefont {S.~R.}\ \bibnamefont
  {White}}\ and\ \bibinfo {author} {\bibfnamefont {R.~M.}\ \bibnamefont
  {Noack}},\ }\href {\doibase 10.1103/PhysRevLett.68.3487} {\bibfield
  {journal} {\bibinfo  {journal} {Phys. Rev. Lett.}\ }\textbf {\bibinfo
  {volume} {68}},\ \bibinfo {pages} {3487} (\bibinfo {year}
  {1992})}\BibitemShut {NoStop}%
\bibitem [{\citenamefont {Schollw{\"o}ck}(2005)}]{schollwoeck2005}%
  \BibitemOpen
  \bibfield  {author} {\bibinfo {author} {\bibfnamefont {U.}~\bibnamefont
  {Schollw{\"o}ck}},\ }\href {\doibase 10.1103/RevModPhys.77.259} {\bibfield
  {journal} {\bibinfo  {journal} {Rev. Mod. Phys.}\ }\textbf {\bibinfo {volume}
  {77}},\ \bibinfo {pages} {259} (\bibinfo {year} {2005})}\BibitemShut
  {NoStop}%
\bibitem [{\citenamefont {Schollw\"ock}(2011)}]{schollwoeck2011}%
  \BibitemOpen
  \bibfield  {author} {\bibinfo {author} {\bibfnamefont {U.}~\bibnamefont
  {Schollw\"ock}},\ }\href {\doibase 10.1016/j.aop.2010.09.012} {\bibfield
  {journal} {\bibinfo  {journal} {Annals of Physics}\ }\textbf {\bibinfo
  {volume} {326}},\ \bibinfo {pages} {96} (\bibinfo {year} {2011})}\BibitemShut
  {NoStop}%
\bibitem [{\citenamefont {Sierra}\ and\ \citenamefont
  {Martin-Delgado}(1998)}]{sierra1998}%
  \BibitemOpen
  \bibfield  {author} {\bibinfo {author} {\bibfnamefont {G.}~\bibnamefont
  {Sierra}}\ and\ \bibinfo {author} {\bibfnamefont {M.~A.}\ \bibnamefont
  {Martin-Delgado}},\ }\href@noop {} {\bibfield  {journal} {\bibinfo  {journal}
  {Preprint}\ } (\bibinfo {year} {1998})},\ \Eprint
  {http://arxiv.org/abs/cond-mat/9811170} {arXiv:cond-mat/9811170} \BibitemShut
  {NoStop}%
\bibitem [{\citenamefont {Nishino}\ and\ \citenamefont
  {Okunishi}(1998)}]{nishino1998}%
  \BibitemOpen
  \bibfield  {author} {\bibinfo {author} {\bibfnamefont {T.}~\bibnamefont
  {Nishino}}\ and\ \bibinfo {author} {\bibfnamefont {K.}~\bibnamefont
  {Okunishi}},\ }\href {\doibase 10.1143/JPSJ.67.3066} {\bibfield  {journal}
  {\bibinfo  {journal} {Journal of the Physical Society of Japan}\ }\textbf
  {\bibinfo {volume} {67}},\ \bibinfo {pages} {3066} (\bibinfo {year}
  {1998})}\BibitemShut {NoStop}%
\bibitem [{\citenamefont {Nishino}\ \emph {et~al.}(2000)\citenamefont
  {Nishino}, \citenamefont {Okunushi}, \citenamefont {Hieida}, \citenamefont
  {Maeshima},\ and\ \citenamefont {Akutsu}}]{nishino2000}%
  \BibitemOpen
  \bibfield  {author} {\bibinfo {author} {\bibfnamefont {T.}~\bibnamefont
  {Nishino}}, \bibinfo {author} {\bibfnamefont {K.}~\bibnamefont {Okunushi}},
  \bibinfo {author} {\bibfnamefont {Y.}~\bibnamefont {Hieida}}, \bibinfo
  {author} {\bibfnamefont {N.}~\bibnamefont {Maeshima}}, \ and\ \bibinfo
  {author} {\bibfnamefont {Y.}~\bibnamefont {Akutsu}},\ }\href {\doibase
  10.1016/S0550-3213(00)00133-4} {\bibfield  {journal} {\bibinfo  {journal}
  {Nucl. Phys. B}\ }\textbf {\bibinfo {volume} {575}},\ \bibinfo {pages} {504}
  (\bibinfo {year} {2000})}\BibitemShut {NoStop}%
\bibitem [{\citenamefont {Nishio}\ \emph {et~al.}(2004)\citenamefont {Nishio},
  \citenamefont {Maeshima}, \citenamefont {Gendiar},\ and\ \citenamefont
  {Nishino}}]{nishio2004}%
  \BibitemOpen
  \bibfield  {author} {\bibinfo {author} {\bibfnamefont {Y.}~\bibnamefont
  {Nishio}}, \bibinfo {author} {\bibfnamefont {N.}~\bibnamefont {Maeshima}},
  \bibinfo {author} {\bibfnamefont {A.}~\bibnamefont {Gendiar}}, \ and\
  \bibinfo {author} {\bibfnamefont {T.}~\bibnamefont {Nishino}},\ }\href@noop
  {} {\bibfield  {journal} {\bibinfo  {journal} {Preprint}\ } (\bibinfo {year}
  {2004})},\ \Eprint {http://arxiv.org/abs/cond-mat/0401115}
  {arXiv:cond-mat/0401115} \BibitemShut {NoStop}%
\bibitem [{\citenamefont {Verstraete}\ and\ \citenamefont
  {Cirac}(2004)}]{verstraete2004}%
  \BibitemOpen
  \bibfield  {author} {\bibinfo {author} {\bibfnamefont {F.}~\bibnamefont
  {Verstraete}}\ and\ \bibinfo {author} {\bibfnamefont {J.~I.}\ \bibnamefont
  {Cirac}},\ }\href@noop {} {\bibfield  {journal} {\bibinfo  {journal}
  {Preprint}\ } (\bibinfo {year} {2004})},\ \Eprint
  {http://arxiv.org/abs/cond-mat/0407066} {arXiv:cond-mat/0407066} \BibitemShut
  {NoStop}%
\bibitem [{\citenamefont {Jordan}\ \emph {et~al.}(2008)\citenamefont {Jordan},
  \citenamefont {Or\'{u}s}, \citenamefont {Vidal}, \citenamefont {Verstraete},\
  and\ \citenamefont {Cirac}}]{jordan2008}%
  \BibitemOpen
  \bibfield  {author} {\bibinfo {author} {\bibfnamefont {J.}~\bibnamefont
  {Jordan}}, \bibinfo {author} {\bibfnamefont {R.}~\bibnamefont {Or\'{u}s}},
  \bibinfo {author} {\bibfnamefont {G.}~\bibnamefont {Vidal}}, \bibinfo
  {author} {\bibfnamefont {F.}~\bibnamefont {Verstraete}}, \ and\ \bibinfo
  {author} {\bibfnamefont {J.~I.}\ \bibnamefont {Cirac}},\ }\href {\doibase
  10.1103/PhysRevLett.101.250602} {\bibfield  {journal} {\bibinfo  {journal}
  {Phys. Rev. Lett.}\ }\textbf {\bibinfo {volume} {101}},\ \bibinfo {pages}
  {250602} (\bibinfo {year} {2008})}\BibitemShut {NoStop}%
\bibitem [{\citenamefont {Vidal}(2007{\natexlab{a}})}]{vidal2007-1}%
  \BibitemOpen
  \bibfield  {author} {\bibinfo {author} {\bibfnamefont {G.}~\bibnamefont
  {Vidal}},\ }\href {\doibase 10.1103/PhysRevLett.99.220405} {\bibfield
  {journal} {\bibinfo  {journal} {Phys. Rev. Lett.}\ }\textbf {\bibinfo
  {volume} {99}},\ \bibinfo {pages} {220405} (\bibinfo {year}
  {2007}{\natexlab{a}})}\BibitemShut {NoStop}%
\bibitem [{\citenamefont {Vidal}(2008)}]{vidal2008}%
  \BibitemOpen
  \bibfield  {author} {\bibinfo {author} {\bibfnamefont {G.}~\bibnamefont
  {Vidal}},\ }\href {\doibase 10.1103/PhysRevLett.101.110501} {\bibfield
  {journal} {\bibinfo  {journal} {Phys. Rev. Lett.}\ }\textbf {\bibinfo
  {volume} {101}},\ \bibinfo {pages} {110501} (\bibinfo {year}
  {2008})}\BibitemShut {NoStop}%
\bibitem [{\citenamefont {Or{\'u}s}(2014{\natexlab{a}})}]{orus2014}%
  \BibitemOpen
  \bibfield  {author} {\bibinfo {author} {\bibfnamefont {R.}~\bibnamefont
  {Or{\'u}s}},\ }\href {\doibase https://doi.org/10.1016/j.aop.2014.06.013}
  {\bibfield  {journal} {\bibinfo  {journal} {Annals of Physics}\ }\textbf
  {\bibinfo {volume} {349}},\ \bibinfo {pages} {117 } (\bibinfo {year}
  {2014}{\natexlab{a}})}\BibitemShut {NoStop}%
\bibitem [{\citenamefont {Or{\'u}s}(2014{\natexlab{b}})}]{orus2014-1}%
  \BibitemOpen
  \bibfield  {author} {\bibinfo {author} {\bibfnamefont {R.}~\bibnamefont
  {Or{\'u}s}},\ }\href {\doibase 10.1140/epjb/e2014-50502-9} {\bibfield
  {journal} {\bibinfo  {journal} {The European Physical Journal B}\ }\textbf
  {\bibinfo {volume} {87}},\ \bibinfo {pages} {280} (\bibinfo {year}
  {2014}{\natexlab{b}})}\BibitemShut {NoStop}%
\bibitem [{\citenamefont {Bridgeman}\ and\ \citenamefont
  {Chubb}(2017)}]{bridgeman2017}%
  \BibitemOpen
  \bibfield  {author} {\bibinfo {author} {\bibfnamefont {J.~C.}\ \bibnamefont
  {Bridgeman}}\ and\ \bibinfo {author} {\bibfnamefont {C.~T.}\ \bibnamefont
  {Chubb}},\ }\href {\doibase 10.1088/1751-8121/aa6dc3} {\bibfield  {journal}
  {\bibinfo  {journal} {J. Phys. A: Math. Theor.}\ }\textbf {\bibinfo {volume}
  {50}},\ \bibinfo {pages} {223001} (\bibinfo {year} {2017})},\ \Eprint
  {http://arxiv.org/abs/arXiv:1603.03039} {arXiv:1603.03039} \BibitemShut
  {NoStop}%
\bibitem [{\citenamefont {Kraus}\ \emph {et~al.}(2010)\citenamefont {Kraus},
  \citenamefont {Schuch}, \citenamefont {Verstraete},\ and\ \citenamefont
  {Cirac}}]{kraus2010}%
  \BibitemOpen
  \bibfield  {author} {\bibinfo {author} {\bibfnamefont {C.~V.}\ \bibnamefont
  {Kraus}}, \bibinfo {author} {\bibfnamefont {N.}~\bibnamefont {Schuch}},
  \bibinfo {author} {\bibfnamefont {F.}~\bibnamefont {Verstraete}}, \ and\
  \bibinfo {author} {\bibfnamefont {J.~I.}\ \bibnamefont {Cirac}},\ }\href
  {\doibase 10.1103/PhysRevA.81.052338} {\bibfield  {journal} {\bibinfo
  {journal} {Phys. Rev. A}\ }\textbf {\bibinfo {volume} {81}},\ \bibinfo
  {pages} {052338} (\bibinfo {year} {2010})}\BibitemShut {NoStop}%
\bibitem [{\citenamefont {Schuch}\ \emph {et~al.}(2012)\citenamefont {Schuch},
  \citenamefont {Wolf},\ and\ \citenamefont {Cirac}}]{schuch2012}%
  \BibitemOpen
  \bibfield  {author} {\bibinfo {author} {\bibfnamefont {N.}~\bibnamefont
  {Schuch}}, \bibinfo {author} {\bibfnamefont {M.~M.}\ \bibnamefont {Wolf}}, \
  and\ \bibinfo {author} {\bibfnamefont {J.~I.}\ \bibnamefont {Cirac}},\
  }\href@noop {} {\bibfield  {journal} {\bibinfo  {journal} {Preprint}\ }
  (\bibinfo {year} {2012})},\ \Eprint {http://arxiv.org/abs/1201.3945}
  {arXiv:1201.3945} \BibitemShut {NoStop}%
\bibitem [{\citenamefont {Dubail}\ and\ \citenamefont
  {Read}(2015)}]{dubail2015}%
  \BibitemOpen
  \bibfield  {author} {\bibinfo {author} {\bibfnamefont {J.}~\bibnamefont
  {Dubail}}\ and\ \bibinfo {author} {\bibfnamefont {N.}~\bibnamefont {Read}},\
  }\href {\doibase 10.1103/PhysRevB.92.205307} {\bibfield  {journal} {\bibinfo
  {journal} {Phys. Rev. B}\ }\textbf {\bibinfo {volume} {92}},\ \bibinfo
  {pages} {205307} (\bibinfo {year} {2015})}\BibitemShut {NoStop}%
\bibitem [{\citenamefont {Haegeman}\ \emph {et~al.}(2013)\citenamefont
  {Haegeman}, \citenamefont {Osborne}, \citenamefont {Verschelde},\ and\
  \citenamefont {Verstraete}}]{haegeman2013}%
  \BibitemOpen
  \bibfield  {author} {\bibinfo {author} {\bibfnamefont {J.}~\bibnamefont
  {Haegeman}}, \bibinfo {author} {\bibfnamefont {T.~J.}\ \bibnamefont
  {Osborne}}, \bibinfo {author} {\bibfnamefont {H.}~\bibnamefont {Verschelde}},
  \ and\ \bibinfo {author} {\bibfnamefont {F.}~\bibnamefont {Verstraete}},\
  }\href {\doibase 10.1103/PhysRevLett.110.100402} {\bibfield  {journal}
  {\bibinfo  {journal} {Phys. Rev. Lett.}\ }\textbf {\bibinfo {volume} {110}},\
  \bibinfo {pages} {100402} (\bibinfo {year} {2013})}\BibitemShut {NoStop}%
\bibitem [{\citenamefont {Fishman}\ and\ \citenamefont
  {White}(2015)}]{fishman2015}%
  \BibitemOpen
  \bibfield  {author} {\bibinfo {author} {\bibfnamefont {M.~T.}\ \bibnamefont
  {Fishman}}\ and\ \bibinfo {author} {\bibfnamefont {S.~R.}\ \bibnamefont
  {White}},\ }\href {\doibase 10.1103/PhysRevB.92.075132} {\bibfield  {journal}
  {\bibinfo  {journal} {Phys. Rev. B}\ }\textbf {\bibinfo {volume} {92}},\
  \bibinfo {pages} {075132} (\bibinfo {year} {2015})}\BibitemShut {NoStop}%
\bibitem [{\citenamefont {Evenbly}\ and\ \citenamefont
  {White}(2016)}]{evenbly2016}%
  \BibitemOpen
  \bibfield  {author} {\bibinfo {author} {\bibfnamefont {G.}~\bibnamefont
  {Evenbly}}\ and\ \bibinfo {author} {\bibfnamefont {S.~R.}\ \bibnamefont
  {White}},\ }\href {\doibase 10.1103/PhysRevLett.116.140403} {\bibfield
  {journal} {\bibinfo  {journal} {Phys. Rev. Lett.}\ }\textbf {\bibinfo
  {volume} {116}},\ \bibinfo {pages} {140403} (\bibinfo {year}
  {2016})}\BibitemShut {NoStop}%
\bibitem [{\citenamefont {Haegeman}\ \emph {et~al.}(2018)\citenamefont
  {Haegeman}, \citenamefont {Swingle}, \citenamefont {Walter}, \citenamefont
  {Cotler}, \citenamefont {Evenbly},\ and\ \citenamefont
  {Scholz}}]{haegeman2018}%
  \BibitemOpen
  \bibfield  {author} {\bibinfo {author} {\bibfnamefont {J.}~\bibnamefont
  {Haegeman}}, \bibinfo {author} {\bibfnamefont {B.}~\bibnamefont {Swingle}},
  \bibinfo {author} {\bibfnamefont {M.}~\bibnamefont {Walter}}, \bibinfo
  {author} {\bibfnamefont {J.}~\bibnamefont {Cotler}}, \bibinfo {author}
  {\bibfnamefont {G.}~\bibnamefont {Evenbly}}, \ and\ \bibinfo {author}
  {\bibfnamefont {V.~B.}\ \bibnamefont {Scholz}},\ }\href {\doibase
  10.1103/PhysRevX.8.011003} {\bibfield  {journal} {\bibinfo  {journal} {Phys.
  Rev. X}\ }\textbf {\bibinfo {volume} {8}},\ \bibinfo {pages} {011003}
  (\bibinfo {year} {2018})}\BibitemShut {NoStop}%
\bibitem [{\citenamefont {Haegeman}\ \emph {et~al.}(2011)\citenamefont
  {Haegeman}, \citenamefont {Cirac}, \citenamefont {Osborne}, \citenamefont
  {Pi\ifmmode~\check{z}\else \v{z}\fi{}orn}, \citenamefont {Verschelde},\ and\
  \citenamefont {Verstraete}}]{haegeman2011}%
  \BibitemOpen
  \bibfield  {author} {\bibinfo {author} {\bibfnamefont {J.}~\bibnamefont
  {Haegeman}}, \bibinfo {author} {\bibfnamefont {J.~I.}\ \bibnamefont {Cirac}},
  \bibinfo {author} {\bibfnamefont {T.~J.}\ \bibnamefont {Osborne}}, \bibinfo
  {author} {\bibfnamefont {I.}~\bibnamefont {Pi\ifmmode~\check{z}\else
  \v{z}\fi{}orn}}, \bibinfo {author} {\bibfnamefont {H.}~\bibnamefont
  {Verschelde}}, \ and\ \bibinfo {author} {\bibfnamefont {F.}~\bibnamefont
  {Verstraete}},\ }\href {\doibase 10.1103/PhysRevLett.107.070601} {\bibfield
  {journal} {\bibinfo  {journal} {Phys. Rev. Lett.}\ }\textbf {\bibinfo
  {volume} {107}},\ \bibinfo {pages} {070601} (\bibinfo {year}
  {2011})}\BibitemShut {NoStop}%
\bibitem [{\citenamefont {Haegeman}\ \emph {et~al.}(2016)\citenamefont
  {Haegeman}, \citenamefont {Lubich}, \citenamefont {Oseledets}, \citenamefont
  {Vandereycken},\ and\ \citenamefont {Verstraete}}]{haegeman2016}%
  \BibitemOpen
  \bibfield  {author} {\bibinfo {author} {\bibfnamefont {J.}~\bibnamefont
  {Haegeman}}, \bibinfo {author} {\bibfnamefont {C.}~\bibnamefont {Lubich}},
  \bibinfo {author} {\bibfnamefont {I.}~\bibnamefont {Oseledets}}, \bibinfo
  {author} {\bibfnamefont {B.}~\bibnamefont {Vandereycken}}, \ and\ \bibinfo
  {author} {\bibfnamefont {F.}~\bibnamefont {Verstraete}},\ }\href {\doibase
  10.1103/PhysRevB.94.165116} {\bibfield  {journal} {\bibinfo  {journal} {Phys.
  Rev. B}\ }\textbf {\bibinfo {volume} {94}},\ \bibinfo {pages} {165116}
  (\bibinfo {year} {2016})}\BibitemShut {NoStop}%
\bibitem [{\citenamefont {Wei\ss{}e}\ \emph {et~al.}(2006)\citenamefont
  {Wei\ss{}e}, \citenamefont {Wellein}, \citenamefont {Alvermann},\ and\
  \citenamefont {Fehske}}]{weisse2006}%
  \BibitemOpen
  \bibfield  {author} {\bibinfo {author} {\bibfnamefont {A.}~\bibnamefont
  {Wei\ss{}e}}, \bibinfo {author} {\bibfnamefont {G.}~\bibnamefont {Wellein}},
  \bibinfo {author} {\bibfnamefont {A.}~\bibnamefont {Alvermann}}, \ and\
  \bibinfo {author} {\bibfnamefont {H.}~\bibnamefont {Fehske}},\ }\href
  {\doibase 10.1103/RevModPhys.78.275} {\bibfield  {journal} {\bibinfo
  {journal} {Rev. Mod. Phys.}\ }\textbf {\bibinfo {volume} {78}},\ \bibinfo
  {pages} {275} (\bibinfo {year} {2006})}\BibitemShut {NoStop}%
\bibitem [{\citenamefont {Thouless}\ and\ \citenamefont
  {Kirkpatrick}(1981)}]{thouless1981}%
  \BibitemOpen
  \bibfield  {author} {\bibinfo {author} {\bibfnamefont {D.~J.}\ \bibnamefont
  {Thouless}}\ and\ \bibinfo {author} {\bibfnamefont {S.}~\bibnamefont
  {Kirkpatrick}},\ }\href {\doibase 10.1088/0022-3719/14/3/007} {\bibfield
  {journal} {\bibinfo  {journal} {Journal of Physics C: Solid State Physics}\
  }\textbf {\bibinfo {volume} {14}},\ \bibinfo {pages} {235} (\bibinfo {year}
  {1981})}\BibitemShut {NoStop}%
\bibitem [{\citenamefont {Lewenkopf}\ and\ \citenamefont
  {Mucciolo}(2013)}]{lewenkopf2013}%
  \BibitemOpen
  \bibfield  {author} {\bibinfo {author} {\bibfnamefont {C.~H.}\ \bibnamefont
  {Lewenkopf}}\ and\ \bibinfo {author} {\bibfnamefont {E.~R.}\ \bibnamefont
  {Mucciolo}},\ }\href {\doibase 10.1007/s10825-013-0458-7} {\bibfield
  {journal} {\bibinfo  {journal} {Journal of Computational Electronics}\
  }\textbf {\bibinfo {volume} {12}},\ \bibinfo {pages} {203} (\bibinfo {year}
  {2013})}\BibitemShut {NoStop}%
\bibitem [{\citenamefont {Datta}(2000)}]{datta2000nanoscale}%
  \BibitemOpen
  \bibfield  {author} {\bibinfo {author} {\bibfnamefont {S.}~\bibnamefont
  {Datta}},\ }\href {\doibase 10.1006/spmi.2000.0920} {\bibfield  {journal}
  {\bibinfo  {journal} {Superlattices and microstructures}\ }\textbf {\bibinfo
  {volume} {28}},\ \bibinfo {pages} {253} (\bibinfo {year} {2000})}\BibitemShut
  {NoStop}%
\bibitem [{\citenamefont {Gaury}\ \emph {et~al.}(2014)\citenamefont {Gaury},
  \citenamefont {Weston}, \citenamefont {Santin}, \citenamefont {Houzet},
  \citenamefont {Groth},\ and\ \citenamefont {Waintal}}]{gaury2014}%
  \BibitemOpen
  \bibfield  {author} {\bibinfo {author} {\bibfnamefont {B.}~\bibnamefont
  {Gaury}}, \bibinfo {author} {\bibfnamefont {J.}~\bibnamefont {Weston}},
  \bibinfo {author} {\bibfnamefont {M.}~\bibnamefont {Santin}}, \bibinfo
  {author} {\bibfnamefont {M.}~\bibnamefont {Houzet}}, \bibinfo {author}
  {\bibfnamefont {C.}~\bibnamefont {Groth}}, \ and\ \bibinfo {author}
  {\bibfnamefont {X.}~\bibnamefont {Waintal}},\ }\href {\doibase
  https://doi.org/10.1016/j.physrep.2013.09.001} {\bibfield  {journal}
  {\bibinfo  {journal} {Physics Reports}\ }\textbf {\bibinfo {volume} {534}},\
  \bibinfo {pages} {1 } (\bibinfo {year} {2014})},\ \bibinfo {note} {numerical
  simulations of time-resolved quantum electronics}\BibitemShut {NoStop}%
\bibitem [{\citenamefont {Goedecker}(1999)}]{goedecker1999}%
  \BibitemOpen
  \bibfield  {author} {\bibinfo {author} {\bibfnamefont {S.}~\bibnamefont
  {Goedecker}},\ }\href {\doibase 10.1103/RevModPhys.71.1085} {\bibfield
  {journal} {\bibinfo  {journal} {Rev. Mod. Phys.}\ }\textbf {\bibinfo {volume}
  {71}},\ \bibinfo {pages} {1085} (\bibinfo {year} {1999})}\BibitemShut
  {NoStop}%
\bibitem [{\citenamefont {Eisert}\ \emph {et~al.}(2010)\citenamefont {Eisert},
  \citenamefont {Cramer},\ and\ \citenamefont {Plenio}}]{eisert2010}%
  \BibitemOpen
  \bibfield  {author} {\bibinfo {author} {\bibfnamefont {J.}~\bibnamefont
  {Eisert}}, \bibinfo {author} {\bibfnamefont {M.}~\bibnamefont {Cramer}}, \
  and\ \bibinfo {author} {\bibfnamefont {M.~B.}\ \bibnamefont {Plenio}},\
  }\href {\doibase 10.1103/RevModPhys.82.277} {\bibfield  {journal} {\bibinfo
  {journal} {Rev. Mod. Phys.}\ }\textbf {\bibinfo {volume} {82}},\ \bibinfo
  {pages} {277} (\bibinfo {year} {2010})}\BibitemShut {NoStop}%
\bibitem [{\citenamefont {Hastings}(2007{\natexlab{a}})}]{hastings:arealaw}%
  \BibitemOpen
  \bibfield  {author} {\bibinfo {author} {\bibfnamefont {M.}~\bibnamefont
  {Hastings}},\ }\href@noop {} {\bibfield  {journal} {\bibinfo  {journal} {J.
  Stat. Mech.}\ ,\ \bibinfo {pages} {P08024}} (\bibinfo {year}
  {2007}{\natexlab{a}})},\ \Eprint {http://arxiv.org/abs/arXiv:0705.2024}
  {arXiv:0705.2024} \BibitemShut {NoStop}%
\bibitem [{\citenamefont {Arad}\ \emph {et~al.}(2017)\citenamefont {Arad},
  \citenamefont {Landau}, \citenamefont {Vazirani},\ and\ \citenamefont
  {Vidick}}]{arad:rg-algorithms-and-area-laws}%
  \BibitemOpen
  \bibfield  {author} {\bibinfo {author} {\bibfnamefont {I.}~\bibnamefont
  {Arad}}, \bibinfo {author} {\bibfnamefont {Z.}~\bibnamefont {Landau}},
  \bibinfo {author} {\bibfnamefont {U.}~\bibnamefont {Vazirani}}, \ and\
  \bibinfo {author} {\bibfnamefont {T.}~\bibnamefont {Vidick}},\ }\href@noop {}
  {\bibfield  {journal} {\bibinfo  {journal} {Commun. Math. Phys.}\ }\textbf
  {\bibinfo {volume} {356}},\ \bibinfo {pages} {65} (\bibinfo {year} {2017})},\
  \Eprint {http://arxiv.org/abs/arXiv:1602.08828} {arXiv:1602.08828}
  \BibitemShut {NoStop}%
\bibitem [{\citenamefont {{Huang}}(2014)}]{huang:area-law}%
  \BibitemOpen
  \bibfield  {author} {\bibinfo {author} {\bibfnamefont {Y.}~\bibnamefont
  {{Huang}}},\ }\href@noop {} {\  (\bibinfo {year} {2014})},\ \Eprint
  {http://arxiv.org/abs/arXiv:1403.0327} {arXiv:1403.0327} \BibitemShut
  {NoStop}%
\bibitem [{\citenamefont {White}(1998)}]{white1998}%
  \BibitemOpen
  \bibfield  {author} {\bibinfo {author} {\bibfnamefont {S.~R.}\ \bibnamefont
  {White}},\ }\href {\doibase https://doi.org/10.1016/S0370-1573(98)00010-6}
  {\bibfield  {journal} {\bibinfo  {journal} {Physics Reports}\ }\textbf
  {\bibinfo {volume} {301}},\ \bibinfo {pages} {187 } (\bibinfo {year}
  {1998})}\BibitemShut {NoStop}%
\bibitem [{\citenamefont {Verstraete}\ and\ \citenamefont
  {Cirac}(2006)}]{verstraete2006}%
  \BibitemOpen
  \bibfield  {author} {\bibinfo {author} {\bibfnamefont {F.}~\bibnamefont
  {Verstraete}}\ and\ \bibinfo {author} {\bibfnamefont {J.~I.}\ \bibnamefont
  {Cirac}},\ }\href {\doibase 10.1103/PhysRevB.73.094423} {\bibfield  {journal}
  {\bibinfo  {journal} {Phys. Rev. B}\ }\textbf {\bibinfo {volume} {73}},\
  \bibinfo {pages} {094423} (\bibinfo {year} {2006})}\BibitemShut {NoStop}%
\bibitem [{\citenamefont {Hastings}(2007{\natexlab{b}})}]{hastings2007}%
  \BibitemOpen
  \bibfield  {author} {\bibinfo {author} {\bibfnamefont {M.~B.}\ \bibnamefont
  {Hastings}},\ }\href {\doibase 10.1103/PhysRevB.76.035114} {\bibfield
  {journal} {\bibinfo  {journal} {Phys. Rev. B}\ }\textbf {\bibinfo {volume}
  {76}},\ \bibinfo {pages} {035114} (\bibinfo {year}
  {2007}{\natexlab{b}})}\BibitemShut {NoStop}%
\bibitem [{\citenamefont {Schuch}\ \emph {et~al.}(2008)\citenamefont {Schuch},
  \citenamefont {Wolf}, \citenamefont {Verstraete},\ and\ \citenamefont
  {Cirac}}]{schuch2008}%
  \BibitemOpen
  \bibfield  {author} {\bibinfo {author} {\bibfnamefont {N.}~\bibnamefont
  {Schuch}}, \bibinfo {author} {\bibfnamefont {M.~M.}\ \bibnamefont {Wolf}},
  \bibinfo {author} {\bibfnamefont {F.}~\bibnamefont {Verstraete}}, \ and\
  \bibinfo {author} {\bibfnamefont {J.~I.}\ \bibnamefont {Cirac}},\ }\href
  {\doibase 10.1103/PhysRevLett.100.030504} {\bibfield  {journal} {\bibinfo
  {journal} {Phys. Rev. Lett.}\ }\textbf {\bibinfo {volume} {100}},\ \bibinfo
  {pages} {030504} (\bibinfo {year} {2008})}\BibitemShut {NoStop}%
\bibitem [{\citenamefont {Botero}\ and\ \citenamefont
  {Reznik}(2003)}]{botero2003}%
  \BibitemOpen
  \bibfield  {author} {\bibinfo {author} {\bibfnamefont {A.}~\bibnamefont
  {Botero}}\ and\ \bibinfo {author} {\bibfnamefont {B.}~\bibnamefont
  {Reznik}},\ }\href {\doibase 10.1103/PhysRevA.67.052311} {\bibfield
  {journal} {\bibinfo  {journal} {Phys. Rev. A}\ }\textbf {\bibinfo {volume}
  {67}},\ \bibinfo {pages} {052311} (\bibinfo {year} {2003})}\BibitemShut
  {NoStop}%
\bibitem [{\citenamefont {Vidal}\ \emph {et~al.}(2003)\citenamefont {Vidal},
  \citenamefont {Latorre}, \citenamefont {Rico},\ and\ \citenamefont
  {Kitaev}}]{vidal2003}%
  \BibitemOpen
  \bibfield  {author} {\bibinfo {author} {\bibfnamefont {G.}~\bibnamefont
  {Vidal}}, \bibinfo {author} {\bibfnamefont {J.~I.}\ \bibnamefont {Latorre}},
  \bibinfo {author} {\bibfnamefont {E.}~\bibnamefont {Rico}}, \ and\ \bibinfo
  {author} {\bibfnamefont {A.}~\bibnamefont {Kitaev}},\ }\href {\doibase
  10.1103/PhysRevLett.90.227902} {\bibfield  {journal} {\bibinfo  {journal}
  {Phys. Rev. Lett.}\ }\textbf {\bibinfo {volume} {90}},\ \bibinfo {pages}
  {227902} (\bibinfo {year} {2003})}\BibitemShut {NoStop}%
\bibitem [{\citenamefont {Peschel}(2003)}]{peschel2003}%
  \BibitemOpen
  \bibfield  {author} {\bibinfo {author} {\bibfnamefont {I.}~\bibnamefont
  {Peschel}},\ }\href {http://stacks.iop.org/0305-4470/36/i=14/a=101}
  {\bibfield  {journal} {\bibinfo  {journal} {Journal of Physics A:
  Mathematical and General}\ }\textbf {\bibinfo {volume} {36}},\ \bibinfo
  {pages} {L205} (\bibinfo {year} {2003})}\BibitemShut {NoStop}%
\bibitem [{\citenamefont {Botero}\ and\ \citenamefont
  {Reznik}(2004)}]{botero2004}%
  \BibitemOpen
  \bibfield  {author} {\bibinfo {author} {\bibfnamefont {A.}~\bibnamefont
  {Botero}}\ and\ \bibinfo {author} {\bibfnamefont {B.}~\bibnamefont
  {Reznik}},\ }\href {\doibase https://doi.org/10.1016/j.physleta.2004.08.037}
  {\bibfield  {journal} {\bibinfo  {journal} {Physics Letters A}\ }\textbf
  {\bibinfo {volume} {331}},\ \bibinfo {pages} {39 } (\bibinfo {year}
  {2004})}\BibitemShut {NoStop}%
\bibitem [{\citenamefont {Bravyi}\ and\ \citenamefont
  {Gosset}(2017)}]{Bravyi2017}%
  \BibitemOpen
  \bibfield  {author} {\bibinfo {author} {\bibfnamefont {S.}~\bibnamefont
  {Bravyi}}\ and\ \bibinfo {author} {\bibfnamefont {D.}~\bibnamefont
  {Gosset}},\ }\href {\doibase 10.1007/s00220-017-2976-9} {\bibfield  {journal}
  {\bibinfo  {journal} {Communications in Mathematical Physics}\ }\textbf
  {\bibinfo {volume} {356}},\ \bibinfo {pages} {451} (\bibinfo {year}
  {2017})}\BibitemShut {NoStop}%
\bibitem [{\citenamefont {Bravyi}(2004)}]{bravyi2004}%
  \BibitemOpen
  \bibfield  {author} {\bibinfo {author} {\bibfnamefont {S.}~\bibnamefont
  {Bravyi}},\ }\href@noop {} {\bibfield  {journal} {\bibinfo  {journal}
  {Preprint}\ } (\bibinfo {year} {2004})},\ \Eprint
  {http://arxiv.org/abs/quant-ph/0404180} {arXiv:quant-ph/0404180} \BibitemShut
  {NoStop}%
\bibitem [{\citenamefont {Wimmer}(2012)}]{Wimmer2012}%
  \BibitemOpen
  \bibfield  {author} {\bibinfo {author} {\bibfnamefont {M.}~\bibnamefont
  {Wimmer}},\ }\href {\doibase 10.1145/2331130.2331138} {\bibfield  {journal}
  {\bibinfo  {journal} {ACM Transactions on Mathematical Software (TOMS)}\
  }\textbf {\bibinfo {volume} {38}},\ \bibinfo {pages} {30} (\bibinfo {year}
  {2012})}\BibitemShut {NoStop}%
\bibitem [{\citenamefont {L\"owdin}(1955)}]{lowdin1955}%
  \BibitemOpen
  \bibfield  {author} {\bibinfo {author} {\bibfnamefont {P.-O.}\ \bibnamefont
  {L\"owdin}},\ }\href {\doibase 10.1103/PhysRev.97.1474} {\bibfield  {journal}
  {\bibinfo  {journal} {Phys. Rev.}\ }\textbf {\bibinfo {volume} {97}},\
  \bibinfo {pages} {1474} (\bibinfo {year} {1955})}\BibitemShut {NoStop}%
\bibitem [{\citenamefont {Giedke}\ and\ \citenamefont
  {Cirac}(2002)}]{Giedke2002}%
  \BibitemOpen
  \bibfield  {author} {\bibinfo {author} {\bibfnamefont {G.}~\bibnamefont
  {Giedke}}\ and\ \bibinfo {author} {\bibfnamefont {J.~I.}\ \bibnamefont
  {Cirac}},\ }\href@noop {} {\bibfield  {journal} {\bibinfo  {journal} {Phys.\
  Rev.\ A}\ }\textbf {\bibinfo {volume} {66}},\ \bibinfo {pages} {032316}
  (\bibinfo {year} {2002})},\ \Eprint {http://arxiv.org/abs/quant-ph/0204085}
  {quant-ph/0204085} \BibitemShut {NoStop}%
\bibitem [{\citenamefont {Eisert}\ \emph {et~al.}(2002)\citenamefont {Eisert},
  \citenamefont {Scheel},\ and\ \citenamefont {Plenio}}]{Eisert2002}%
  \BibitemOpen
  \bibfield  {author} {\bibinfo {author} {\bibfnamefont {J.}~\bibnamefont
  {Eisert}}, \bibinfo {author} {\bibfnamefont {S.}~\bibnamefont {Scheel}}, \
  and\ \bibinfo {author} {\bibfnamefont {M.~B.}\ \bibnamefont {Plenio}},\
  }\href@noop {} {\bibfield  {journal} {\bibinfo  {journal} {Phys. Rev. Lett.}\
  }\textbf {\bibinfo {volume} {89}},\ \bibinfo {pages} {137903} (\bibinfo
  {year} {2002})},\ \Eprint {http://arxiv.org/abs/quant-ph/0204052}
  {quant-ph/0204052} \BibitemShut {NoStop}%
\bibitem [{\citenamefont {Fiur{\'a\v s}ek}(2002)}]{fiurasek2002}%
  \BibitemOpen
  \bibfield  {author} {\bibinfo {author} {\bibfnamefont {J.}~\bibnamefont
  {Fiur{\'a\v s}ek}},\ }\href@noop {} {\bibfield  {journal} {\bibinfo
  {journal} {Phys. Rev. Lett.}\ }\textbf {\bibinfo {volume} {89}},\ \bibinfo
  {pages} {137904} (\bibinfo {year} {2002})},\ \Eprint
  {http://arxiv.org/abs/quant-ph/0204069} {quant-ph/0204069} \BibitemShut
  {NoStop}%
\bibitem [{Note1()}]{Note1}%
  \BibitemOpen
  \bibinfo {note} {This can also be seen without calculation by noting that
  this amounts to a postselected teleportation protocol (i.e.\ attaching a
  maximally entangled state and projecting onto the maximally entangled
  state).}\BibitemShut {Stop}%
\bibitem [{\citenamefont {White}(2005)}]{white2005}%
  \BibitemOpen
  \bibfield  {author} {\bibinfo {author} {\bibfnamefont {S.~R.}\ \bibnamefont
  {White}},\ }\href {\doibase 10.1103/PhysRevB.72.180403} {\bibfield  {journal}
  {\bibinfo  {journal} {Phys. Rev. B}\ }\textbf {\bibinfo {volume} {72}},\
  \bibinfo {pages} {180403} (\bibinfo {year} {2005})}\BibitemShut {NoStop}%
\bibitem [{Note2()}]{Note2}%
  \BibitemOpen
  \bibinfo {note} {This can be shown by observing that Eq.~\protect \textup
  {\hbox {\mathsurround \z@ \protect \normalfont (\ignorespaces \ref
  {eq:time-evol-cm}\unskip \@@italiccorr )}} is equivalent to $c_k(t)=\DOTSB
  \sum@ \slimits@ O(t)_{kl} c_l(0)$ and thus $\protect \mathaccentV {dot}05Fc_k
  = 4\DOTSB \sum@ \slimits@ H_{kl} c_l$, and on the other hand comparing it
  with the evolution $\protect \mathaccentV {dot}05Fc_k = \protect \mathrm
  {i}[\protect \mathcal H,c_k]$ of $c_k$ in the Heisenberg
  picture.}\BibitemShut {Stop}%
\bibitem [{\citenamefont {Kraus}\ and\ \citenamefont
  {Cirac}(2010)}]{kraus2010-ghf}%
  \BibitemOpen
  \bibfield  {author} {\bibinfo {author} {\bibfnamefont {C.~V.}\ \bibnamefont
  {Kraus}}\ and\ \bibinfo {author} {\bibfnamefont {J.~I.}\ \bibnamefont
  {Cirac}},\ }\href {\doibase 10.1088/1367-2630/12/11/113004} {\bibfield
  {journal} {\bibinfo  {journal} {New Journal of Physics}\ }\textbf {\bibinfo
  {volume} {12}},\ \bibinfo {pages} {113004} (\bibinfo {year}
  {2010})}\BibitemShut {NoStop}%
\bibitem [{\citenamefont {McCulloch}(2008)}]{mcculloch2008infinite}%
  \BibitemOpen
  \bibfield  {author} {\bibinfo {author} {\bibfnamefont {I.~P.}\ \bibnamefont
  {McCulloch}},\ }\href@noop {} {\bibfield  {journal} {\bibinfo  {journal}
  {Preprint}\ } (\bibinfo {year} {2008})},\ \Eprint
  {http://arxiv.org/abs/0804.2509} {arXiv:0804.2509} \BibitemShut {NoStop}%
\bibitem [{\citenamefont {Crosswhite}\ \emph {et~al.}(2008)\citenamefont
  {Crosswhite}, \citenamefont {Doherty},\ and\ \citenamefont
  {Vidal}}]{crosswhite2008}%
  \BibitemOpen
  \bibfield  {author} {\bibinfo {author} {\bibfnamefont {G.~M.}\ \bibnamefont
  {Crosswhite}}, \bibinfo {author} {\bibfnamefont {A.~C.}\ \bibnamefont
  {Doherty}}, \ and\ \bibinfo {author} {\bibfnamefont {G.}~\bibnamefont
  {Vidal}},\ }\href {\doibase 10.1103/PhysRevB.78.035116} {\bibfield  {journal}
  {\bibinfo  {journal} {Phys. Rev. B}\ }\textbf {\bibinfo {volume} {78}},\
  \bibinfo {pages} {035116} (\bibinfo {year} {2008})}\BibitemShut {NoStop}%
\bibitem [{\citenamefont {Vanderstraeten}\ \emph {et~al.}(2019)\citenamefont
  {Vanderstraeten}, \citenamefont {Haegeman},\ and\ \citenamefont
  {Verstraete}}]{vanderstraeten2019}%
  \BibitemOpen
  \bibfield  {author} {\bibinfo {author} {\bibfnamefont {L.}~\bibnamefont
  {Vanderstraeten}}, \bibinfo {author} {\bibfnamefont {J.}~\bibnamefont
  {Haegeman}}, \ and\ \bibinfo {author} {\bibfnamefont {F.}~\bibnamefont
  {Verstraete}},\ }\href {\doibase 10.21468/SciPostPhysLectNotes.7} {\bibfield
  {journal} {\bibinfo  {journal} {SciPost Phys. Lect. Notes}\ ,\ \bibinfo
  {pages} {7}} (\bibinfo {year} {2019})}\BibitemShut {NoStop}%
\bibitem [{\citenamefont {Vidal}(2007{\natexlab{b}})}]{vidal2007tebd}%
  \BibitemOpen
  \bibfield  {author} {\bibinfo {author} {\bibfnamefont {G.}~\bibnamefont
  {Vidal}},\ }\href {\doibase 10.1103/PhysRevLett.98.070201} {\bibfield
  {journal} {\bibinfo  {journal} {Phys. Rev. Lett.}\ }\textbf {\bibinfo
  {volume} {98}},\ \bibinfo {pages} {070201} (\bibinfo {year}
  {2007}{\natexlab{b}})}\BibitemShut {NoStop}%
\bibitem [{\citenamefont {Anderson}(1961)}]{anderson1961}%
  \BibitemOpen
  \bibfield  {author} {\bibinfo {author} {\bibfnamefont {P.~W.}\ \bibnamefont
  {Anderson}},\ }\href {\doibase 10.1103/PhysRev.124.41} {\bibfield  {journal}
  {\bibinfo  {journal} {Phys. Rev.}\ }\textbf {\bibinfo {volume} {124}},\
  \bibinfo {pages} {41} (\bibinfo {year} {1961})}\BibitemShut {NoStop}%
\bibitem [{\citenamefont {Toulouse}(1970)}]{toulouse1970}%
  \BibitemOpen
  \bibfield  {author} {\bibinfo {author} {\bibfnamefont {G.}~\bibnamefont
  {Toulouse}},\ }\href {\doibase 10.1103/PhysRevB.2.270} {\bibfield  {journal}
  {\bibinfo  {journal} {Phys. Rev. B}\ }\textbf {\bibinfo {volume} {2}},\
  \bibinfo {pages} {270} (\bibinfo {year} {1970})}\BibitemShut {NoStop}%
\bibitem [{\citenamefont {Schlottmann}(1980)}]{schlottmann1980}%
  \BibitemOpen
  \bibfield  {author} {\bibinfo {author} {\bibfnamefont {P.}~\bibnamefont
  {Schlottmann}},\ }\href {\doibase 10.1103/PhysRevB.22.613} {\bibfield
  {journal} {\bibinfo  {journal} {Phys. Rev. B}\ }\textbf {\bibinfo {volume}
  {22}},\ \bibinfo {pages} {613} (\bibinfo {year} {1980})}\BibitemShut
  {NoStop}%
\bibitem [{\citenamefont {Filyov}\ \emph {et~al.}(1981)\citenamefont {Filyov},
  \citenamefont {Tzvelik},\ and\ \citenamefont {Wiegmann}}]{filyov1981}%
  \BibitemOpen
  \bibfield  {author} {\bibinfo {author} {\bibfnamefont {V.}~\bibnamefont
  {Filyov}}, \bibinfo {author} {\bibfnamefont {A.}~\bibnamefont {Tzvelik}}, \
  and\ \bibinfo {author} {\bibfnamefont {P.}~\bibnamefont {Wiegmann}},\ }\href
  {\doibase https://doi.org/10.1016/0375-9601(81)90055-4} {\bibfield  {journal}
  {\bibinfo  {journal} {Physics Letters A}\ }\textbf {\bibinfo {volume} {81}},\
  \bibinfo {pages} {175 } (\bibinfo {year} {1981})}\BibitemShut {NoStop}%
\bibitem [{\citenamefont {Ghosh}\ \emph {et~al.}(2014)\citenamefont {Ghosh},
  \citenamefont {Ribeiro},\ and\ \citenamefont {Haque}}]{ghosh2014}%
  \BibitemOpen
  \bibfield  {author} {\bibinfo {author} {\bibfnamefont {S.}~\bibnamefont
  {Ghosh}}, \bibinfo {author} {\bibfnamefont {P.}~\bibnamefont {Ribeiro}}, \
  and\ \bibinfo {author} {\bibfnamefont {M.}~\bibnamefont {Haque}},\ }\href
  {http://stacks.iop.org/1742-5468/2014/i=4/a=P04011} {\bibfield  {journal}
  {\bibinfo  {journal} {Journal of Statistical Mechanics: Theory and
  Experiment}\ }\textbf {\bibinfo {volume} {2014}},\ \bibinfo {pages} {P04011}
  (\bibinfo {year} {2014})}\BibitemShut {NoStop}%
\bibitem [{\citenamefont {Wolf}(2006)}]{wolf2006}%
  \BibitemOpen
  \bibfield  {author} {\bibinfo {author} {\bibfnamefont {M.~M.}\ \bibnamefont
  {Wolf}},\ }\href {\doibase 10.1103/PhysRevLett.96.010404} {\bibfield
  {journal} {\bibinfo  {journal} {Phys. Rev. Lett.}\ }\textbf {\bibinfo
  {volume} {96}},\ \bibinfo {pages} {010404} (\bibinfo {year}
  {2006})}\BibitemShut {NoStop}%
\bibitem [{\citenamefont {Gioev}\ and\ \citenamefont
  {Klich}(2006)}]{gioev2006}%
  \BibitemOpen
  \bibfield  {author} {\bibinfo {author} {\bibfnamefont {D.}~\bibnamefont
  {Gioev}}\ and\ \bibinfo {author} {\bibfnamefont {I.}~\bibnamefont {Klich}},\
  }\href {\doibase 10.1103/PhysRevLett.96.100503} {\bibfield  {journal}
  {\bibinfo  {journal} {Phys. Rev. Lett.}\ }\textbf {\bibinfo {volume} {96}},\
  \bibinfo {pages} {100503} (\bibinfo {year} {2006})}\BibitemShut {NoStop}%
\bibitem [{\citenamefont {Barthel}\ \emph {et~al.}(2006)\citenamefont
  {Barthel}, \citenamefont {Chung},\ and\ \citenamefont
  {Schollw\"ock}}]{barthel2006}%
  \BibitemOpen
  \bibfield  {author} {\bibinfo {author} {\bibfnamefont {T.}~\bibnamefont
  {Barthel}}, \bibinfo {author} {\bibfnamefont {M.-C.}\ \bibnamefont {Chung}},
  \ and\ \bibinfo {author} {\bibfnamefont {U.}~\bibnamefont {Schollw\"ock}},\
  }\href {\doibase 10.1103/PhysRevA.74.022329} {\bibfield  {journal} {\bibinfo
  {journal} {Phys. Rev. A}\ }\textbf {\bibinfo {volume} {74}},\ \bibinfo
  {pages} {022329} (\bibinfo {year} {2006})}\BibitemShut {NoStop}%
\bibitem [{\citenamefont {Li}\ \emph {et~al.}(2006)\citenamefont {Li},
  \citenamefont {Ding}, \citenamefont {Yu}, \citenamefont {Roscilde},\ and\
  \citenamefont {Haas}}]{li2006}%
  \BibitemOpen
  \bibfield  {author} {\bibinfo {author} {\bibfnamefont {W.}~\bibnamefont
  {Li}}, \bibinfo {author} {\bibfnamefont {L.}~\bibnamefont {Ding}}, \bibinfo
  {author} {\bibfnamefont {R.}~\bibnamefont {Yu}}, \bibinfo {author}
  {\bibfnamefont {T.}~\bibnamefont {Roscilde}}, \ and\ \bibinfo {author}
  {\bibfnamefont {S.}~\bibnamefont {Haas}},\ }\href {\doibase
  10.1103/PhysRevB.74.073103} {\bibfield  {journal} {\bibinfo  {journal} {Phys.
  Rev. B}\ }\textbf {\bibinfo {volume} {74}},\ \bibinfo {pages} {073103}
  (\bibinfo {year} {2006})}\BibitemShut {NoStop}%
\bibitem [{Note3()}]{Note3}%
  \BibitemOpen
  \bibinfo {note} {See Ref.~\protect \rev@citealpnum {landau2015polynomial} for
  an MPS-based variational method where convergence can be guaranteed in
  certain cases.}\BibitemShut {Stop}%
\bibitem [{\citenamefont {Nazarov}\ \emph {et~al.}(2009)\citenamefont
  {Nazarov}, \citenamefont {Nazarov},\ and\ \citenamefont
  {Blanter}}]{nazarov2009quantum}%
  \BibitemOpen
  \bibfield  {author} {\bibinfo {author} {\bibfnamefont {Y.~V.}\ \bibnamefont
  {Nazarov}}, \bibinfo {author} {\bibfnamefont {Y.}~\bibnamefont {Nazarov}}, \
  and\ \bibinfo {author} {\bibfnamefont {Y.~M.}\ \bibnamefont {Blanter}},\
  }\href@noop {} {\emph {\bibinfo {title} {Quantum transport: introduction to
  nanoscience}}}\ (\bibinfo  {publisher} {Cambridge university press},\
  \bibinfo {year} {2009})\BibitemShut {NoStop}%
\bibitem [{\citenamefont {Landau}\ \emph {et~al.}(2015)\citenamefont {Landau},
  \citenamefont {Vazirani},\ and\ \citenamefont
  {Vidick}}]{landau2015polynomial}%
  \BibitemOpen
  \bibfield  {author} {\bibinfo {author} {\bibfnamefont {Z.}~\bibnamefont
  {Landau}}, \bibinfo {author} {\bibfnamefont {U.}~\bibnamefont {Vazirani}}, \
  and\ \bibinfo {author} {\bibfnamefont {T.}~\bibnamefont {Vidick}},\ }\href
  {\doibase 10.1038/nphys3345} {\bibfield  {journal} {\bibinfo  {journal}
  {Nature Physics}\ }\textbf {\bibinfo {volume} {11}},\ \bibinfo {pages} {566}
  (\bibinfo {year} {2015})}\BibitemShut {NoStop}%
\end{thebibliography}%

\end{document}